\documentclass[twocolumn,secnumarabic, amssymb, floatfix,
amsmath,tightenlines,showpacs,superscriptaddress,
preprintnumbers,aps,prb]{revtex4-1}

\usepackage{times}
\usepackage{color}
\usepackage{graphicx}
\usepackage{dcolumn}
\usepackage{amssymb}
\usepackage{amsmath}
\usepackage{amsfonts}
\usepackage{docs}%
\usepackage{bm}%
\usepackage[colorlinks=true,linkcolor=blue,pagecolor=blue,
filecolor=blue,menucolor=blue,urlcolor=blue,
citecolor=blue,anchorcolor=blue]{hyperref}%

\newcommand{\etal}{\emph{et al.}}
\newcommand{\sns}{$S$/$N$/$S $ }
\newcommand{\sfs}{$S$/$F$/$S $ }
\newcommand{\sffs}{$S$/$F$/$F$/$S $ }
\newcommand{\sfffs}{$S$/$F$/$F$/$F$/$S $ }
\newcommand{\sffn}{$S$/$F$/$F$/$N $ }

\newcommand{\ff}{$F$/$F $ }
\newcommand{\fs}{$S$/$F $ }
\newcommand{\f}{$F $ }

\begin{document}

\title{
Proximity Induced Vortices
and Long-Range Triplet Supercurrents in
Ferromagnetic Josephson Junctions and Spin Valves}

\author{Mohammad Alidoust }
\email{phymalidoust@gmail.com}
\affiliation{Department of Physics,
University of Basel, Klingelbergstrasse 82, CH-4056 Basel, Switzerland}
\affiliation{Department of Physics,
Faculty of Sciences, University of Isfahan, Hezar Jerib Avenue,
Isfahan 81746-73441, Iran}
\author{Klaus Halterman}
\email{klaus.halterman@navy.mil}
\affiliation{Michelson Lab, Physics
Division, Naval Air Warfare Center, China Lake, California 93555,
USA}

\date{\today}

\begin{abstract}
Using a spin-parameterized quasiclassical Keldysh-Usadel
technique,
we theoretically study
 supercurrent transport in several types of diffusive
 ferromagnetic($F$)/superconducting($S$) configurations
 with differing magnetization textures.
We separate out the even- and odd-frequency components of the
supercurrent within the low proximity limit and
identify the relative contributions from the
singlet and triplet channels.
We first consider inhomogeneous one-dimensional Josephson
structures consisting of a uniform bilayer magnetic \sffs structure
and
a trilayer \sfffs
configuration, in which case the
outer
$F$ layers can
have either a uniform or conical texture relative
to the
central uniform $F$ layer.
Our results demonstrate that for supercurrents
flowing {\it perpendicular} to the \ff interfaces,
incorporating a conical texture yields the most effective way to observe
the signatures of the long-ranged spin-triplet supercurrents.
We also consider three different types of finite-sized
two-dimensional
magnetic structures subjected to an applied magnetic
field normal to the junction plane:
a \sfs junction with uniform magnetization
texture, and
two \sffs configurations with differing \ff bilayer arrangements.
In one case, the \ff
interface is parallel with the \fs junction interfaces while in
the other case, the \ff junction is oriented perpendicular to
the \fs interfaces.
We then discuss the proximity vortices and corresponding
spatial maps of
currents inside the junctions.
For the uniform \sfs junction, we analytically calculate the magnetic field
induced supercurrent and pair potential in both the narrow and wide junction regimes,
thus providing insight into the variations in the Fraunhofer diffraction patterns and
proximity vortices when transitioning from a wide junction to a narrow one.
Our extensive computations demonstrate that the induced long-range spin-triplet
supercurrents can
deeply penetrate uniform \ff bilayers
when
spin-singlet supercurrents flow {\it parallel} to the \ff interfaces.
This is in stark contrast to configurations
where a spin-singlet supercurrent flows
perpendicular to the \ff interfaces. We pinpoint the origin of
the induced triplet and singlet correlations through
spatial profiles of the decomposed
total supercurrents. We find that the penetration of the long-range spin-triplet
supercurrents associated with supercurrents flowing parallel to the \ff interfaces,
are more pronounced when the thickness of the \f strips are unequal.
Lastly, if one of the $S$ terminals is
replaced with a finite-sized
normal metal, we demonstrate that the corresponding
experimentally accessible \sffn spin valve presents an effective platform
in which the predicted long-range effects can be effectively generated
and probed.

\end{abstract}

\pacs{74.50.+r, 74.25.Ha, 74.78.Na, 74.50.+r, 74.45.+c, 74.78.FK,
72.80.Vp, 68.65.Pq, 81.05.ue}

\maketitle

\section{introduction}\label{sec:introduction} 

The interaction between the different order parameters in proximity
coupled nanostructures comprised of ferromagnets ($F$) and
superconductors ($S$) has attracted considerable attention from
numerous scientific disciplines in both the theoretical and
experimental communities.
\cite{zutic,golubov1,buzdin1,bergeret1,efetov1,eschrigh1,al1,al2,al3}
The interplay between ferromagnetism and superconductivity
at low temperatures has constituted a unique arena for researchers
in condensed matter studying superconducting spintronics in the
clean, diffusive and non-equilibrium regimes.
\cite{Bobkov,brataas1,Ioffe,Barash,Cottet,Crouzy,Fominov,
Radovic1,Sellier,Pugach2,Jin,eschrigh3}
Interest in superconducting electronics involving $S/F$ hybrids has
substantially increased during the past decade due to considerable
advances in nanofabrication techniques. This consequently has led to
more possibilities for $S/F$ heterostructures playing a practical
role in nanoscale systems including, quantum computers and
ultra-sensitive detectors.
\cite{al2,alidoust2,alidoust1,Zdravkov,Bakurskiy,
Bakurskiy,makhlin,efetov1,eschrigh1,golubov1,golubov1,bergeret1,
buzdin1,Baibich,Grunberg,Ioffe}
Several interesting and important effects have been found and
studied both theoretically and experimentally, such as $0$-$\pi$
transitions \cite{ryaz0,Bulaevskii,buzdin2,kh_0pi}, and the
existence of triplet correlations
\cite{bergeret1,bergeret2,Asano2,efetov1,Lofwander1,
Lofwander2,Kontos,Sosnin,halterman1,Birge,Keizer}.

When a single quantization axis can be defined throughout the
system, such as in a simple $S/F$ bilayer with a uniformly magnetized
ferromagnet, the Cooper pair wavefunction is composed of singlet and
opposite-spin triplet components.
\cite{bergeret1,bergeret2,Asano2} These components have zero
spin projection along the quantization axis, which is the same
direction as the magnetization.
These two types of superconducting
correlations oscillate and strongly decay inside the $F$ layer over
length scales determined by $\xi_F$. In the diffusive regime studied here,
$\xi_F=\sqrt{D/h}$,
where $D$ and $h$ represent the diffusion constant and exchange
field magnitudes, respectively. In the ballistic regime, $\xi_F=\hbar v_F/2h$,
where $v_F$ is the fermi velocity. Due to these relatively
small length scales,
the zero-spin triplet correlations are often referred to as short-ranged.
\cite{bergeret1,bergeret2,halterman1,efetov1} However, if the
magnetization of the $F$ layer possesses an inhomogeneous pattern,
equal-spin triplet correlations can be generated.
\cite{bergeret1,efetov1,Lofwander1,Lofwander2,Kontos,Sosnin} These types of correlations
have non-zero spin projection along the quantization
axis. The equal-spin correlations penetrate into a uniform diffusive $F$
media over a length scale that is the same as singlets in a normal
metal. \cite{Houzet3,rob2,Asano2,alidoust_missner}
For instance, it has been theoretically shown that in the
diffusive regime, a particular trilayer \sfffs Josephson junctions with
non-collinear magnetizations may host triplet supercurrents that
are manifested in a slowly decaying critical current as a function of junction
thickness \cite{Houzet3}.
It has also been demonstrated that to
reveal the long-ranged nature of proximity triplet supercurrents in
the {\it diffusive} regime,
a simple uniformly magnetized \sffs junction may not possess the
requisite magnetic inhomogeneity, and
consequently a counterpart layered \sfffs
junction is necessary \cite{Houzet1}.
In contrast to the diffusive regime, it has recently been shown
in the {\it ballistic} regime
that it is possible to generate
long-range odd-frequency triplet correlations in \sffs Josephson
junctions containing two uniform $F$ layers with misaligned
magnetization orientations and differing thicknesses ($d_{F1}\neq
d_{F2}$).\cite{Trifunovic2,Trifunovic3,pugach_paral}
The signature of these long-ranged triplet correlations
are theoretically predicted to be revealed in the second
harmonic term of the Josephson current
\cite{Trifunovic2,Trifunovic3,Hikino}.
The  
signatures of the equal-spin triplet correlations have been observed in
experiments as well \cite{rob1,rob2,rob4,Birge,Keizer}.
When the magnet is fully spin polarized, as in half-metallic systems, 
these type of triplet
correlations can be produced when there are  spin-active interfaces present.
\cite{Keizer,Lofwander1,halterman2,eschrigh3}
A
significant thrust of these works is the
formulation of simple and optimal conditions to detect the odd-frequency
pairings in \fs heterostructures. To this end, spin-valve $S/F/F$
heterostructures have recently attracted
interest from both the theoretical and experimental
communities
\cite{Oboznov,Ovsyannikov,Karminskaya,kh_sv,buzdin3,Beasley,
Fominov2,Zdravkov,exper1,exper2,exper3,exper4,exper5,rob4,
exper6,exper7,alidoust_missner}.
The advantages of such spin valves are their
less complicated experimental implementation
and greater control of their magnetization state
compared to
layered magnetic Josephson junctions.

In this paper, we make use of a spin parametrization scheme
for the Green's function, the Usadel equation, and associated
boundary conditions. 
This method provides a suitable
framework for separating the {\it supercurrent}
into spin singlet, opposite-spin triplet, and equal-spin triplet components, using the
spin parametrization technique outlined for a generic three dimensional system.
Our model allows for investigations into a broad range of realistic finite-size 
ferromagnet/superconductor hybrids
with arbitrary magnetization patterns subject to an external magnetic field\cite{alidoust_nfrh1, alidoust_nfrh2,al1}.
The spin decomposed supercurrent  accurately pinpoints
the contribution from different superconducting pairings, their influence upon the
total supercurrent, and their spatial variations
within the magnetic regions.\cite{al2}

We first consider three types of
{\it one-dimensional} Josephson junctions and
study the critical supercurrent spin decomposed components for differing \f layer
thicknesses.
Our results demonstrate that
in the low proximity regime of the diffusive limit,
the most effective
way to observe signatures of the long-ranged spin-triplet supercurrents
(where the supercurrent flows perpendicular to \ff interfaces)
involves the
 use of
inhomogeneous magnetic structures comprised of combinations of rotating exchange
interactions (e.g. conical texture in Holmium [$Ho$]) and uniform ferromagnets.
We find that the supercurrent spin decomposed component
corresponding to the rotating component of magnetization is long-ranged
in such a situation and dominates
the behavior of total supercurrent.
Trilayer \sfffs structures with
uniform ferromagnets\cite{nazarov,Houzet3} are shown to weakly display
long-range spin-triplet signatures in this {\it low proximity} limit.

Next, we consider three different types of finite-sized {\it two-dimensional}
magnetic Josephson junctions subject to an applied magnetic field.\cite{al1}
Our general
analytical and numerical
framework permits
the study of
magnetization textures with
highly intricate patterns.\cite{alidoust_nfrh1, alidoust_nfrh2}
Our methodology also allows for rather general
geometric parameters,
including arbitrary ratios of the side lengths
describing the ferromagnet strips.
We
first consider a \sfs Josephson junction with a uniform magnetization
texture, thus extending the results of a normal \sns Josephson junction.\cite{Cuevas_frh2}
In doing this, we employ simplifying
approximations that permit explicit analytical solutions to the
anomalous Green's function.
This consequently leads to tractable and transparent
analytical
expressions for the spatial dependence to the current density and pair potential.
In particular, we implement the so-called wide and narrow junction
limits, which results in  considerable simplifications to the
Usadel equations.
In the wide-junction limit, the Fraunhofer diffraction pattern appears with
$\Phi_0$ (the magnetic flux quantum) periodicity in the critical supercurrent as a function
of 
external magnetic flux,
whereas a narrow-junction transitions
from an oscillating Fraunhofer pattern to a
monotonically
decaying one, similar to its normal metal counter part\cite{Cuevas_frh2}. 
Associated with these signatures of the supercurrent is
the appearance of arrays of proximity
vortices\cite{Cuevas_frh1,Cuevas_frh2,alidoust_nfrh1,alidoust_nfrh2,Ledermann}
which
provide useful information
regarding  the Fraunhofer response of the
supercurrent to an external magnetic field.
By initially considering uniformly magnetized structures
with a single $F$ layer,
the nature of the proximity vortices and
current flow mappings in
more complicated magnetically inhomogeneous junctions discussed below
are better understood
in addition to the $0$-$\pi$ transition influences on the critical
supercurrent responses\cite{alidoust_nfrh1}.

To explore the possibility of induced long-range triplet effects,
additional magnetic inhomogeneity is introduced by the addition of another
ferromagnet layer, thus
establishing double magnet
\sffs Josephson junctions.
These types of structures
comprise
the main focus of the paper.
Two types of \sffs configurations are considered: In one
case, the \ff interface is parallel to the
interfaces of the $S$ leads, while in the other case, the \ff junction is
oriented perpendicular to them.
In either scenario,
when an external magnetic field is present, it is
applied normal to the junction plane. Our findings demonstrate
that a diffusive \sffs Josephson junction in the {\it low
proximity} limit can generate long-ranged triplet supercurrents depending on
the direction of charge supercurrent with respect to the \ff interface orientation.
In particular, if charge
supercurrent flows parallel with the \ff interface,
spin-triplet components generated in one ferromagnetic
wire deeply penetrate the adjacent ferromagnet
with relative orthogonal magnetizations.
For these types of structures, we find that the long-ranged effect
manifests itself when the thickness of the ferromagnetic strips are
unequal.\cite{al1}
With the goal of demonstrating the generality of the introduced scenario above,
isolating the predicted equal-spin triplet component to the
supercurrents flowing parallel to \ff interfaces, and motivated by recent experiments involving
$S/F/F$ spin-valves
\cite{exper1,exper2,exper3,exper4,exper5,Zdravkov,rob4,exper6,exper7},
we turn our attention to
$S/F/F/N$ spin-valves subject to an external
magnetic field ($N$ denotes a normal metal layer).
Our
results show that indeed for certain geometric and material
parameters, diffusive $S/F/F/N$ spin valves can isolate purely equal-spin
odd-triplet correlations arising from the Meissner response, following
the parallel transport scenario above, even in the \textit{low proximity limit}. The
supercurrent moving parallel to \ff contact, in this case is long-ranged,
extends considerably into the
$N$ layer, and can be experimentally probed through direct local
measurements of the current inside the relatively thick normal layer.
We find that an equal-spin triplet supercurrent appears when the
thickness of the two $F$ layers are unequal: $d_{F1}\gg d_{F2}$, and
vanishes when $d_{F1}= d_{F2}$,
consistent with the behavior of
the ferromagnet Josephson junctions mentioned above. Therefore, our
extensive study demonstrates the generality of our proposed scenario to
effectively generate long-ranged supercurrents independent of
geometry implemented.

The paper is organized as follows: We present a
succinct review of the theoretical framework,
spin-parametrization scheme, and parameters employed in Sec.~\ref{sec:theor_gener}.
In Sec.~\ref{sec:1D}, the
one-dimensional spin-parameterized Green's function
is discussed, and in Sec.~\ref{subsec:1D-spin}
we present
the
approach taken to
evaluate the
corresponding
decomposed supercurrents.
In Sec.~\ref{subsec:1D-results},
we discuss the critical
supercurrent, $0$-$\pi$ transitions and equal-spin triplet components
of the supercurrent
for one-dimensional $S$/$F$/$F$/$S$, $S$/$F$/$F$/$F$/$S$,
and $S$/$Ho$/$F$/$Ho$/$S$ structures.
Next, in Sec.~\ref{sec:2D} we expand our investigations into two-dimensional
hybrid junctions.
In Sec.~\ref{subsec:2D-param},  the
pertinent technical
points and parameters used to study the proposed
heterostructures theoretically are presented, and which are chosen to be
aligned with realistic experimental conditions.
In Sec.~\ref{subsec:2D-frnhfr}, we
consider $i)$ the wide-junction limit, and  $ii)$
the narrow-junction limit of a uniform \sfs junction.
Corresponding
analytical expressions
are given for the pair potential and
supercurrent response
when the system is
subject to an
applied magnetic field.
In part $iii)$ we compliment our
analytical expressions with a full numerical
treatment
that does not resort
to the previous simplifying assumptions.
In Sec.~\ref{subsec:2D-spin},
we present the spin-parameterized Usadel equation, supplementary boundary
conditions, and
separate out the contributions
from the odd and
even frequency components of the net
supercurrent
describing these two-dimensional systems.
In Sec.~\ref{subsec:2D-sffs-pral}, we study
one of the main structures, a magnetic
\sffs junction, where the
double layer \ff  interfaces are aligned with the interfaces of the $S$ terminals.
In Sec.~\ref{subsec:2D-sffs-perp}, the remaining structure is discussed, where
the \ff interfaces are
orthogonal to the interfaces of the $S$ terminals.
For both configurations,
we study the
pair potential, charge supercurrent, and its odd or even frequency
decomposition. We show also how  an external
magnetic flux can induce vortex phenomena and
modifications to the singlet and triplet correlations responsible for supercurrent transport.
We also study the influence of ferromagnetic
strip thicknesses on the long-range
spin-triplet contributions to the charge supercurrent.
Finally, in Sec.~\ref{subsec:2D-snpvlv}, we study the long-range spin-triplet
supercurrents in $S/F/F/N$ valves.
These results are then compared with those
obtained for their \sffs counterparts. The experimental
implications of our findings for these structures are also discussed.
Finally, we summarize our
findings in Sec.~\ref{sec:conclusion} and give concluding remarks.

\section{General approach and formalism }\label{sec:theor_gener}
Here we first outline the
theoretical approach
for
generic
three-dimensional systems.
The corresponding
reduced one-dimensional and two-dimensional
cases are presented 
in the subsequent sections.

\subsection{Theoretical methods}\label{subsec:theor-model}

The coupling between an $s$-wave superconductor and a ferromagnet
leads to proximity-induced triplet correlations in addition to the usual
singlet pairings.\cite{bergeret1,bergeret2} The corresponding
coherent superconducting quasiparticles inside a
diffusive medium can be
described by the Usadel
equations, \cite{Usadel} which are a set of coupled complex three-dimensional partial
differential equations. A  general three-dimensional
quasiclassical model for such diffusive ferromagnet/superconductor
heterostructures subject to an external magnetic field is given by the
following Usadel equation; \cite{morten,Usadel,bergeret1}
\begin{eqnarray}\label{eq:full_Usadel}
&&D[\check{\partial},\check{G}(x,y,z)[\hat{\partial},\check{G}(x,y,z)]]+i[
\varepsilon \hat{\rho}_{3}+\nonumber\\&&
\text{diag}[\vec{h}(x,y,z)\cdot\underline\sigma,(\vec{h}(x,y,z)\cdot\underline\sigma)^{T}],\check{G}(x,y,z)]=0,
\end{eqnarray}
in which $\hat{\rho}_{3}$ and $\underline{\sigma}$ are $4\times 4$
and $2\times 2$ Pauli matrices, respectively, and
we denote the
diffusive constant of the medium by $D$.
Here,
the exchange field of a ferromagnetic region,
$\vec{h}(x,y,z)=\big(h^x(x,y,z),h^y(x,y,z),h^z(x,y,z)\big)$, can take
arbitrary directions in configuration space. We have defined the 4$\times$4
version of partial derivative, $\check{\partial}$, by
$\hat{\partial}\equiv\vec{\nabla} \hat{1}-ie
\vec{A}(x,y,z)\hat{\rho}_{3} $ in which $\vec{A}(x,y,z)$ stands for a
vector potential producing the applied external magnetic field
$\vec{H}(x,y,z)$ and
$[\hat{\partial},\hat{G}(x,y,z)]=\vec{\nabla}\hat{G}(x,y,z)-ie[\vec{A}(x,y,z)\hat{\rho}_{3},\hat{G}(x,y,z)]$.
We have denoted the quasiparticles' energy by $\varepsilon$ which is
measured from the fermi surface $\varepsilon_F$.

In the low proximity limit, the normal and anomalous components of
the
Green's function can be approximately written as,
$\underline{F}^{no}(x,y,z)\simeq \underline{1}$ and $\underline{F}(x,y,z)\ll
\underline{1}$, respectively. In this regime therefore, the advanced
component of the Green's function can be directly expressed as:
\begin{align}
\hat{G}^{A}(x,y,z,\varepsilon)\approx\begin{pmatrix}
-\underline{1} & -\underline{F}(x,y,z,-\varepsilon)\\
\underline{F}^\ast(x,y,z,\varepsilon) & \underline{1}\\
\end{pmatrix},
\end{align}
where the underline notation reflects 2$\times$2 matrices. Thus,
the advanced component, $\hat{G}^{A}(x,y,z)$, of total Green's function
$\check{G}$ can be written as:
\begin{align}\label{Advanced Gree}
\hat{G}^{A}(x,y,z)=\begin{pmatrix}
-1 & 0 & -f_{\uparrow\uparrow}(-\varepsilon) & -f_{-}(-\varepsilon) \\
0 & -1  &-f_{+}(-\varepsilon)  & -f_{\downarrow\downarrow}(-\varepsilon) \\
 f_{\uparrow\uparrow}^{\ast}(\varepsilon)  & f_{-}^{\ast}(\varepsilon)  & 1  &  0  \\
f_{+}^{\ast}(\varepsilon)& f_{\downarrow\downarrow}^{\ast}(\varepsilon)  & 0  &  1  \\
\end{pmatrix}.
\end{align}
In general, when a system is in a nonequilibrium state, the
Usadel equation must be supplemented by the appropriate distribution
functions.\cite{Belzig_solid_state} In this paper,
however, we assume  equilibrium conditions for our systems under
consideration, and hence
the three blocks comprising the total Green's function are
related to each other in the following way:
$\hat{G}^{A}(x,y,z)=-(\tau_3\hat{G}^R(x,y,z)\tau_3)^{\dag}$, and
$\hat{G}^{K}(x,y,z)=\tanh(\beta\varepsilon)(\hat{G}^{R}(x,y,z)-\hat{G}^{A}(x,y,z))$, where
$\tau_3$ is the Pauli matrix, and $\beta\equiv k_BT/2$.

The resulting nonlinear complex partial differential equations should
be supplemented by appropriate boundary conditions to properly
capture the
electronic and transport characteristics of $S$/$F$ hybrid
structures.
We employ the Kupriyanov-Lukichev boundary conditions at
the \fs interfaces\cite{cite:zaitsev} and control the induced
proximity correlations using the parameter $\zeta$ as the barrier
resistance:
\begin{eqnarray}\label{eq:bc}
    &&\zeta(\check{G}(x,y,z)\check{\partial}\check{G}(x,y,z))\cdot\hat{\boldsymbol{n}}=
    [\check{G}_{\text{BCS}}(\theta),\check{G}(x,y,z)]+ \nonumber\\&& i (G_S/G_T)
[\text{diag}(\underline{\tau_3}, \underline{\tau_3}), \check{G}(x,y,z)],
\end{eqnarray}
where $\hat{\boldsymbol{n}}$ is a unit vector denoting the
perpendicular direction to an interface.
The parameters $G_S$ and
$G_T$ introduce spin-activity at the $S$/$F$ interfaces.
\cite{alidoust1}
The solution for a bulk
even-frequency $s$-wave superconductor $\hat{G}_{\text{BCS}}^{R}$
reads,\cite{morten}
\begin{align}
\hat{G}^{R}_{\text{BCS}}(\theta)=\left(
                                  \begin{array}{cc}
                                    \mathbf{1}\cosh\vartheta(\varepsilon) & i\tau_2e^{i\theta}\sinh\vartheta(\varepsilon) \\
                                    i\tau_2e^{-i\theta}\sinh\vartheta(\varepsilon) & -\mathbf{1}\cosh\vartheta(\varepsilon) \\
                                  \end{array}
                                \right),
\end{align}
where,
\begin{equation}
\nonumber
\vartheta(\varepsilon)=\text{arctanh}(\frac{\mid\Delta\mid}{\varepsilon}).
\end{equation}
Here we have represented the macroscopic phase of the bulk
superconductor by $\theta$. To have more compact expressions,
we define the following piecewise functions:
\begin{eqnarray}
&&\nonumber s(\varepsilon)\equiv e^{i\theta}\sinh\vartheta(\varepsilon)=\\&&\nonumber-\Delta\left\{\frac{\text{sgn}(\varepsilon)}{\sqrt{\varepsilon^2-\Delta^2}}\Theta(\varepsilon^2-\Delta^2)-\frac{i}{\sqrt{\Delta^2-\varepsilon^2}}\Theta(\Delta^2-\varepsilon^2)\right\},\\
&&\nonumber
c(\varepsilon)\equiv\cosh\vartheta(\varepsilon)=\\&&\nonumber\frac{\mid\varepsilon\mid}{\sqrt{\varepsilon^2-\Delta^2}}\Theta(\varepsilon^2-\Delta^2)-\frac{i\varepsilon}{\sqrt{\Delta^2-\varepsilon^2}}\Theta(\Delta^2-\varepsilon^2),
\end{eqnarray}
where
$\Theta(x)$ denotes the usual step function.

In the situations where an external magnetic field is applied,
it is directed
along the $z$-axis.
We also
use the Coulomb gauge $\vec{\nabla}\cdot\vec{A}(x,y,z)=0$ throughout our
calculations for the vector potential. In a magnetic junction the
vector potential is composed of two parts: $a$) a
part due to the magnetic field
associated with the exchange interaction in the ferromagnetic layer
$\vec{h}(x,y,z)$, and $b)$ a part due to the external magnetic field $\vec{H}(x,y,z)$.
The contribution due to the exchange interaction
measured in experiments reveals itself
as a shift in the observed magnetic interference patterns \cite{rob1,Keizer}.
However, as  this has found
good agreement with experiments,
one can safely neglect the part of $\vec{A}(x,y,z)$
arising from the exchange interaction.\cite{rob1,alidoust_nfrh1}
Thus, we assume that
the external magnetic flux contribution dominates, and the vector
potential can be determined entirely by the external magnetic field
$\vec{H}(x,y,z)$. In this paper, we consider the regime where the junction
width $W_{F}$ is smaller than the Josephson penetration length
$\lambda_{J}$ \cite{Cuevas_frh1,Cuevas_frh2}. Therefore,
screening of the magnetic field by
Josephson currents can be safely ignored.
\cite{Angers}
In general, the external magnetic field strongly influences
the macroscopic phases of the superconducting leads. To avoid
such effects
in our considered systems, we assume that the external magnetic
field passes only across the nonsuperconducting sandwiched strips.
These assumptions lead to results that are
in very good agreement
with those found in experiments.
\cite{Cuevas_frh1,Cuevas_frh2,alidoust_nfrh1,Angers,Chiodi,Clem1}
This will be discussed in more detail in Sec.~\ref{sec:2D}.

One of the most important quantities in the context of quantum
transport through
Josephson junction systems is the charge supercurrent,
which provides valuable
information about the superconducting properties of the system
and the associated favorable
experimental conditions under which to detect them.
Under equilibrium
conditions, the vector current density can be expressed by the
Keldysh block as follows:
\begin{equation}
\label{eq:currentdensity}
\vec{J}(x,y,z) = J_{0} \int_{-\infty}^{+\infty}\hspace{-.2cm}
d\varepsilon\text{Tr}\Big\{\rho_{3}
\big(\check{G}(x,y,z)[\check{\partial},\check{G}(x,y,z)]\big)^{K}\Big\},
\end{equation}
where $J_{0}  =  N_{0} e D/4$,
$N_{0}$ is the number of states at
the Fermi surface,
and $e$ is the electron charge.
The vector current
density,
$\vec{J}(x,y,z)$,
provides a local spatial
map and measure of the
charge supercurrent
flow through the system.
To obtain the total
Josephson charge current flowing along a
particular direction inside the
junction,
it is necessary  to perform an additional integration of Eq.~(\ref{eq:currentdensity})
over the direction perpendicular
to the transport direction.
For instance, the total charge current
flowing along the $x$ direction can be obtained from,
\begin{equation}
I_x=I_{0}\int_{W_F} dy \vec{J}(x,y,z)\cdot \hat{x},
\end{equation}
in which $W_F$ is the junction width (see e.g., Fig.~\ref{fig:model1}).
Likewise, the  charge supercurrent flow in the $y$
direction can be obtained via integration of $J_y(x,y,z)$ over
the $x$ coordinate. Another physically relevant quantity which gives
additional insight into the local behavior of the
singlet correlations throughout the Josephson structure is the spatial maps of pair
potential, ${U}_{\text{pair}}(x,y,z)$. This pair correlation function is
defined using the Keldysh block of the total Green's function,
$\check{G}(x,y,z)$\cite{morten};
\begin{align}\label{eq:full_pair}
U_{\text{pair}}(x,y,z)=U_{0}\text{Tr}\Big\{\frac{\hat{\rho}_{1}-i\hat{\rho}_{2}}{2}\hat{\tau}_{3}\int_{-\infty}^{+\infty}
d\varepsilon \check{\text{G}}^{K}(x,y,z)\Big\},
\end{align}
where we normalize the pair
potential as, $\tilde{U}_{\text{pair}}(x,y,z)=U_{\text{pair}}(x,y,z)/U_{0}$, where
$U_{0}=-N_{0}\lambda/8$, and $\lambda$ is a constant inside the
superconducting regions.
We note that
although the pair potential must vanish
outside of the intrinsically superconducting regions,
the pair amplitude, $\tilde{U}_{\text{pair}}(x,y,z)$, is generally nonzero
in the ferromagnetic regions due to
the proximity effect.

\subsection{Spin-parametrization and parameters}\label{subsec:theor-spin}

Due to the possible appearance of triplet pairings in such
hybrid structures \cite{bergeret1}, we assume a fixed quantization
axis and employ a spin-parametrization scheme. In this
scheme, the Green's function is decomposed
into the even and odd frequency components, taking
the spin-quantization axis to be oriented along $z$ direction.
Since we consider the low proximity limit, the anomalous
component of the  Green's function takes the following form in terms of
even- ($\mathbb{S}$) and odd- ($\mathbb{T}$) frequency parts:
\begin{eqnarray} \label{eq:decomp}
\underline{F}(x,y,z,\varepsilon)= i\Big
[\mathbb{S}(x,y,z,\varepsilon)+\vec{\tau}\cdot\vec{\mathbb{T}}(x,y,z,\varepsilon)\Big
]\tau_y,
\end{eqnarray}
where $\vec{\tau}=(\tau_x,\tau_y,\tau_z)$ is a vector comprised of
Pauli matrices and $\vec{\mathbb{T}}(x,y,z,\varepsilon)=
(\mathbb{T}_x,\mathbb{T}_y,\mathbb{T}_z )$.
Thus, the probability
of finding odd-frequency triplet superconducting correlations with
zero spin projection along the $z$-axis is $|\mathbb{T}_z|^2$.\cite{Champel1,Eschrig2}
Likewise, if $\mathbb{T}_x$ or
$\mathbb{T}_y$ are finite, there exists triplet correlations with
$\pm 1$ spin projections along the spin quantization axis.
\cite{Lofwander1,Lofwander2,efetov2,Champel1,Eschrig2}

We have normalized all lengths by the superconducting coherence length $\xi_S$.
The
quasiparticles' energy $\varepsilon$
and the exchange energy intensity
are normalized by the zero temperature
superconducting order parameter, $\Delta_0$.
We also
assume a low temperature of
$T=0.05T_c$,  where $T_c$
is the critical temperature
of
the bulk superconducting banks.
We use natural units, with
 $\hbar=k_B=1$, where $k_B$ is the Boltzmann constant and define $\beta=k_BT/2$.
 \begin{figure}[t]
\includegraphics[width=8.50cm,height=6.80cm]{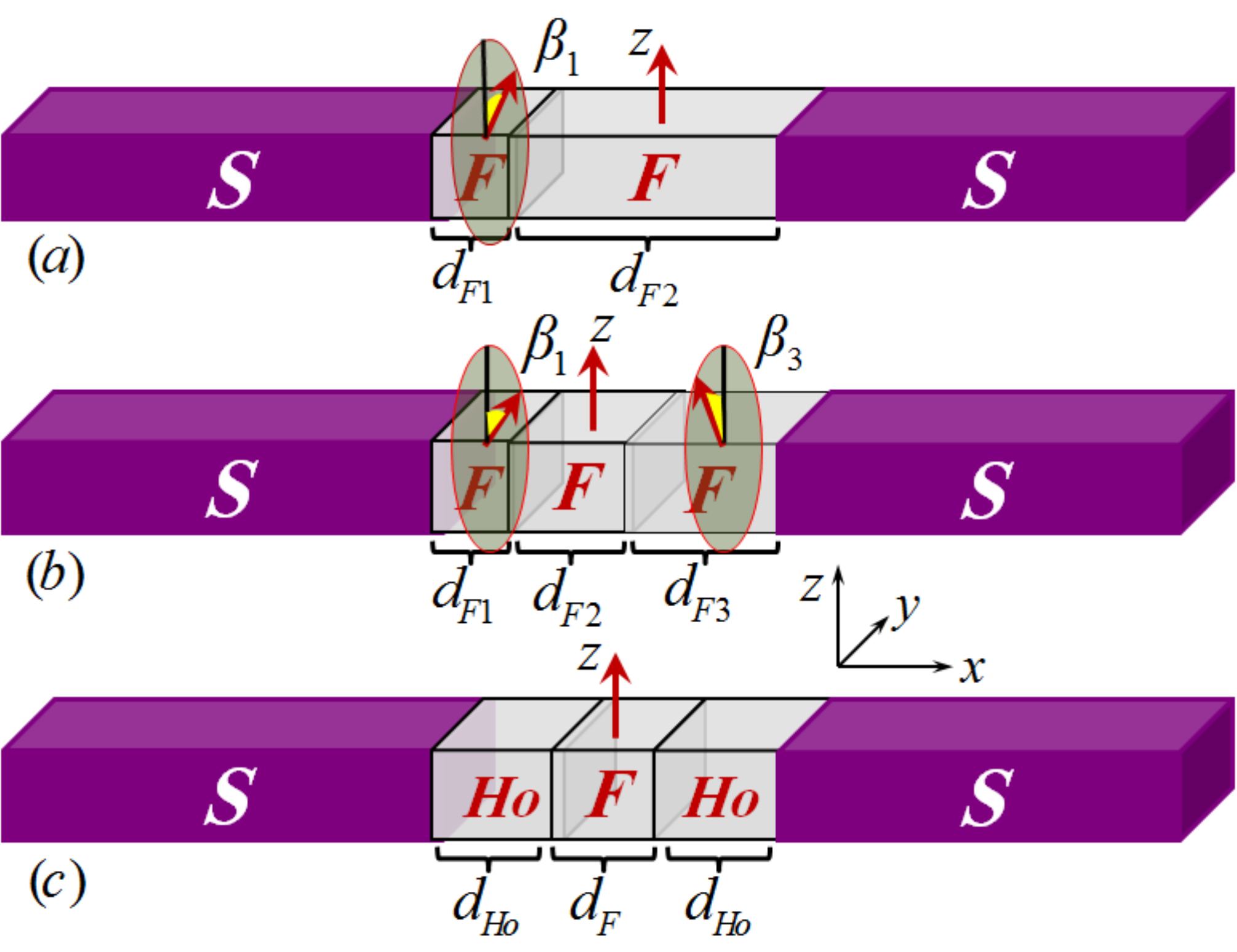}
\caption{\label{fig:model1_1}(Color online) Schematic of three
experimental sets up. (a) simple \sffs Josephson junction with
ferromagnetic layers' widths $d_{F1}$, $d_{F2}$ and magnetization
orientations $\vec{h}_1=h_0(0,\sin\beta_1,\cos\beta_1)$,
$\vec{h}_2=h_0(0,0,1)$, respectively. (b) trilayer \sfffs junction
where the magnetization of middle $F$ layer, with width $d_{F2}$, is
fixed in the $z$ direction. The other layers have $d_{F1}$, $d_{F3}$
widths with magnetization orientations
$\vec{h}_{1,3}=h_0(0,\sin\beta_{1,3},\cos\beta_{1,3})$,
respectively. (c) trilayer of ferromagnets where the outer $F$
layers have Holmium-like magnetization patterns namely,
$\vec{h}_{Ho}=h_0(\cos\alpha,\sin\alpha\sin\gamma
x/a,\sin\alpha\cos\gamma x/a)$. The length of middle $F$ layer is
labeled by $d_{F}$ while the two outer Holmium-like $F$ layers are
assumed to have equal lengths $d_{Ho}$. The $x$ axis is normal to
the junction interfaces and the systems have infinite sizes in the $y$ and $z$ directions.}
\end{figure}
We
consider weak exchange field strengths of $|\vec{h}|=5 \Delta_0$
corresponding to that found in ferromagnet alloys\cite{Zdravkov}
such as, e.g., ${\rm Cu}_{x}{\rm Ni}_y$. A barrier resistance of
$\zeta=4.0$ ensures sufficiently opaque \fs
interfaces leading to
appropriate solutions to the Usadel equations within
the low proximity limit,
$\underline{F}(x,y,z,\varepsilon)\ll \underline{1}$.
Having now outlined the
Keldysh-Usadel quasiclassical formalism and spin-parametrization framework employed in this work, we
now proceed to present our analytical and numerical findings in the
next sections.

\section{One-dimensional hybrid structures}\label{sec:1D}
In this section, we consider the effectively one-dimensional hybrid structures
sketched in Fig.~\ref{fig:model1_1}. We first derive
the Usadel equations and boundary conditions. We then decompose
the supercurrent
into its singlet and triplet components using the three-dimensional
spin-parametrization scheme discussed in Sec.~\ref{subsec:theor-spin}. Utilizing this singlet-triplet
decomposition, we analyze and characterize the supercurrent behavior based on
the individual components involved.
\subsection{Spin-parameterized supercurrent}\label{subsec:1D-spin}

Upon decomposing the Green's function via Eq.~(\ref{eq:decomp}),
the Usadel equation, Eq.~(\ref{eq:full_Usadel}), transforms into the
following eight coupled complex partial differential equations for one-dimensional systems,
\begin{widetext}
\begin{subequations}\label{Linearized Usadel Eq.}
\begin{eqnarray}
&&D\left\{\mp\partial_x^{2} \mathbb{T}_{x}(-\varepsilon)+i\partial_x^{2} \mathbb{T}_{y}(-\varepsilon)\right\}+i\left\{-2 \varepsilon (\mp \mathbb{T}_{x}(-\varepsilon)+i \mathbb{T}_{y}(-\varepsilon)) \mp 2\mathbb{S}(-\varepsilon) (h_{x} \mp i h_{y}) \right\}=0\label{a}\\
&&D\left\{\mp\partial_x^2\mathbb{S}(-\varepsilon)+\partial_x^2\mathbb{T}_z(-\varepsilon)\right\}+i\left\{\mp2\mathbb{T}_x(-\varepsilon) h_{x} \mp 2 \mathbb{T}_{y}(-\varepsilon) h_{y} - 2 (\mp\mathbb{S}(-\varepsilon)+\mathbb{T}_z(-\varepsilon)) (\varepsilon \pm h_{z}) \right\}=0\label{b}\\
&&D\left\{\mp\partial_x^{2}\mathbb{T}_{x}^{\ast}(\varepsilon)-i\partial_x^{2}\mathbb{T}_{y}^{\ast}(\varepsilon)\right\}+i\left\{\pm 2 (h_{x} \pm i h_{y}) \mathbb{S}^{\ast}(\varepsilon) - 2 \varepsilon (\mp\mathbb{T}_{x}^{\ast}(\varepsilon)-i\mathbb{T}_{y}^{\ast}(\varepsilon)) \right\}=0\label{c}\\
&&D\left\{\mp\partial_x^{2}\mathbb{S}^{\ast}(\varepsilon)+\partial_x^2\mathbb{T}_{z}^{\ast}(\varepsilon)\right\}+i\left\{
2 (-\varepsilon \pm
h_{z})(\mp\mathbb{S}^{\ast}(\varepsilon)+\mathbb{T}_{z}^{\ast}(\varepsilon))
\pm 2 h_{x} \mathbb{T}_{x}^{\ast}(\varepsilon) \pm 2 h_{y}
\mathbb{T}_{y}^{\ast}(\varepsilon)\right\}=0\label{d}
\end{eqnarray}
\end{subequations}
\end{widetext}
The Kupriyanov-Lukichev boundary conditions at the left $S$/$F$
interface, Eq. (\ref{eq:bc}), are transformed in the same way,
leading to the following differential equations:
\begin{subequations} \label{bc_F1}
\begin{eqnarray}
&&(\zeta\partial_x  - c^{\ast}(\varepsilon))(\mp \mathbb{T}_{x}(-\varepsilon)+i \mathbb{T}_{y}(-\varepsilon))=0,\\
&&(\zeta\partial_x  - c^{\ast}(\varepsilon))(\mp\mathbb{S}(-\varepsilon)+\mathbb{T}_z(-\varepsilon)) =\mp s^{\ast}(\varepsilon), \\
&&(\zeta\partial_x  - c^{\ast}(\varepsilon))(\mp\mathbb{T}_{x}^{\ast}(\varepsilon)-i\mathbb{T}_{y}^{\ast}(\varepsilon))=0,\\
&&(\zeta\partial_x  -
c^{\ast}(\varepsilon))(\mp\mathbb{S}^{\ast}(\varepsilon)+\mathbb{T}_{z}^{\ast}(\varepsilon))
=\pm s^{\ast}(\varepsilon).
\end{eqnarray}
\end{subequations}
Similarly, the
Kupriyanov-Lukichev boundary conditions at the right $F$/$S$
interface are also transformed as,
\begin{subequations}\label{bc_F2}
\begin{eqnarray}
&&(\zeta\partial_x  + c^{\ast}(\varepsilon))(\mp \mathbb{T}_{x}(-\varepsilon)+i \mathbb{T}_{y}(-\varepsilon))=0,\\
&&(\zeta\partial_x  + c^{\ast}(\varepsilon))(\mp\mathbb{S}(-\varepsilon)+\mathbb{T}_z(-\varepsilon)) =\pm s^{\ast}(\varepsilon), \\
&&(\zeta\partial_x  + c^{\ast}(\varepsilon))(\mp\mathbb{T}_{x}^{\ast}(\varepsilon)-i\mathbb{T}_{y}^{\ast}(\varepsilon))=0,\\
&&(\zeta\partial_x  +
c^{\ast}(\varepsilon))(\mp\mathbb{S}^{\ast}(\varepsilon)+\mathbb{T}_{z}^{\ast}(\varepsilon))
=\mp s^{\ast}(\varepsilon) .
\end{eqnarray}
\end{subequations}
By solving this coupled set of complex differential equations with
the boundary conditions [Eqs. (\ref{bc_F1}) and (\ref{bc_F2})], the
relevant physical quantities can be obtained.
We consider the $x$-axis be normal to the interfaces, as shown in
Fig. \ref{fig:model1}. The decomposition introduced above leads to
the following expression for the supercurrent density within the
junction [Eq. (\ref{eq:currentdensity})]:
\begin{eqnarray}\label{eq:current}
&&\nonumber J(x) =J_0 \int_{-\infty}^{\infty} d\varepsilon
\left\{\right.\\\nonumber && \left.
\mathbb{S}(\varepsilon)\partial_x\mathbb{S}^{\ast}(-\varepsilon) -
\mathbb{S}(-\varepsilon)\partial_x\mathbb{S}^{\ast}(\varepsilon)+
\mathbb{S}(\varepsilon)^{\ast}\partial_x\mathbb{S}(-\varepsilon)
-\right.\\\nonumber && \left.
\mathbb{S}(-\varepsilon)^{\ast}\partial_x\mathbb{S}(\varepsilon)
-\partial_x\mathbb{T}_x(-\varepsilon)\mathbb{T}_x^{\ast}(\varepsilon)
+\partial_x\mathbb{T}_x(\varepsilon)\mathbb{T}_x^{\ast}(-\varepsilon)
-\right.\\\nonumber && \left.
\partial_x\mathbb{T}_x^{\ast}(-\varepsilon)\mathbb{T}_x(\varepsilon)
+\partial_x\mathbb{T}_x^{\ast}(\varepsilon)\mathbb{T}_x(-\varepsilon)
-\partial_x\mathbb{T}_y(-\varepsilon)\mathbb{T}_y^{\ast}(\varepsilon)
+\right.\\\nonumber && \left.
\partial_x\mathbb{T}_y(\varepsilon)\mathbb{T}_y^{\ast}(-\varepsilon)
-\partial_x\mathbb{T}_y^{\ast}(-\varepsilon)\mathbb{T}_y(\varepsilon)
+\partial_x\mathbb{T}_y^{\ast}(\varepsilon)\mathbb{T}_y(-\varepsilon)-\right.\\\nonumber
&& \left.
\partial_x\mathbb{T}_z(-\varepsilon)\mathbb{T}_z^{\ast}(\varepsilon)
+\partial_x\mathbb{T}_z(\varepsilon)\mathbb{T}_z^{\ast}(-\varepsilon)
-
\partial_x\mathbb{T}_z^{\ast}(-\varepsilon)\mathbb{T}_z(\varepsilon)+
\right.\\ &&
\left.\partial_x\mathbb{T}_z^{\ast}(\varepsilon)\mathbb{T}_z(-\varepsilon)
\right\}\tanh(\varepsilon\beta).
\end{eqnarray}

The current through the junction can be easily obtained by
integration of the current density along the $y$ and $z$ directions over the
$F$ junction widths $W$ and $L$, respectively
(corresponding to the cross section of the wire).
We assume that our system is very wide in
the $y$ direction, so that  the one-dimensional approximation is
valid, and therefore the current density remains constant in the $y$ and
$z$ directions. It is convenient to define the normalization constant, $I_0\equiv
LWJ_0$, for the supercurrent, $I(x)$. To extract the
contributions to the total supercurrent from the even frequency
singlet, and odd frequency triplet correlations, we have also
decomposed the supercurrent accordingly into four components:
\begin{subequations}\label{eq:components}
\begin{eqnarray}
&&\nonumber I_{S0}(x) =I_0 \int_{-\infty}^{\infty} d\varepsilon
\left\{
\mathbb{S}(\varepsilon)\partial_x\mathbb{S}^{\ast}(-\varepsilon) -
\mathbb{S}(-\varepsilon)\partial_x\mathbb{S}^{\ast}(\varepsilon)+\right.\\
&& \left.
\mathbb{S}(\varepsilon)^{\ast}\partial_x\mathbb{S}(-\varepsilon)
-\mathbb{S}(-\varepsilon)^{\ast}\partial_x\mathbb{S}(\varepsilon)\right\}\tanh(\varepsilon\beta),\\\nonumber
&&\nonumber I_{Sx}(x) =I_0 \int_{-\infty}^{\infty} d\varepsilon
\left\{
-\partial_x\mathbb{T}_x(-\varepsilon)\mathbb{T}_x^{\ast}(\varepsilon)
+\partial_x\mathbb{T}_x(\varepsilon)\mathbb{T}_x^{\ast}(-\varepsilon)
\right.\\ && \left.
-\partial_x\mathbb{T}_x^{\ast}(-\varepsilon)\mathbb{T}_x(\varepsilon)
+\partial_x\mathbb{T}_x^{\ast}(\varepsilon)\mathbb{T}_x(-\varepsilon)\right\}\tanh(\varepsilon\beta),\\
&&\nonumber I_{Sy}(x) =I_0 \int_{-\infty}^{\infty} d\varepsilon
\left\{-\partial_x\mathbb{T}_y(-\varepsilon)\mathbb{T}_y^{\ast}(\varepsilon)
+
\partial_x\mathbb{T}_y(\varepsilon)\mathbb{T}_y^{\ast}(-\varepsilon)\right.\\ && \left.
-\partial_x\mathbb{T}_y^{\ast}(-\varepsilon)\mathbb{T}_y(\varepsilon)
+\partial_x\mathbb{T}_y^{\ast}(\varepsilon)\mathbb{T}_y(-\varepsilon)\right\}\tanh(\varepsilon\beta),\\
&&\nonumber I_{Sz}(x) =I_0 \int_{-\infty}^{\infty} d\varepsilon
\left\{\partial_x\mathbb{T}_z(-\varepsilon)\mathbb{T}_z^{\ast}(\varepsilon)
+\partial_x\mathbb{T}_z(\varepsilon)\mathbb{T}_z^{\ast}(-\varepsilon)
\right.\\ &&
\left.-\partial_x\mathbb{T}_z^{\ast}(-\varepsilon)\mathbb{T}_z(\varepsilon)+
\partial_x\mathbb{T}_z^{\ast}(\varepsilon)\mathbb{T}_z(-\varepsilon)
\right\}\tanh(\varepsilon\beta),
\end{eqnarray}
\end{subequations}
where the total supercurrent is thus the sum of decomposed terms,
namely,
\begin{equation}
I_{tot}(x)=I_{S0}(x)+I_{Sx}(x)+I_{Sy}(x)+I_{Sz}(x).
\end{equation}
This decomposition allows for pinpointing the exact behavior of the
even- and odd-frequency supercurrent components.

\subsubsection{Results and discussions} \label{subsec:1D-results}

Various
analytical or numerical
schemes with varying approximations have been employed to investigate
the
structures with magnetization patterns shown in
Fig.~\ref{fig:model1_1}
\cite{alidoust1,alidoust_missner,alidoust2,Houzet3,Houzet1,rob2,Crouzy}. In the analytical
treatments \cite{Houzet3,robinson3}, limiting approximations were
employed.
For example, to study the noncollinear \sfffs structures
\cite{Houzet1,Houzet3}, transparent boundaries are employed at the
\fs interfaces together with the assumption that the anomalous Green's
function varies enough slowly through the magnetic trilayer to warrant
its Taylor expansion.
Our
full numerical
results involve no such approximations, and hence
reveals cases where the inclusion of such effects may be important.
One of the main aspects that our numerical approach reveals is the
crucial role that each of the different types of superconducting
correlations play in the total supercurrent, given by
Eq.~(\ref{eq:currentdensity}). The supercurrent is composed of
different components of even-frequency singlet and odd-frequency
triplet correlations $\mathbb{S}$, $\vec{\mathbb{T}}$, respectively. Such
a decomposition is often neglected in Josephson structures that
involve intricate magnetic textures.

For comparison purposes, we first consider the simpler $S/F/F/S$
heterostructure\cite{Crouzy}
shown in Fig.~\ref{fig:model1_1}(a). Most of the results have
$d_{F1}\neq d_{F2}$, and in all cases shown, the magnetization of
the right $F$ layer is fixed along the $z$-axis while the left $F$
layer magnetization rotates in the $yz$ plane, that is,
$\vec{h}_1=h_0(0,\sin\beta_1,\cos\beta_1)$. The magnetization
orientation of the left $F$ is thus characterized by the angle
$\beta_1$, since the magnetization is entirely in-plane.
\begin{figure*}[]
\includegraphics[width=15cm,height=6cm]{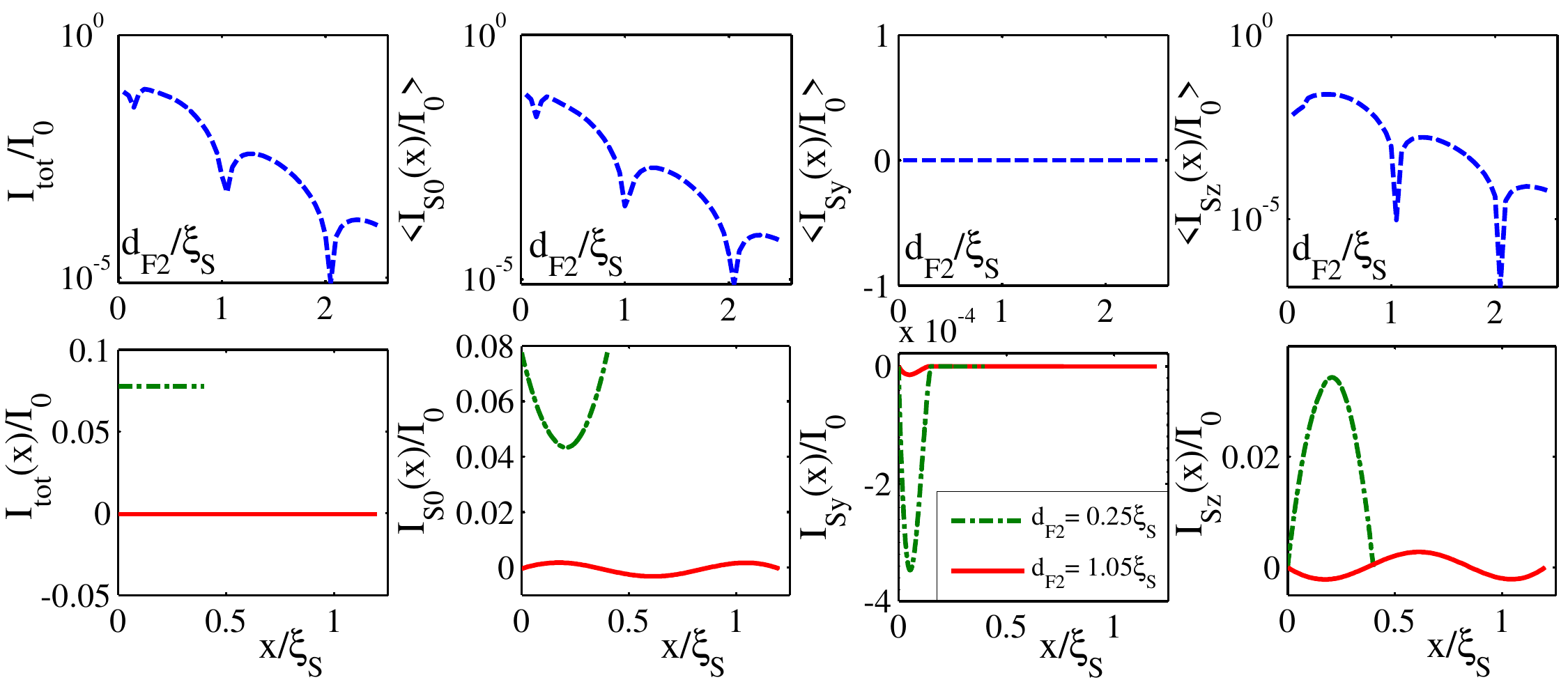}
\caption{\label{fig:s_f_f_s}(Color online) \textit{Top row}:
Critical supercurrent and its components as a function of $d_{F2}$,
the width of the right magnetic layer in a $S/F/F/S$ junction (see
Fig.~\ref{fig:model1_1}(a)). The width of the first magnetic layer is
fixed at $d_{F1}=0.15\xi_S$. The current components are calculated
by performing a spatial average  over the right $F$
layer ($\langle I_{S0}(x)/I_0\rangle, \langle I_{Sy}(x)/I_0\rangle,
\langle I_{Sz}(x)/I_0\rangle$). \textit{Bottom row}: Spatial
behavior of the critical Josephson current and its components as a
function of position $x$  inside the junction for two typical values
of $d_{F2}=0.25\xi_S$, and $d=1.05\xi_S$. The magnetization orientation in
the left $F$ layer is fixed at a representative value of
$\beta_1=0.2\pi$, while the magnetization in the right $F$ layer
always points along $z$, corresponding to $\beta_2=0$.}
\end{figure*}
The top row
of Fig.~\ref{fig:s_f_f_s} illustrates the total critical
supercurrent $I_{tot}$ and its decomposed components ($I_{S0}$,
$I_{Sx}$, $I_{Sy}$, $I_{Sz}$) versus the thickness of right $F$
layer, $d_{F2}$.
The current components generally vary with position $x$, so in order
to display an overall view of their behavior as a function of
magnetization orientation, we spatially average [denoted by
$\langle...\rangle$] each component  over $d_{F2}$.
The thickness of the left $F$ layer is set typically at
$d_{F1}=0.15\xi_S$, and its magnetization has two components $h_y$,
and $h_z$, using a representative angle of $\beta_{1}=0.2\pi$ (these
values are chosen to in part support our comparison purposes in
Subsec. \ref{subsec:2D-snpvlv}). To show the fine features of the
$0$-$\pi$ transition profiles, we have used a logarithmic scale for
the magnitude of the critical supercurrent and its decomposed
components. The
critical current (far left panel) undergoes multiple $0$-$\pi$
transitions when varying $d_{F2}$. The decomposed current components
$\langle I_{S0}\rangle$ and $\langle I_{Sz}\rangle$
also show the same behavior as seen in the remaining panels.
Next, in the bottom row
of Fig.~\ref{fig:s_f_f_s}, we
plot the maximum
current and its components as a function of position for
two representative values of the right $F$ layer's
thickness ($d_{F2}=0.25\xi_S$, and $d_{F2}=1.05\xi_S$). The current components, in contrast to
the total current, often vary inside the magnetic layers: $I_{S0}$,
and $I_{Sz}$ are shown to propagate within the two $F$ layers. The
spin-1 triplet component, $I_{Sy}$, is however localized within the
left $F$ layer where $\beta_2=0.2\pi$ and thus $h_y\neq 0$.
We have investigated a wide range of parameter sets, involving
$\beta_1$, $d_{F1}$, and the superconducting phase differences.
The
$I_{Sy}$ component, does not propagate into the
right $F$ region where the exchange field
is directed along $z$,
demonstrating consistency with previous
studies\cite{Houzet1}.
As discussed in the introduction, recent theoretical works showed
that signatures of the triplet supercurrent may be detected by the
appearance of a second harmonic in the supercurrent in ballistic
$S/F/F/S$ Josephson junction, provided that $d_{F1} \gg d_{F2}$.
The higher harmonics were shown to decay exponentially
(faster than the first harmonic) when varying the system parameters such 
as the thickness of the magnetic layers,  and exchange field intensities, 
in the full proximity limit of the diffusive regime.\cite{buzdin1}
Therefore, in the low proximity limit we consider in our manuscript,
the higher harmonics are absent.\cite{buzdin1} 
It has been suggested that a trilayer\cite{Houzet1,Houzet3} of
uniform magnetic materials with noncollinear magnetizations can
reveal the signatures of long-ranged spin-triplet correlations where the two outer layers
produce nonzero spin projections which can be detected in the
middle $F$ layer
with orthogonal magnetization.
To elucidate the source of the
long-range triplet behavior
in these
types of
trilayer configurations\cite{Houzet3},
we investigate next
the details of the individual
components comprising  the total supercurrent.

We therefore
consider
a \sfffs trilayer structure, as  depicted in
Fig.~\ref{fig:model1_1}(b).
The magnetization of the central $F$ layer
is pinned in the $z$ orientation, coinciding with the
spin-quantization axis.
The relative in-plane magnetization
directions in the surrounding left and right $F$ layers are
described simply by the angles $\beta_1$, and $\beta_2$, respectively. We
denote the thicknesses of the left, middle, and right $F$ layers by
$d_{F1}$, $d_{F2}$, and $d_{F3}$, respectively. In the top set of
\begin{figure*}[]
\includegraphics[width=15cm,height=7.0cm]{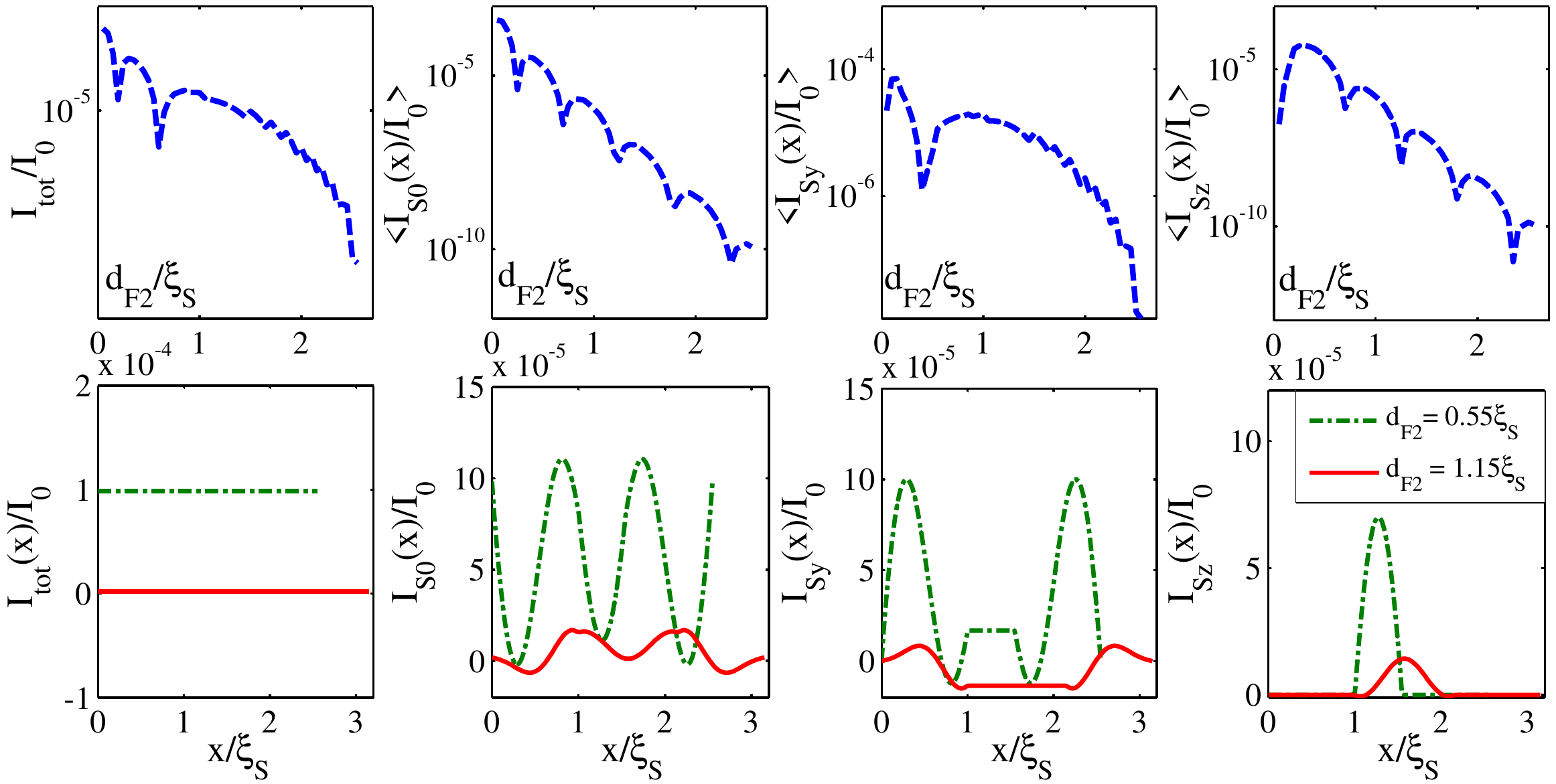}
\caption{\label{fig:s_f_f_f_s}(Color online) \textit{Top row}:
Critical supercurrent and its components against the thickness of
the middle $F$ layer $d_{F2}$ in the $S/F/F/F/S$ structure shown in Fig.~\ref{fig:model1_1}(b).
The thickness of the outer $F$ layers, $d_{F1}$ and
$d_{F3}$, are equal and fixed at $\xi_S$.
The components are
spatially averaged over the middle $F$ layer thickness ($\langle
I_{S0}(x)/I_0\rangle, \langle I_{Sy}(x)/I_0\rangle, \langle
I_{Sz}(x)/I_0\rangle$).
\textit{Bottom row}: Critical current as a
function of position inside the junction $x$ for two different
thicknesses of the middle $F$ layer, $d_{F2}=0.55\xi_S$, and $d_{F2}=1.15\xi_S$. The
magnetization of the
middle $F$ layer is fixed along the $z$ direction,
$\beta_2=0$, while the magnetization is  oriented towards the
$y$ direction in the outer $F$ layers ($\beta_1=\beta_3=\pi/2$).}
\end{figure*}
panels in Fig.~\ref{fig:s_f_f_f_s}, the spatially averaged
supercurrent and its singlet and triplet constituents are shown
as a function of $d_{F2}$.
We here show the results for  $d_{F1}=d_{F3}=\xi_S$, which
is
representative of the numerous equal-width cases
investigated numerically.
The
bottom set of panels illustrate the spatial behavior of the
supercurrent and its components. To isolate the spin-1 triplet
contribution, $I_{Sy}$, to the supercurrent in the middle $F$ layer,
we set $\beta_1=\beta_3=\pi/2$. In this case, the magnetization of
the two outer $F$ layers are strictly along $y$ and orthogonal to
the exchange field direction of middle $F$. Such a magnetization
configuration
has been suggested as
optimal
for detecting the signatures of the
spin triplet supercurrents \cite{Houzet3}.
As seen in the figure, the critical supercurrent versus the middle
$F$ layer thickness shows multiple $0$-$\pi$ transitions,
corresponding to points where the current nearly vanishes and then
eventually changes sign. The averaged components of the total supercurrent, $I_{S0}$ and
$I_{Sz}$ demonstrate short-range signatures as exhibited 
by the multiple cusps compared to the equal-spin triplet
component $I_{Sy}$. The total critical current behavior is dominated
by the triplet term, $I_{Sy}$, which undergoes fewer sign changes
than the other components, and consequently fewer $0$-$\pi$
transitions, when changing $d_{F2}$.
Thus for the regime considered here, the supercurrent does not
exhibit a very slow monotonic decay as a function of the central magnetic
junction thickness, as reflected in the absence of long-ranged
behavior in $I_{Sy}$ vs $d_{F2}$.

To further explore the behavior of the current throughout the junction, we
next examine (bottom row, Fig.~\ref{fig:s_f_f_f_s}) the spatial
dependence to the total supercurrent and its singlet and triplet
components for representative values of unequal middle $F$ layer
thicknesses, $d_{F2}=0.55\xi_S$, and $d_{F2}=1.15\xi_S$.
We immediately observe from the left panel that as expected, the
maximum total supercurrent is a constant in all parts of the
junction, reflecting  conservation of current there. The triplet
component with zero-spin projection, $I_{Sz}$, is localized in the
middle $F$ layer where the magnetization is directed along $z$. In
contrast, the singlet component $I_{S0}$ oscillates throughout the
junction while the triplet component, $I_{Sy}$,  propagates
\textit{without} decay in the middle ferromagnet, which has its
magnetization direction orthogonal to the spin orientation of
$I_{Sy}$.
Thus, we observe that
$I_{Sy}$ is long-ranged in the middle $F$ layer. Although
$I_{Sy}$ is spatially constant throughout the middle $F$ layer, it
changes sign, depending on $d_{F2}$. This characteristic is  seen in the top row of
Fig.~\ref{fig:s_f_f_f_s}.
We have investigated
with our full numerical
method
several different
geometrical parameter sets,
including, e.g., much smaller $d_{F1}$, and $d_{F3}$,
as well as other
$\beta_1$, and $\beta_3$. We
typically found that
 the results presented in the top row of
Fig.~\ref{fig:s_f_f_f_s} are quite
representative of the
singlet and triplet supercurrent behavior
for the
low proximity regime.

In a recent experiment\cite{rob1} involving
$Ho/F/Ho$
trilayers with inhomogeneous magnetization patterns,
a long-ranged Josephson supercurrent through
the ferromagnet
was detected.
We here
fully characterize the  long-ranged triplet
correlations in such $S/Ho/F/Ho/S$ Josephson junctions (see
Fig.~\ref{fig:model1_1}(c)). The sandwiched central $F$ layer
represents a material with uniform magnetization, e.g., Cobalt,
while the outer two $Ho$ layers represent ferromagnets with a conical
magnetization texture, such as that found in Holmium ($Ho$). As with
the previous structures, we focus the study on the supercurrent
behavior as a function of the central $F$ layer thickness. We find
that the supercurrent decays uniformly \textit{without} any sign
change when varying the middle $F$ layer thickness, in agreement
with other works\cite{rob2}. However, the precise
underlying role of the
singlet and triplet components in the overall supercurrent behavior
has  been lacking.
\begin{figure*}[]
\includegraphics[width=18.0cm,height=7.0cm]{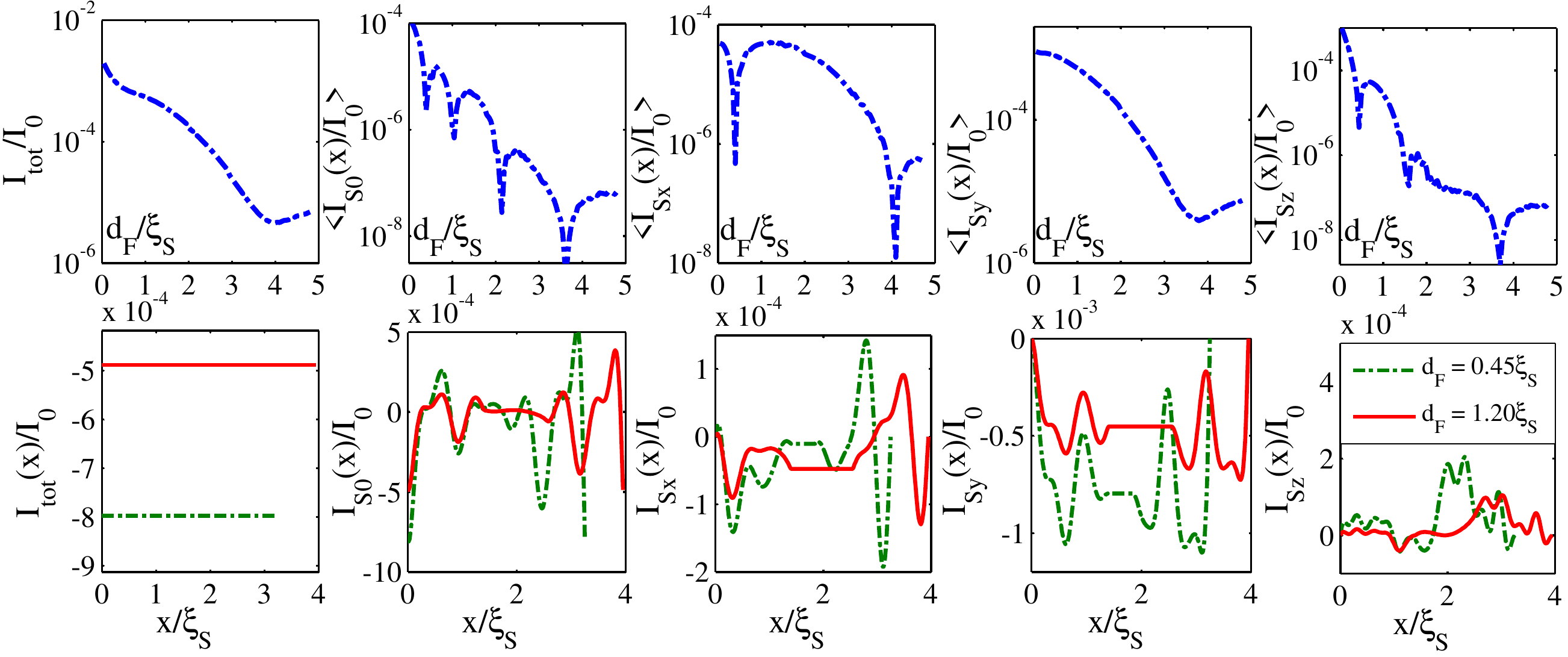}
\caption{\label{fig:s_ho_f_ho_s}(Color online) \textit{Top row}:
Critical current and its components versus middle $F$ layer
thickness, $d_{F}$ in $S/Ho/F/Ho/S$ structure where the outer $F$
layers' magnetization pattern is  Holmium ($Ho$)-like .
The configuration is  illustrated
schematically in Fig.~\ref{fig:model1_1}(c).
The thicknesses of the outer $F$ layers are fixed at
$d_{Ho}=1.4\xi_S$, whereas the magnetization is fixed along the $z$
direction in the middle $F$ layer, $\beta_2=0$. The components of
the critical supercurrents are calculated by spatially averaging
over the thickness of the middle $F$ layer ($\langle
I_{S0}(x)/I_0\rangle, \langle I_{Sy}(x)/I_0\rangle, \langle
I_{Sz}(x)/I_0\rangle$). \textit{Bottom row}: Critical supercurrent
against position inside the magnetic junction $x$ where the
thickness of middle $F$ layer has two values $d_{F}=0.45\xi_S,
1.2\xi_S$.}
\end{figure*}
In Fig.~\ref{fig:s_ho_f_ho_s}, we exhibit the total current and
its decomposition for a variety of parameters appropriate for the
inhomogeneous conical magnetic junction.
The middle $F$ layer is magnetized along
$z$, while the conical magnetization patterns of the Holmium layers
have the adopted form:
\begin{align}
\label{ho_eq} \vec{h}_{Ho}=h_0\left(\cos\alpha,\sin\alpha\sin(\gamma
x/a),\sin\alpha\cos(\gamma x/a)\right),
\end{align}
with the following material parameters: $a=0.02d$ is the distance of interatomic layers,
$d$ denotes the junction
thickness,
$\alpha=4\pi/9$
is the apex angle, and $\gamma=\pi/6$ is the rotation angle of the
cone structure, consistent with experimental values \cite{Sosnin}.
The magnitude of the exchange field is unchanged throughout the $F$
regions. The top set of panels in Fig.~\ref{fig:s_ho_f_ho_s} shows
the critical total current and its components as a function of the
middle uniform $F$ layer thickness $d_{F}$. The components
$I_{S0}$, $I_{Sy}$, and $I_{Sz}$ are averaged over the middle $F$
layer [denoted by $\langle ...\rangle$]. We assume the two $Ho$
layers have identical magnetization patterns \cite{rob2}, and  both
thicknesses equal $d_{Ho}=1.4\xi_S$. The first panel on the left
shows the total critical supercurrent versus $d_F$
and exhibits
the expected decay over a few coherence lengths. Examining the
signatures of the other components in the top panels, we see that
the opposite-spin singlet and triplet components, $I_{S0}$ and
$I_{Sz}$ respectively, demonstrate well defined oscillatory behavior
and corresponding sign changes as a function of $d_{F}$.
We can also conclude that the net
supercurrent arises mainly from the
spin-1 projection of the triplet components, $I_{Sy}$, which is
long-ranged in the middle $F$ with an exchange field direction
orthogonal to the spin-orientation of $I_{Sy}$. For the Holmium
magnetization profile, $h_x$ does not vary in space, and therefore
$I_{Sx}$ behaves similarly to the $I_{Sy}$ component in the
$S/F/F/F/S$ structure (see Fig.~\ref{fig:s_f_f_f_s}). Although
$I_{Sx}$ undergoes fewer sign changes when varying $d_F$, the
$0$-$\pi$ transitions are clearly present for this component. The
triplet component, $I_{Sy}$, the
main contributor to the total current,
does not switch directions when
increasing the middle $F$ layer thickness $d_F$, and its magnitude
often dominates the other components. Its nearly monotonic decay
can be traced back to the corresponding component of the
magnetization profile: in the $Ho$ layer, $h_y$ rotates sinusoidally
as a function of position, generating long-ranged odd-frequency
correlations that are not subject to the spin-splitting effects of
the magnet responsible also for the oscillatory behavior of
superconducting correlations. We have also  found  consistency with
previous studies\cite{rob2}, where  the monotonic decay
of the supercurrent appears when $d_{Ho}$ is large enough to contain
at least one spiral period.
On the other hand,
the sign changing behavior emerges for small $d_{Ho}$, so that
the $Ho$ layers effectively mimics a uniform ferromagnet.
This aspect was revisited in a recent work employing
a lattice model.\cite{annett}
The bottom row of Fig.~\ref{fig:s_ho_f_ho_s} illustrates the
decomposed components of the maximum total supercurrent as a
function of position throughout the ferromagnet regions.
Two representative thicknesses of the middle $F$ layer are
considered: $d_{F}=0.4\xi_S$, and $d_{F}=1.2\xi_S$. Similar to the
$S/F/F/F/S$ junction above, the triplet components with spin projection
$m=\pm 1$ on the $z$-axis ($I_{Sx}$ and $I_{Sy}$), are constant over
the entire middle $F$ region and therefore can be classified as long-ranged.
To summarize this section,
we studied the behavior of the critical 
supercurrent through low-proximity one-dimensional
structures shown in Fig.~\ref{fig:model1_1}.  
By directly decomposing the {\it supercurrent} using the
spin-parametrization technique given in the theoretical methods section,
we numerically studied the
origins of the supercurrent behavior
in terms of its  short-ranged
and
long-ranged components.
Our results showed that \sffs
structures do not support any long-ranged supercurrent components,
reaffirming the findings of Ref.~\onlinecite{Houzet1},
while \sfffs junctions host long-ranged supercurrent
components\cite{Houzet3} that are more prominent
in the inhomogeneous $S$/$Ho$/$F$/$Ho$/$S$ structures as experimentally observed in Ref. \onlinecite{rob1} and verified theoretically by numerical studied of full proximity regime in Ref. \onlinecite{rob2}.
We showed that the long-ranged supercurrent component
corresponds to the rotating component of the magnetization texture, and is the
main contributor to
the  total supercurrent.
The numerical results presented
in this section shall be used for later comparisons when we
categorize structures into two classes based on the supercurrent
direction with respect to the \ff interface orientation: parallel
or perpendicular.
The one-dimensional structures in Fig.~\ref{fig:model1_1}  belong  
to the latter class.
We now direct our attention to two-dimensional hybrids,
including the possibility of an applied magnetic field.
The
singlet-triplet decompositions discussed above
shall be employed to pinpoint exactly the spatial behavior of the associated
components of the total charge supercurrent.

\section{two-dimensional hybrid structures}\label{sec:2D}

In this section we present the main results of the paper.
We first
consider a two-dimensional magnetic \sfs system subject to an
external magnetic field for two regimes:
the wide junction $W_F$ $\gg$ $d_F$, and the narrow junction
$W_F$ $\ll$ $d_F$ regimes.
The magnetic strips are
sandwiched between two $s$-wave superconducting reservoirs, where
the exchange field in the $F$ strips is spatially uniform.
This permits
analytical solutions that are capable of
accurately predicting
the behavior of relevant
physical quantities such as charge and spin supercurrents, as
well as the pair
potential.
The
analytical results are complemented with full numerical investigations,
including
studies of the
dependence
of the critical charge supercurrent on
the external magnetic
field,
and the corresponding
appearance of proximity vortices.
The current density
spatial map is also investigated,
giving a global view of the distribution of supercurrents throughout the junction.
We next consider two kinds of finite-sized magnetic \sffs Josephson junctions
subject to an external magnetic field.
For these systems, analytical routes
are scarce, and we must in general resort
 to numerical approaches.
 In one case, we assume the double layer
 magnetic $F$ strips
 comprising
the \ff junction are parallel with the $S$
interfaces. In the other case, however, we
assume that the $F$ strips are perpendicular to the $S$
interfaces.
Using the
spin-parametrization introduced in Sec.~\ref{subsec:theor-spin}, we can then
study the even and odd frequency
components of the total charge Josephson current inside
the proposed structures.

\subsection{Technical assumptions and parameters}\label{subsec:2D-param}

\begin{figure}[b]
\includegraphics[width=7.50cm,height=4.5cm]{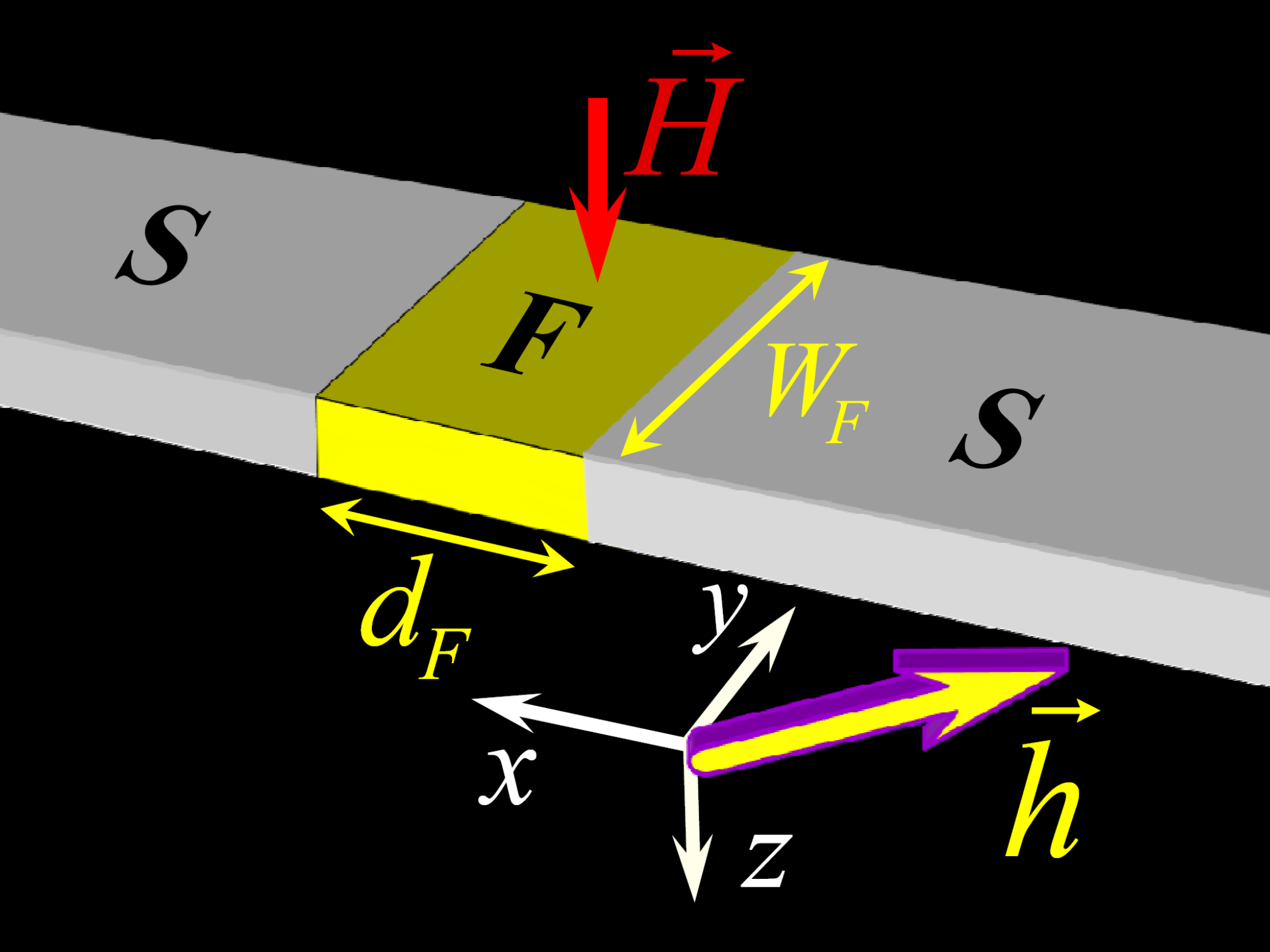}
\caption{\label{fig:model1} Proposed  setup of the Josephson \sfs
junction subject to an external magnetic field $\vec{H}$. The
magnetic wire ($F$)
with rectangular dimensions,  $d_F$ and $W_F$,
is sandwiched between two $s$-wave superconducting ($S$)
leads. The
exchange field of the
magnetic layer, $\vec{h}=(h^x,h^y,h^z)$, can take
arbitrary orientations.
The external
magnetic field is oriented along the $z$ direction. The
two-dimensional system resides  in the $xy$ plane so that the
interfaces are along the $y$ direction and $x$ axis is normal to the
junction. }
\end{figure}

In this subsection, we discuss the  assumptions used in our
calculations, along with the parameters and notations used throughout.
As was previously mentioned,
 the external magnetic field is confined
within the non-superconducting regions\cite{Cuevas_frh2,Cuevas_frh1,alidoust_nfrh1}
(see Fig.~\ref{fig:model1}).
We restrict
the magnetic field to be
oriented
perpendicular to the junction plane, which for our coordinate
system corresponds to
the $z$-axis. Thus,
supercurrent flow resides in the $xy$ plane. The
situation where the external magnetic field is parallel to the
current direction has been studied both
experimentally and theoretically.\cite{Birge}
We can therefore represent the magnetic field as,
$\vec{H}(x)=\vec{H}\Theta(x)\Theta(d_F-x)$, where $d_F$ is the
junction length and $\vec{H}=(0,0,H_z)$. This assumption
also ensures that the macroscopic phases of the two
superconducting electrodes are unaffected by the external magnetic
field. This widely used assumption
has demonstrated good
qualitative agreement with
experimental measurements.
\cite{Cuevas_frh1,Cuevas_frh2,alidoust_nfrh1,Angers}
The
Josephson junctions investigated in this work
therefore are assumed to have negligible magnetic field
screening.\cite{Angers,Chiodi}
If on the contrary, the magnetic field
is not restricted to the $F$ regions,
it becomes necessary to solve a set of
partial differential equations, Eq.~(\ref{eq:full_Usadel}),
self-consistently
in tandem with
Maxwell's equations and the superconducting
order parameter $\Delta(x,y,T)$.
Hence,
a suitable
choice for the vector potential $\vec{A}$
that we use
satisfying the
Coulomb gauge,
$\vec{\nabla}\cdot\vec{A}=0$,
is, $\vec{A}(x)=(-yH_z(x),0,0)$.
In normalizing our equations,
we write the external magnetic flux as
$\Phi=\tilde{\Phi}/\Phi_0$,
where
$\tilde{\Phi}\equiv W_Fd_FH_z$, $W_F$ is the junction width,
and
$\Phi_0=\hbar/2e$ is
the magnetic flux quantum.

\subsection{Uniform \sfs heterostructures}\label{subsec:2D-frnhfr}
We here consider a two-dimensional ferromagnetic \sfs Josephson
junction where the magnetization of the
magnetic strip
is homogeneous.
Although our theoretical approach allows
for completely general patterns in the magnetization texture,
we restrict our focus here to a specific case where the exchange
field has only one component along the $z$ direction,
$\vec{h}(x,y,z)=(0,0,h^z)$,
thus permitting analytical
solutions to the Usadel equation.
The two-dimensional junction resides in the $xy$
plane so that the \fs interfaces are parallel with the $y$ axis
(see Fig.~\ref{fig:model1}).
The corresponding
system of coupled partial differential equations [Eq.~(\ref{eq:full_Usadel})],
now reduces to a smaller set of decoupled partial
differential equations.
We are then able to derive
analytical expressions for the anomalous component of the
Green's function and therefore the charge supercurrent and pair potential.

Using this simplified
system of decoupled partial differential
equations, we consider two regimes: In the first case, we
assume the junction
width $W_F \gg d_F$, thus, terms involving
the ratio $d_F/W_F$ can be dropped,
leading to further simplifications.
In the second regime, the junction width
$W_F\ll d_F$,
corresponding to a
narrow magnetic nanowire.
To be complete, we also
implement a
full numerical investigation,
without the simplifying assumptions above
and with arbitrary values of ratio $W_F/d_F$.
This requires numerical solutions to a
complex system of partial
differential equations [see Eqs.~(\ref{eq:full_Usadel})].
Several checks on the numerics were performed, including
reproducing
previous results involving nonmagnetic \sns Josephson junctions,
where the exchange field of the $F$ layer is equal to zero.\cite{Cuevas_frh1,Cuevas_frh2}

The full Usadel equations in the presence of an external magnetic field
$\vec{H}$, and corresponding vector potential, $\vec{A}$, are written:
\begin{subequations} \label{eq:usadel_vector_pot}
\begin{eqnarray}
&&\nonumber\vec{\nabla}^{2}f_{\pm}(-\varepsilon)-2ie\vec{\nabla}\cdot\vec{A}f_{\pm}(-\varepsilon)-4ie\vec{A}\cdot\vec{\nabla}f_{\pm}(-\varepsilon)-\\&&
4e^{2}A^{2}f_{\pm}(-\varepsilon)-\frac{2i(\varepsilon\mp
h^z)}{D}f^{R}_{\pm}(-\varepsilon)=0,\\
&&\nonumber\vec{\nabla}^{2}f^{*}_{\pm}(\varepsilon)+2ie\vec{\nabla}\cdot\vec{A}f^{*}_{\pm}(\varepsilon)+4ie\vec{A}\cdot\vec{\nabla}f^{*}_{\pm}(\varepsilon)-\\&&
4e^{2}A^{2}f^{*}_{\pm}(\varepsilon)-\frac{2i(\varepsilon\pm
h^z)}{D}f^{*}_{\pm}(\varepsilon)=0,
\end{eqnarray}
\end{subequations}
where $\vec{\nabla}\equiv (\partial_x,\partial_y,\partial_z)$. The
above decoupled partial differential equations appear only for a
magnetic junction where the magnetization has one component
$\vec{h}=(0,0,h^z)$. If we now expand
the boundary conditions given
by Eq.~(\ref{eq:bc}), at the left \fs
interface, we find,
\begin{subequations}\label{eq:bcc1}
\begin{eqnarray}
&& \Big\{\zeta(\vec{\nabla} - 2ie\vec{A}) -
c^{\ast}(\varepsilon)\pm
i\frac{G_{S}}{G_{T}}\Big\}f_{\mp}(-\varepsilon) =\mp
s^{\ast}(\varepsilon)e^{-i\theta_l}, \\
&&\Big\{\zeta(\vec{\nabla} + 2ie\vec{A}) - c^{\ast}(\varepsilon)\mp
i\frac{G_{S}}{G_{T}}\Big\}f^{\ast}_{\mp}(\varepsilon) =\pm
s^{\ast}(\varepsilon)e^{i\theta_l},
\end{eqnarray}
\end{subequations}
while the boundary conditions at the right interface take the following
form:
\begin{subequations}\label{eq:bcc2}
\begin{eqnarray}
&& \Big\{\zeta(\vec{\nabla} -
2ie\vec{A})+c^{\ast}(\varepsilon)\mp i\frac{G_{S}}{G_{T}}\Big\}
f_{\mp}(-\varepsilon) =\pm
s^{\ast}(\varepsilon)e^{-i\theta_r},\\
&&\Big\{\zeta(\vec{\nabla} + 2ie\vec{A})+c^{\ast}(\varepsilon)\pm
i\frac{G_{S}}{G_{T}}\Big\}f^{\ast}_{\mp}(\varepsilon)=\mp
s^{\ast}(\varepsilon)e^{i\theta_r}.
\end{eqnarray}
\end{subequations}
The simplifying geometric approximations mentioned above
can now be applied to the above equations,
while adhering to the requirement that
the corresponding regimes
are experimentally accessible.

\subsubsection{Wide junction limit, $W_F$ $\gg$ $d_F$: Analytical results}\label{ssubsec:2D-frnhfr-W}
If we assume that the width of junction is much larger than its
length, terms involving $d_F/W_F$ in Eq.~(\ref{eq:usadel_vector_pot})
can be neglected,
yielding the following
decoupled Usadel
equations:
\begin{subequations}\label{eq:linearized Usadel1}
\begin{eqnarray}
 &&\nonumber\partial^{2}_{x}f_{\pm}(-\varepsilon)+4i\Phi
y\partial_{x}f_{\pm}(-\varepsilon)
-4\Phi^2y^2f_{\pm}(-\varepsilon)\\&& -2i\frac{\epsilon_\mp
}{\epsilon_{T}}f_{\pm}(-\varepsilon)=0,\\
&&\nonumber\partial^{2}_{x}f^\ast_{\pm}(\varepsilon)+4i\Phi
y\partial_{x}f^{\ast}_{\pm}(\varepsilon)
-4\Phi^2y^2f^{^\ast}_{\pm}(\varepsilon)\\&& -2i\frac{\epsilon_\pm
}{\epsilon_{T}}f^{^\ast}_{\pm}(\varepsilon)=0.
\end{eqnarray}
\end{subequations}
Here, we define $\epsilon_{\pm}\equiv\varepsilon\pm h^z(x,y,z)$, and
the Thouless energy $\epsilon_T=D/d_F^2$.
Note that all
partial derivatives in Eqs.~(\ref{eq:linearized Usadel1})
are solely
with respect to the $x$ coordinate. In other
words, the original two-dimensional problem is now
reduced to a quasi one-dimensional one.
These uncoupled
differential equations can be solved analytically thus permitting
additional insight into the transport properties of ferromagnetic Josephson junctions.
The Kupriyanov-Lukichev boundary conditions, Eq.~(\ref{eq:bc}),
at the left \fs interface located at $x=0$ reduces to:
\begin{subequations}\label{B.C. 1}
\begin{eqnarray}
&&\Big\{\zeta(\partial_{x}+2iy\Phi)+c(-\varepsilon) \mp
i\frac{G_{S}}{G_{T}}
\Big\}f_{\pm}(-\varepsilon)=\mp s(-\varepsilon)e^{i\theta_l},\\
&&\Big\{\zeta(\partial_{x}-2iy\Phi)+c(-\varepsilon) \pm
i\frac{G_{S}}{G_{T}}\Big\}f^{\ast}_{\pm}(\varepsilon)=\pm
s(-\varepsilon)e^{-i\theta_l}.
\end{eqnarray}
\end{subequations}
Similarly, the boundary conditions at the right \fs interface
located at $x=d_F$ can
be written as follows:
\begin{subequations}\label{B.C. 2a}
\begin{eqnarray}
&&\Big\{\zeta(\partial_{x}+2iy\Phi)-c(-\varepsilon) \pm
i\frac{G_{S}}{G_{T}}\Big\}f_{\pm}(-\varepsilon)=\pm
s(-\varepsilon)e^{-i\theta_r},\\
&&\Big\{\zeta(\partial_{x}-2iy\Phi)-c(-\varepsilon) \mp
i\frac{G_{S}}{G_{T}}\Big\}f^{\ast}_{\pm}(\varepsilon)=\mp
s(-\varepsilon)e^{i\theta_r}.
\end{eqnarray}
\end{subequations}
The macroscopic phases of the
left and right superconducting terminals
are labeled $\theta_l$ and $\theta_r$, respectively.
The magnetic
strips are assumed isolated in the $y$ direction so that physically no
current passes through the boundaries at $y=0$, and $y= W_F$.
Thus, to ensure
that the supercurrent does not pass through the vacuum boundaries in
the $y$ direction, we have the following conditions:
\begin{subequations}\label{B.C. 2}
\begin{eqnarray}
&&\partial_y f_{\pm}(\pm\varepsilon)=0,\\&&
\partial_y
f^{\ast}_{\pm}(\pm\varepsilon)=0.
\end{eqnarray}
\end{subequations}

With the solutions to Eq.~(\ref{eq:linearized Usadel1})
at hand, we are now in a position to
calculate the current density and the pair potential for a given magnetic flux.
The current density, given by Eq.~(\ref{eq:currentdensity}),
can thus be expressed as:
\begin{eqnarray}\label{eq:current1}
&&\nonumber\vec{J}(x,y)=J_{0}\int_{-\infty}^{\infty}d\varepsilon\tanh(\varepsilon
\beta)\Big\{\Big.
f_{-}(-\varepsilon)\vec{\nabla}f^{*}_{+}(\varepsilon)\\&&\nonumber+
f_{+}(-\varepsilon)\vec{\nabla}f^{*}_{-}(\varepsilon)-
f_{+}(\varepsilon)\vec{\nabla}f^{*}_{-}(-\varepsilon)-f_{-}(\varepsilon)\\&&\nonumber\vec{\nabla}f^{*}_{+}(-\varepsilon)+
f^{*}_{-}(-\varepsilon)\vec{\nabla}f_{+}(\varepsilon)+f^{*}_{+}(-\varepsilon)\vec{\nabla}f_{-}(\varepsilon)\\&&\nonumber-
f^{*}_{+}(\varepsilon)\vec{\nabla}f_{-}(-\varepsilon)-f^{*}_{-}(\varepsilon)\vec{\nabla}f_{+}(-\varepsilon)+4ie\vec{A}\big
[\Big.
\\&&\nonumber f^{*}_{-}(\varepsilon)f_{+}(-\varepsilon)-f^{*}_{+}(-\varepsilon)f_{-}(\varepsilon)-f^{*}_{-}(-\varepsilon)f_{+}(\varepsilon)\\&&\big.+f^{*}_{+}(\varepsilon)f_{-}(-\varepsilon)\big ]\Big\}.
\end{eqnarray}
The normalized pair potential, Eq.~(\ref{eq:full_pair})
now reads:
\begin{eqnarray}\label{eq:Linearized pair_potential}
\nonumber \tilde{U}_{\text{pair}}=\int_{-\infty}^{\infty} \big[
f_{+}(\varepsilon)-f_{+}(-\varepsilon)+\big.\\ \big.
f_{-}(-\varepsilon)-f_{-}(\varepsilon)\big]\tanh(\varepsilon
\beta)d\varepsilon.
\end{eqnarray}
If we solve the Usadel equations Eqs.~(\ref{eq:linearized Usadel1})
using the boundary conditions (\ref{B.C. 1}) and (\ref{B.C. 2}), we arrive at the following
solutions to the anomalous component of the Green's function:
\begin{eqnarray}
f_{\pm}(-\varepsilon)=\mp\frac{\mathcal{N}^{\pm}}{\mathcal{D}^{\pm}},
\end{eqnarray}
where the numerators $\mathcal{N}^{\pm}$ and denominators
$\mathcal{D}^{\pm}$ are given by
\begin{eqnarray}
&&\nonumber\mathcal{N}^{\pm}=s^{\ast}(\varepsilon) e^{-\frac{1}{2} i
(4 x y \Phi +\phi )} \Big\{e^{2 i y \Phi } \big[ \zeta \lambda_{\mp}
\cosh (x \lambda_{\mp})+\Big.\big.
\\&&\nonumber\big(c^{\ast}(\varepsilon)\pm i \frac{G_{S}}{G_{T}}\big) \sinh
(x \lambda_{\mp})\big] +e^{i \phi } \big[\zeta \lambda_{\mp}\times
   \big. \nonumber\\&&\nonumber\Big. \big. \cosh (\lambda_{\mp}-x \lambda_{\mp})+\big(c^{\ast}(\varepsilon)\pm  i \frac{G_{S}}{G_{T}}\big) \sinh (\lambda_{\mp}-x
   \lambda_{\mp})\big]\Big\},\nonumber
         \end{eqnarray}
\begin{eqnarray}
   &&
\mathcal{D}^{\pm}=\sinh (\lambda_{\mp}) \Big\{\zeta ^2
\lambda_{\mp}^2+\big(c^{\ast}(\varepsilon)\pm i
\frac{G_{S}}{G_{T}}\big)^2\Big\}+\nonumber\\&&\nonumber 2 \zeta
   \lambda_{\mp} \big(c^{\ast}(\varepsilon)\pm i \frac{G_{S}}{G_{T}}\big) \cosh
   (\lambda_{\mp}).\nonumber
      \end{eqnarray}
Similar solutions can be found for $f^{*}_{\pm}(-\varepsilon)$.
In order to simplify notation, we have defined $\lambda_{\pm}=2i(\epsilon\pm h^z)/\epsilon_T$,
and the macroscopic phase difference of the
superconducting terminals is denoted by $\phi=\theta_l-\theta_r$.
The charge current density expressed by Eq.~(\ref{eq:current1})
involves eight terms, $f_{\pm}(\pm\varepsilon)$, and
$f_{\pm}^{\ast}(\pm\varepsilon)$,
which should be derived to
obtain an analytical expression for the supercurrent flow.
In our calculations thus far,
the interfaces are assumed spin-active, namely,
$G_S/G_T\neq 0$.
To maintain
tractable analytic solutions, we drop the
$c(\varepsilon)$-terms
in addition to the spin-active contributions,
which is appropriate
for
experimental conditions involving
highly impure superconducting terminals.
These widely used
approximations lead to
intuitive and physically relevant analytical solutions.\cite{bergeret1,buzdin1,alidoust2}
Substituting the
solutions to the Usadel equations into Eq.~(\ref{eq:current1}), we
find the charge supercurrent density in the $x$
direction:
\begin{equation}\label{eq:current_wide_junc}
I(\phi)=J_0\int_{-\infty}^{\infty}d\varepsilon\frac{2 i
\tanh(\varepsilon \beta) \mathbb{N}}{\zeta ^2 \lambda_{-}
\lambda_{+}}\int_{-1/2}^{1/2}dy\sin (\phi -2 y \Phi ),
    \end{equation}
where,
\begin{eqnarray}
&&\nonumber\mathbb{N}={s^{\ast}}^{2}(-\varepsilon) \Big(\lambda_{+}
\csc (\lambda_{-})+\lambda_{-} \csc (\lambda_{+})\Big)+\\\nonumber
&&{s^{\ast}}^{2}(\varepsilon) \Big(\lambda_{+}
\text{csch}(\lambda_{-})+\lambda_{-} \text{csch}(\lambda_{+})\Big).
\end{eqnarray}
Integrating the junction width over the $y$-direction,
we end up with the total charge supercurrent
across the junction:
\begin{equation}\label{eq:fraunhofer_anal}
I(\phi,\Phi)=I_{0}\sin\phi\frac{\sin\Phi}{\Phi},
    \end{equation}
where we have extracted the phase and flux dependent terms and absorbed the
remaining coefficients into
$I_0$.
The maximum charge supercurrent occurs when the
superconducting phase difference equals $|\phi|=\pi/2$.
From Eq.~(\ref{eq:fraunhofer_anal}),
we see immediately that
the critical charge current exhibits the
well-known Fraunhofer interference diffraction pattern as
a function of the
externally applied flux $\Phi$.
We also recover the results of a normal \sns junction \cite{Cuevas_frh2},
where $h^z=0$.
Thus, our analytical
expressions for wide \sfs Josephson junctions
experiencing
perpendicularly directed external magnetic flux yields
the same critical current response
as a normal \sns junction\cite{Cuevas_frh2}).
Here, however, there are
additional, experimentally tunable physical quantities
which can
cause sign changes in $I_0(h^z,G_S/G_T)$, and consequently $I(\phi, \Phi)$.
When presenting a global view of the current density, it is
illustrative to examine a spatial map of its behavior.
By utilizing Eq.~(\ref{eq:currentdensity}), it is possible
to calculate the charge current density
throughout the $xy$ plane for a
wide magnetic \sfs Josephson
junction.

If we now insert the recently obtained solutions to the Usadel equations
into the pair potential
equation, Eq.~(\ref{eq:Linearized pair_potential}), we arrive at the
following analytical formula which provides a spatial map of
$U_{\text{pair}}$
for a wide junction:
\begin{eqnarray}\label{pair_analytical}
\nonumber
&&\tilde{U}_{\text{pair}}=\int_{-\infty}^{+\infty}d\varepsilon\Big\{e^{-\frac{i}{2}(\phi+4xy\Phi)}\big(
\big[s^\ast(-\varepsilon)\lambda_{+}\cos(x\lambda_{-})\times\big.\big.\Big.\\\nonumber
&&\csc(\lambda_{-})+s^\ast(-\varepsilon)\lambda_{-}\cos(x\lambda_{+})\csc(\lambda_{+})+
s^\ast(\varepsilon)\lambda_{+}\times\\\nonumber
&&\cosh(x\lambda_{-})\text{csch}(\lambda_{-})+
s^\ast(\varepsilon)\lambda_{+}\cosh(x\lambda_{+})\text{csch}(\lambda_{+})\big]e^{2iy\Phi}\\\nonumber
&&+\big[
s^\ast(-\varepsilon)\lambda_{+}\cos(\lambda_{-}-x\lambda_{-})\csc(\lambda_{-})
+ s^\ast(-\varepsilon)\lambda_{-}\times\big.\\\nonumber
&&\cos(\lambda_{+}-x\lambda_{+})\csc(\lambda_{+})+
s^\ast(\varepsilon)\lambda_{+}\cosh(\lambda_{-}-x\lambda_{-})\times\\
&&\big.\big.\Big.\text{csch}(\lambda_{-})+
s^\ast(\varepsilon)\lambda_{-}\cosh(\lambda_{+}-x\lambda_{+})\text{csch}(\lambda_{+})\big]e^{i\phi}
 \big) \Big\}\tanh(\varepsilon \beta).
\end{eqnarray}
If we restrict the proximity pair potential profile above by considering a fixed
$x$ position, corresponding to the middle of the
junction ($x=1/2$), we arrive at an expression which is
now $x$ independent:
\begin{eqnarray} \label{eq:pair_analytic}
\nonumber
\tilde{U}_{\text{pair}}&=&\int_{-\infty}^{+\infty}d\varepsilon\frac{\cos\big([\phi-2y\Phi]/2\big)}{\zeta\lambda_{-}\lambda_{+}}\Big\{
s^{\ast}(-\varepsilon)\lambda_{+}\csc(\lambda_{-}/2)\Big.\\\nonumber
&+& s^{\ast}(-\varepsilon)\lambda_{-}\csc(\lambda_{+}/2)
+s^{\ast}(\varepsilon)\lambda_{+}\text{csch}(\lambda_{-}/2)\\
&+&\Big.
s^{\ast}(\varepsilon)\lambda_{-}\text{csch}(\lambda_{+}/2)\Big\}\tanh(\varepsilon \beta).
\end{eqnarray}
As seen, the $\cos\big([\phi-2y\Phi]/2\big)$ term is zero at
$\phi-2y\Phi=m\pi$, for $m$ an odd integer.
Therefore, the
zeros of the proximity pair potential at the middle of magnetic strip are located
at $y=(\phi-m\pi)/2\Phi$ so that $-1/2<y<1/2$.
Setting $h^z=0$,
recovers the
nonmagnetic \sns junction result for the proximity pair potential.\cite{Cuevas_frh2}
Comparing with the current density, Eq.~(\ref{eq:current_wide_junc}), we see
that the current density and proximity pair potential both vanish at the same locations,
however the current density vanishes at additional positions corresponding to
$\sin\big(\phi-2y\Phi\big)=0$, or, when $\phi-2y\Phi=m\pi$ for  $m=0,
\pm 1, \pm 2, \pm 3...$ . The origin of these extra zeroes in the
current density arises from the cancellation of
counter-propagating currents from the orbital motion of the
quasiparticles.
These paths are
visualized using spatial mappings, presented below.

\subsubsection{Narrow junction limit, $W_F$ $\ll$ $d_F$: Analytical results}\label{ssubsec:2D-frnhfr-N}

The next useful regime that
leads to analytical results is that
corresponding to a
narrow junction, that is, $W_F \ll d_F$. In this case, we
assume that the width of the ferromagnetic layer $W_F \lesssim \xi_{H}$,
where
$\xi_{H}$ is the characteristic length describing
the Green's
function oscillations in the ferromagnetic layer.
This assumption
permits averaging the relevant equations
over the junction width.
Therefore, making the
substitutions
$\langle-Hy\hat{x}\rangle_{y}=0$ and $\langle
H^{2}y^{2}\rangle_{y}=H^{2}/12$, where $\langle \ldots \rangle_y$ denotes
spatial averaging over the
$y$ direction, results in the modified Usadel equations:
\begin{subequations}\label{averaged Usadel}
\begin{eqnarray}
&&\partial^{2}_{x}f_{\pm}(-\varepsilon)-2(\frac{\Gamma_{H}+i\epsilon_{\pm}}{\epsilon_{T}})f_{\pm}(-\varepsilon)=0,\\
&&\partial^{2}_{x}f^{\ast}_{\pm}(\varepsilon)-2(\frac{\Gamma_{H}+i\epsilon_{\pm}}{\epsilon_{T}})f^{\ast}_{\pm}(\varepsilon)=0,
\end{eqnarray}
\end{subequations}
where we define $\Gamma_{H}\equiv D\pi^{2}H^{2}W_{F}^{2}/6$, and thus,
$\Gamma_{H}/\epsilon_{T}=\Phi^{2}/6$. The quantity, $\Gamma_{H}$, is known as the
magnetic depairing energy \cite{Cuevas_frh2}. As can be seen, the
above assumptions have considerably simplified the Usadel equations,
and we are now able to efficiently
solve the differential equations and derive
analytical expressions for useful physical quantities.
After some straightforward calculations, we arrive at the following solutions to the
Usadel equations for the anomalous Green's function:
\begin{eqnarray}\label{eq:f_narrow}
f_{\pm}(-\varepsilon)=\mp\frac{\mathcal{N}^{\pm}}{\mathcal{D}^{\pm}},
\end{eqnarray}
where the numerators $\mathcal{N}^{\pm}$ and denominators
$\mathcal{D}^{\pm}$ are now defined by the following expressions;
\begin{eqnarray}
&&\nonumber\mathcal{N}^{\pm}=s^{\ast}(\varepsilon) e^{-\frac{i \phi
}{2}} \Big\{\big(c^{\ast}(\varepsilon)\pm i \frac{G_{S}}{G_{T}}\big)
\times\Big.\\&&\nonumber\big(\sinh (x \lambda^{+}_{\mp})+e^{i \phi }
 \sinh (\lambda^{+}_{\mp}-x \lambda^{+}_{\mp})\big)+\zeta
\lambda^{+}_{\mp} e^{i \phi }\times\\&&\nonumber\Big. \cosh
   (\lambda^{+}_{\mp}-x \lambda^{+}_{\mp})+\zeta  \lambda^{+}_{\mp} \cosh (x
   \lambda^{+}_{\mp})\Big\},
      \end{eqnarray}
      and,
\begin{eqnarray}
&&\nonumber \mathcal{D}^{\pm}=\sinh (\lambda^{+}_{\mp}) \Big\{\zeta
^2 {\lambda^{+}_{\mp}}^2+\big(c^{\ast}(\varepsilon)\pm i
\frac{G_{S}}{G_{T}}\big)^2\Big\}+\\&&\nonumber 2 \zeta
   \lambda^{+}_{\mp} \big(c^{\ast}(\varepsilon)\pm i \frac{G_{S}}{G_{T}}\big) \cosh
   (\lambda^{+}_{\mp}).
\end{eqnarray}
Here, we have introduced a simplified notation,
where we define:
${\lambda^{+}_{\pm}}^2\equiv\frac{\Gamma_{H}+2i\epsilon_{\pm}}{\epsilon_{T}}$
and
${\lambda^{-}_{\pm}}^2\equiv\frac{\Gamma_{H}-2i\epsilon_{\pm}}{\epsilon_{T}}$.
If we substitute these solutions to the anomalous Green's
function, Eq.~(\ref{eq:f_narrow}), into the Josephson current
relation, Eq.~(\ref{eq:current1}), we arrive at the flow of
charge supercurrent through a narrow \sfs junction in the presence of
an external magnetic flux $\Phi$:
\begin{eqnarray}
&&I(\phi)=\int_{-\infty}^{\infty}d\varepsilon\frac{2 i
\tanh(\varepsilon \beta) \mathbb{N} }{\zeta ^2 {\lambda^{-}_{-}}
{\lambda^{-}_{+}}
   {\lambda^{+}_{-}} {\lambda^{+}_{+}}}\sin
(\phi ),\\&& \nonumber \mathbb{N}= -{\lambda^{-}_{-}}
\Big\{s^{\ast}(-\varepsilon)^2 {\lambda^{+}_{-}} {\lambda^{+}_{+}}
\text{csch}
   ({\lambda^{-}_{+}})-\Big.\\&& \nonumber \Big. s^{\ast}(\varepsilon)^2 {\lambda^{-}_{+}} \big({\lambda^{+}_{+}}
   \text{csch}
   ({\lambda^{+}_{-}})+{\lambda^{+}_{-}} \text{csch} ({\lambda^{+}_{+}})\big)\Big\}-\\&& \nonumber \Big. s^{\ast}(-\varepsilon)^2 {\lambda^{-}_{+}} {\lambda^{+}_{-}}
   {\lambda^{+}_{+}} \text{csch} ({\lambda^{-}_{-}}),
    \end{eqnarray}
        \begin{figure}[b]
\includegraphics[width=7cm,height=6.0cm]{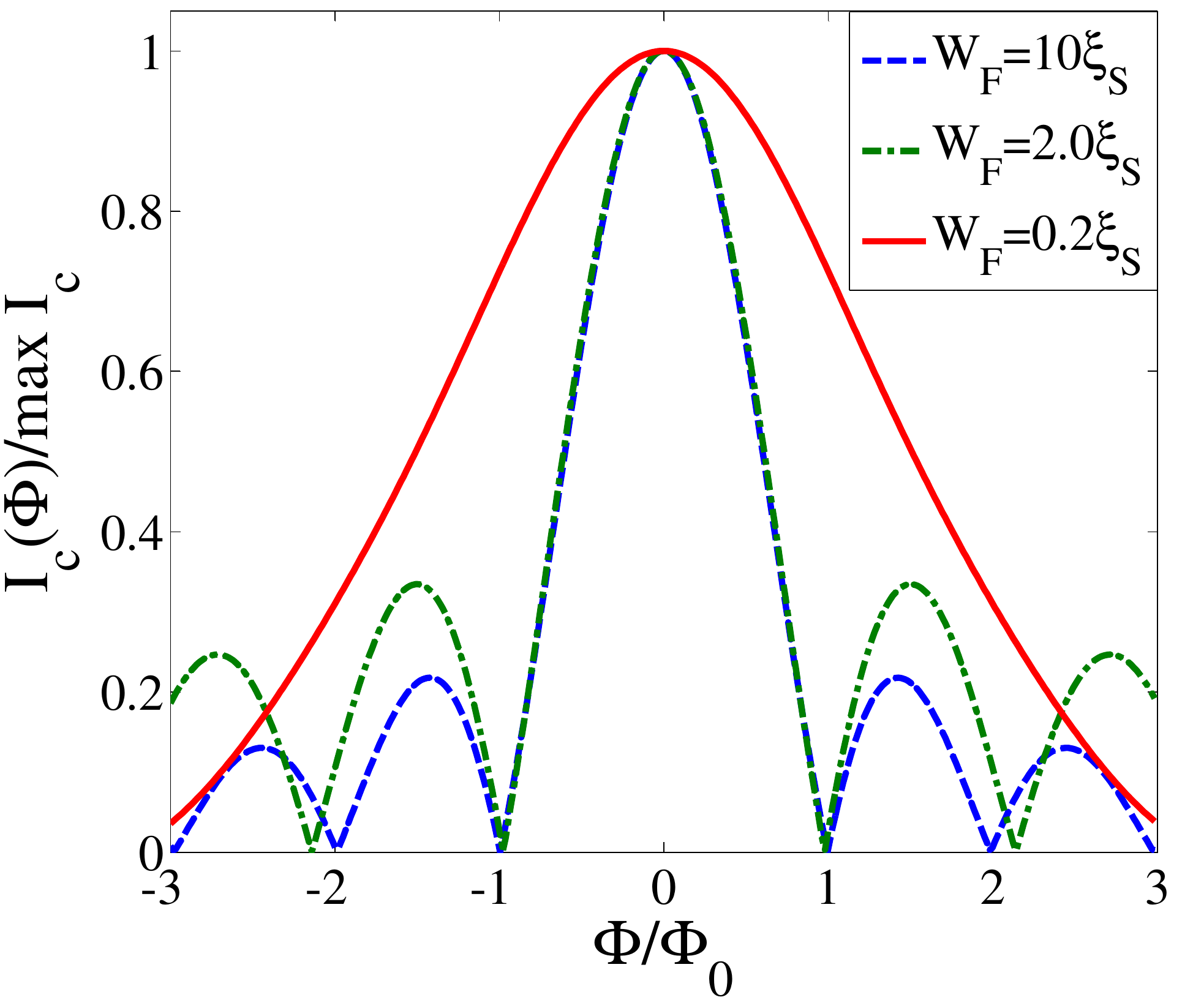}
\caption{\label{fig:magnetic_pattern} Magnetic interference pattern
of an \sfs Josephson junction (see Fig.~\ref{fig:model1}).
The critical supercurrent $I_c(\Phi)$, is shown normalized
by its maximum value, $\max(I_c)$, and plotted
against the external
magnetic flux $\Phi$.
Three values of the magnetic layer width $W_F$
are considered: $10\xi_S$, $2.0\xi_S$, and $0.2\xi_S$, whereas the
length of $F$ is assumed fixed at a representative value of
$d_F=2.0\xi_S$. }
\end{figure}
where the junction characteristics are
obtained via averaged values.
The critical supercurrent thus shows a monotonic decaying behavior
against the external magnetic flux $\Phi$ in the narrow junction
regime. This fact can be understood by noting the role of $\Gamma_H$
in the Usadel equations. In the narrow junction limit,
the external magnetic field breaks the coherence of Cooper pairs and
the Fraunhofer diffraction patterns for the critical current in the
wide junction limit turns to a monotonic decay.
The proximity vortices, which are closely linked to the Fraunhofer
patterns, vanish in this regime due to the narrow size of junction width.
This finding is also
in agreement with the nonmagnetic \sns junction counterpart
\cite{Cuevas_frh1,Cuevas_frh2}. If we substitute the
solutions from Eq.~(\ref{eq:f_narrow}) into the pair potential, Eq.~(\ref{eq:Linearized pair_potential}), we find that the
zeros which appeared in the wide
junction limit have now vanished in the narrow junction regime.
We now proceed to compliment our
simplified analytical results with a more complete
numerical investigation.

\subsubsection{Interference patterns, proximity vortices,
and current densities: Numerical investigations}\label{ssubsec:2D-frnhfr-Arb}
Previously we utilized  various
approximations to simplify
situations and permit explicit analytical solutions.
To investigate
the validity of our
analytic results,
we solve
numerically the Usadel equations, Eq.~(\ref{eq:usadel_vector_pot})
with appropriate boundary conditions found in Eqs.~(\ref{eq:bcc1}) and (\ref{eq:bcc2}).
We now
retain the
$c(\varepsilon)$ terms,
and include a broader range of junction
widths, not only those that are very narrow or very wide,
but also  intermediate widths that are not amenable to an analytical treatment.
We shall  present
results for
the current density and
for  spatial maps of the  pair potential in the junction
subjected to an
external magnetic field.

Figure~\ref{fig:magnetic_pattern} exhibits the results of our
numerical studies for the critical supercurrent response to an
external magnetic flux, $\Phi$, in a \sfs junction with a variety of ratios  $W_F/d_F$.
The
junction length and exchange field intensity are fixed at particular
values of $d_F=2.0\xi_S$ and $|\vec{h}|=5.0\Delta_0$
respectively.
The interface resistance parameter is  set to $\zeta=4.0$.
These choices ensure the validity of the Green's functions
in the low proximity limit:
$\underline{F}(x,y,z,\varepsilon)\ll
0.1\underline{F}^{no}(x,y,z,\varepsilon)$.
To more clearly see the effects
in the scaled plots,
we normalize the
critical charge supercurrent by the maximum of this quantity
max($I_c$) for each case separately.
\begin{figure}[b]
\includegraphics[width=7.50cm,height=6.0cm]{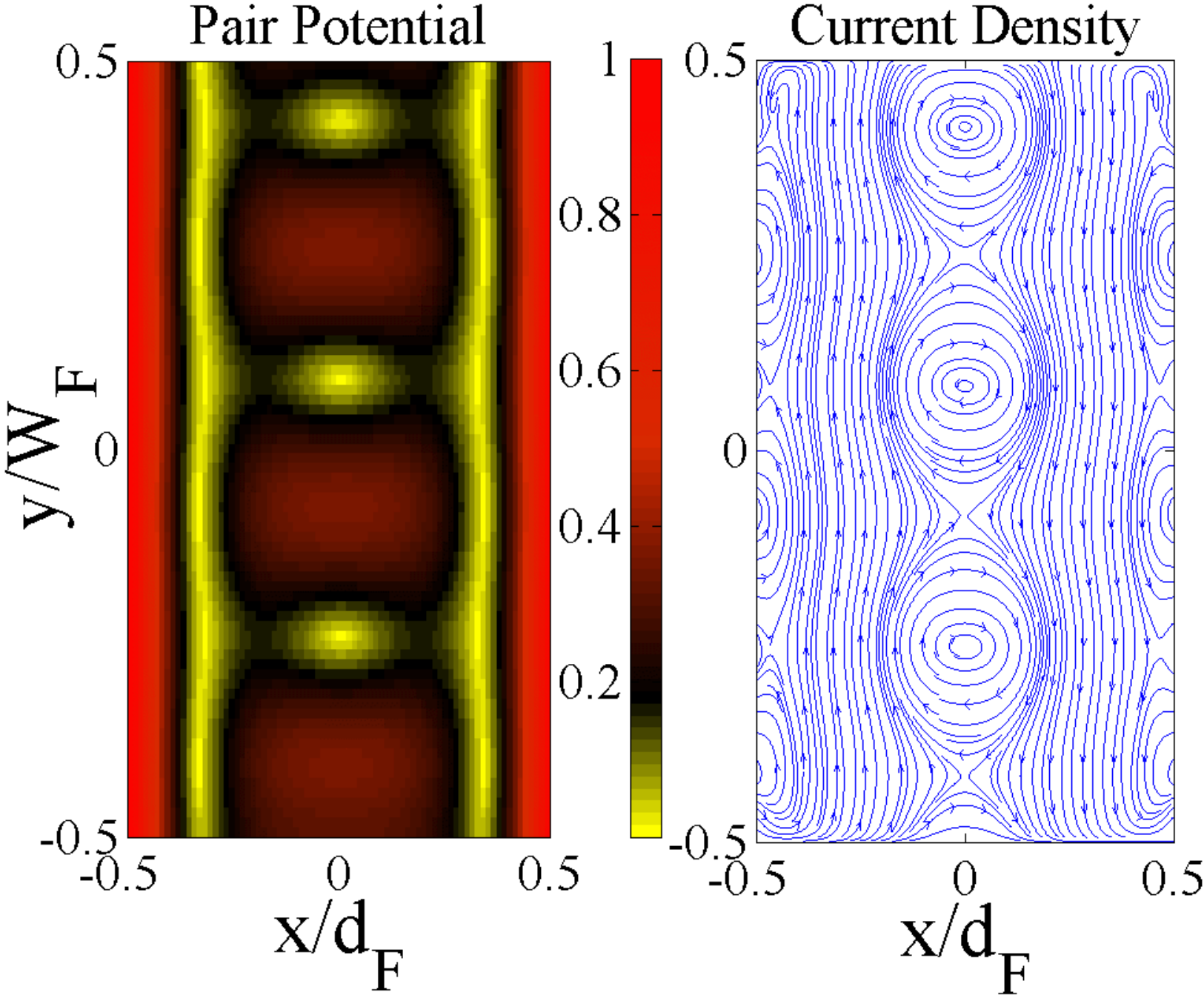}
\caption{\label{fig:pairp_crntdnsty_sfs} Spatial maps of the
pair
potential and current density in the presence of an external
magnetic field applied normally to the plane of an \sfs Josephson
junction (see Fig.~\ref{fig:model1}). We
assume the junction is relatively wide, $W_F=10\xi_S$, while its length
is sufficiently short to allow for current transport across the $x$ axis,
namely,  $d_F=2.0\xi_S$. The external magnetic flux is
$\Phi=3\Phi_0$, and the macroscopic phase difference is
corresponds to $|\phi|=\pi/2$.
}
\end{figure}
The figure clearly reveals a full Fraunhofer pattern for the
case of a wide
junction $W_F=10\xi_S$. This pattern is consistent with the analytic expression
for the
critical current given by Eq.~(\ref{eq:fraunhofer_anal}).
The
critical supercurrent decays with  increasing
magnetic field and
undergoes a series of cusps, indicating that
$I_c$
changes sign.
The observed sign-change reflects aspects of the orbital
motions of the quasiparticles,
while the decaying behavior is indicative of the
pair breaking nature of an external magnetic field as we explained previously.
A relatively
narrower junction  width  of $W_F=2.0\xi_S$ is also investigated.
As shown,
the ideal Fraunhofer pattern is now modified, yet still retains
its trademark signature. This diffraction pattern transitions to a
uniformly decaying behavior for the sufficiently narrow junction with width
$W_F=0.2\xi_S$.
This narrowest junction confines the orbital motion of
the quasiparticles and thus the pair-breaking by the external
field
plays a dominant role
in the critical
supercurrent response.

In Fig.~\ref{fig:pairp_crntdnsty_sfs} we plot the
2D spatial dependence of the
pair
potential and charge current density for a wide junction,
where the larger width
can more effectively demonstrate
the orbital paths as they relate to the pair correlations and supercurrent response.
To be consistent with
Fig.~\ref{fig:magnetic_pattern},
we
set $\Phi=3\Phi_0$, $W_F=10\xi_S$, and $|\phi|=\pi/2$.
As shown,
three zeros in the pair potential appear at the middle of the  junction (at
$x=0$) as
was found analytically in Sec.~\ref{ssubsec:2D-frnhfr-N}.
The intensity of the external magnetic flux
$\Phi$ determines the number of zeros and distance between
neighboring zeros.
The panel on
the right corresponds to a
spatial and vector map of the charge current
density,
revealing the
circulating paths of the quasiparticles. If we
set $|\phi|=\pi/2$ in Eq.~(\ref{eq:pair_analytic}),
with the constraint $-1/2<y<1/2$,
we find that $y=-1/12, 3/12, 5/12$
corresponds precisely to the
zeroes of the pair potential
found from the general numerical treatment, shown in Fig.~\ref{fig:pairp_crntdnsty_sfs}.
Likewise, $y=\pm 1/12$, $y=\pm 3/12$, and $y=\pm
5/12$ gives the zeroes of the current density.
As seen, the additional zeros in the current density corresponds to
locations where the trajectories of
two opposing circular paths
overlap,
thus canceling one another.
The other $y$ values giving zeroes, which correlate
with the
same locations as the pair potential, are at the cores of the
circulating paths.
Such behavior is
reminiscent of
Abrokosov vortices, where the supercurrent circulates around
normal state cores.
The
zeros in the pair potential may thus be viewed as proximity vortices.
\cite{Cuevas_frh2,Suderow,Abrikosov1,Abrikosov2,Abrikosov3}
Abrikosov vortices,
which carry a single magnetic quantum
flux $\Phi_0$, are however intrinsic
to type II superconductors.\cite{Abrikosov1,Abrikosov2,Abrikosov3,Suderow}
One of the criteria
for categorizing the superconducting state of a material is the nature of these
intrinsic vortices.\cite{Tinkham}
Nonetheless, such proximity vortices
are generally geometry-dependent
and  rely on 
the mutual interaction between a
magnetic field with the superconductor,
in contrast to  intrinsic Abrikosov vortices. \cite{Cuevas_frh2}

\subsection{Spin-parametrization}\label{subsec:2D-spin}

The Usadel equations in this section
have dealt solely with the even-frequency
superconducting correlations with spin-zero projection
along the spin-quantization axis.
This is due to the fact that we have only considered
ferromagnetic strips with a uniform magnetization texture.
For
inhomogeneous magnetization textures, the complex
partial differential equations become coupled and increase
in number to
eight in the low proximity limit.
This number is doubled if the full proximity limit
\cite{alidoust1} is considered.
Fortunately, it has been well established that the low proximity limit is
sufficient to capture the essential physical properties of
proximity systems such as the ones proposed in this paper.
To investigate the
behavior of even- and odd-frequency correlations, we employ a
spin-parametrization technique\cite{Lofwander2,Champel1,Champel2}
that has been frequently used to study
the characteristics of magnetic systems.
\cite{Lofwander2,Champel1,Champel2,Hikino,alidoust_missner,
buzdin1,Houzet1,Oboznov,bergeret1,bergeret2}

If we now substitute
this decomposition of the anomalous Green's function into the Usadel
equation, Eq.~(\ref{eq:full_Usadel}), and consider a two-dimensional
system, we end up with the following coupled set of differential
equations in the presence of an external flux $\Phi$:
\begin{widetext}
\begin{subequations}\label{eq:usadel_param_Phi}
\begin{eqnarray}
&&\nonumber D\bigg\{\mp\partial_x^{2} \mathbb{T}_{x}(-\varepsilon)\mp\Big(\frac{d_{F}}{W_{F}}\Big)^{2}\partial_y^{2} \mathbb{T}_{x}(-\varepsilon)+i\partial_x^{2} \mathbb{T}_{y}(-\varepsilon)+i\Big(\frac{d_{F}}{W_{F}}\Big)^{2}\partial_y^{2} \mathbb{T}_{y}(-\varepsilon)+4i\Phi y(\mp\partial_x\mathbb{T}_{x}(-\varepsilon)+i\partial_x\mathbb{T}_{y}(-\varepsilon))\bigg.\\&&\bigg.-4\Phi^2y^{2}(\mp\mathbb{T}_{x}(-\varepsilon)+i\mathbb{T}_{y}(-\varepsilon))\bigg\} +id_{F}^2\Big\{\mathbb{T}_{y}-2 \varepsilon (\mp \mathbb{T}_{x}(-\varepsilon)+i \mathbb{T}_{y}(-\varepsilon)) \mp 2\mathbb{S}(-\varepsilon) (h^{x} \mp i h^{y}) \Big\}=0,\\
&&\nonumber D\bigg\{\mp\partial_x^2\mathbb{S}(-\varepsilon)\mp\Big(\frac{d_{F}}{W_{F}}\Big)^{2}\partial_y^2\mathbb{S}(-\varepsilon)+\partial_x^2\mathbb{T}_z(-\varepsilon)+\Big(\frac{d_{F}}{W_{F}}\Big)^{2}\partial_y^2\mathbb{T}_z(-\varepsilon)+4i\Phi y(\mp \partial_x\mathbb{S}(-\varepsilon)+\partial_x \mathbb{T}_z(-\varepsilon))\bigg. \\&& \bigg. -4\Phi^2y^2(\mp \mathbb{S}(-\varepsilon)+\mathbb{T}_z(-\varepsilon))\bigg\}+id_{F}^2\Big\{\mp2\mathbb{T}_x(-\varepsilon) h^{x} \mp 2 \mathbb{T}_{y}(-\varepsilon) h^{y} - 2 (\mp\mathbb{S}(-\varepsilon)+\mathbb{T}_z(-\varepsilon)) (\varepsilon \pm h^{z}) \Big\}=0,\\
&&\nonumber D\bigg\{\mp\partial_x^{2}\mathbb{T}_{x}^{\ast}(\varepsilon)\mp\Big(\frac{d_{F}}{W_{F}}\Big)^{2}\partial_y^{2}\mathbb{T}_{x}^{\ast}(\varepsilon)-i\partial_x^{2}\mathbb{T}_{y}^{\ast}(\varepsilon)-i\Big(\frac{d_{F}}{W_{F}}\Big)^{2}\partial_y^{2}\mathbb{T}_{y}^{\ast}(\varepsilon)-4i\Phi y (\mp\partial_x \mathbb{T}_{x}^{\ast}(\varepsilon)-i \partial_x \mathbb{T}_{y}^{\ast}(\varepsilon))\bigg. \\&& \bigg. -4\Phi^2y^2(\mp\mathbb{T}_{x}^{\ast}(\varepsilon)-i \mathbb{T}_{y}^{\ast}(\varepsilon)) \bigg\}+id_{F}^2\Big\{\pm 2 (h^{x} \pm i h^{y}) \mathbb{S}^{\ast}(\varepsilon) - 2 \varepsilon (\mp\mathbb{T}_{x}^{\ast}(\varepsilon)-i\mathbb{T}_{y}^{\ast}(\varepsilon)) \Big\}=0,\\
&&\nonumber
D\bigg\{\mp\partial_x^{2}\mathbb{S}^{\ast}(\varepsilon)\mp\Big(\frac{d_{F}}{W_{F}}\Big)^{2}\partial_y^{2}\mathbb{S}^{\ast}(\varepsilon)+\partial_x^2\mathbb{T}_{z}^{\ast}(\varepsilon)+\Big(\frac{d_{F}}{W_{F}}\Big)^{2}\partial_y^2\mathbb{T}_{z}^{\ast}(\varepsilon)-4i\Phi
y(\mp
\partial_x\mathbb{S}^{\ast}(\varepsilon)+\partial_x\mathbb{T}_{z}^{\ast}(\varepsilon))\bigg.\\&&\bigg. -4\Phi^2y^2(\mp
\mathbb{S}^{\ast}(\varepsilon)+\mathbb{T}_{z}^{\ast}(\varepsilon))
\bigg\}+id_{F}^2\Big\{ 2 (-\varepsilon \pm
h^{z})(\mp\mathbb{S}^{\ast}(\varepsilon)+\mathbb{T}_{z}^{\ast}(\varepsilon))
\pm 2 h^{x} \mathbb{T}_{x}^{\ast}(\varepsilon) \pm 2 h^{y}
\mathbb{T}_{y}^{\ast}(\varepsilon)\Big\}=0.
\end{eqnarray}
\end{subequations}
The spin parameterized boundary conditions at the left \fs
interface, Eq. (\ref{eq:bc}), now reads:
\begin{subequations} \label{eq:bc_param1}
\begin{eqnarray}
&&\Big\{\zeta(\partial_x + 2i\Phi y) - c^{\ast}(\varepsilon)\Big\}\Big(\mp \mathbb{T}_{x}(-\varepsilon)+i \mathbb{T}_{y}(-\varepsilon)\Big)=0,\\
&&\Big\{\zeta(\partial_x + 2i\Phi y) - c^{\ast}(\varepsilon)\pm i\frac{G_S}{G_T}\Big\}\Big(\mp\mathbb{S}(-\varepsilon)+\mathbb{T}_z(-\varepsilon)\Big) \pm s^{\ast}(\varepsilon)e^{i\theta_l} =0,\\
&&\Big\{\zeta(\partial_x - 2i\Phi y) - c^{\ast}(\varepsilon)\Big\}\Big(\mp\mathbb{T}_{x}^{\ast}(\varepsilon)-i\mathbb{T}_{y}^{\ast}(\varepsilon)\Big)=0,\\
&&\Big\{\zeta(\partial_x - 2i\Phi y) - c^{\ast}(\varepsilon)\mp i\frac{G_S}{G_T}\Big\}\Big(\mp\mathbb{S}^{\ast}(\varepsilon)+\mathbb{T}_{z}^{\ast}(\varepsilon)\Big) \mp s^{\ast}(\varepsilon)e^{-i\theta_l} =0,
\end{eqnarray}
\end{subequations}
and for the right \fs interface, we have:
\begin{subequations}\label{eq:bc_param2}
\begin{eqnarray}
&&\Big\{\zeta(\partial_x + 2i \Phi y)+c^{\ast}(\varepsilon)\Big\} \Big(\mp \mathbb{T}_{x}(-\varepsilon)+i \mathbb{T}_{y}(-\varepsilon)\Big) =0,\\
&&\Big\{\zeta(\partial_x + 2i \Phi y)+c^{\ast}(\varepsilon)\mp i\frac{G_S}{G_T}\Big\} \Big(\mp\mathbb{S}(-\varepsilon)+\mathbb{T}_z(-\varepsilon)\Big) \mp s^{\ast}(\varepsilon)e^{i\theta_r}=0,\\
&&\Big\{\zeta(\partial_x - 2i \Phi y)+c^{\ast}(\varepsilon)\Big\} \Big(\mp\mathbb{T}_{x}^{\ast}(\varepsilon)-i\mathbb{T}_{y}^{\ast}(\varepsilon)\Big)=0,\\
&&\Big\{\zeta(\partial_x - 2i \Phi y)+c^{\ast}(\varepsilon)\pm i\frac{G_S}{G_T}\Big\}\Big(\mp\mathbb{S}^{\ast}(\varepsilon)+\mathbb{T}_{z}^{\ast}(\varepsilon)\Big) \pm s^{\ast}(\varepsilon)e^{-i\theta_r}=0.
\end{eqnarray}
\end{subequations}
Using Eq.~(\ref{eq:currentdensity}),
the spin-parameterized current
density through the junction (along the $x$ direction)
in the presence
of an external magnetic field is:
\begin{eqnarray}\label{eq:current_param}
\nonumber J_x(x,y) =&&J_0 \int_{-\infty}^{\infty} d\varepsilon
\bigg\{\bigg.
\mathbb{S}(\varepsilon)\partial_x\mathbb{S}^{\ast}(-\varepsilon) -
\mathbb{S}(-\varepsilon)\partial_x\mathbb{S}^{\ast}(\varepsilon)+
\mathbb{S}(\varepsilon)^{\ast}\partial_x\mathbb{S}(-\varepsilon)
-\mathbb{S}(-\varepsilon)^{\ast}\partial_x\mathbb{S}(\varepsilon)-\partial_x\mathbb{T}_x(-\varepsilon)\mathbb{T}_x^{\ast}(\varepsilon)
+\partial_x\mathbb{T}_x(\varepsilon)\mathbb{T}_x^{\ast}(-\varepsilon)\\\nonumber
&& \left. -\right. \left.
\partial_x\mathbb{T}_x^{\ast}(-\varepsilon)\mathbb{T}_x(\varepsilon)
+\partial_x\mathbb{T}_x^{\ast}(\varepsilon)\mathbb{T}_x(-\varepsilon)
-\partial_x\mathbb{T}_y(-\varepsilon)\mathbb{T}_y^{\ast}(\varepsilon)
+\partial_x\mathbb{T}_y(\varepsilon)\mathbb{T}_y^{\ast}(-\varepsilon)
-\partial_x\mathbb{T}_y^{\ast}(-\varepsilon)\mathbb{T}_y(\varepsilon)
+\partial_x\mathbb{T}_y^{\ast}(\varepsilon)\mathbb{T}_y(-\varepsilon)\right.\\\nonumber
&& \left. -\right. \left.
\partial_x\mathbb{T}_z(-\varepsilon)\mathbb{T}_z^{\ast}(\varepsilon)
+\partial_x\mathbb{T}_z(\varepsilon)\mathbb{T}_z^{\ast}(-\varepsilon)
-
\partial_x\mathbb{T}_z^{\ast}(-\varepsilon)\mathbb{T}_z(\varepsilon)+\partial_x\mathbb{T}_z^{\ast}(\varepsilon)\mathbb{T}_z(-\varepsilon)
- 4i\Phi y
\Big[\mathbb{S}(\varepsilon)\mathbb{S}^{\ast}(-\varepsilon)-
\mathbb{S}(-\varepsilon)\mathbb{S}^{\ast}(\varepsilon) \right.\\ &&
+ \mathbb{T}_x(-\varepsilon)\mathbb{T}_x^{\ast}(\varepsilon) -
\mathbb{T}_x(\varepsilon)\mathbb{T}_x^{\ast}(-\varepsilon)+\Big.\bigg.\Big.
\mathbb{T}_y(-\varepsilon)\mathbb{T}_y^{\ast}(\varepsilon) -
\mathbb{T}_y(\varepsilon)\mathbb{T}_y^{\ast}(-\varepsilon)+
\mathbb{T}_z(-\varepsilon)\mathbb{T}_z^{\ast}(\varepsilon) -
\mathbb{T}_z(\varepsilon)\mathbb{T}_z^{\ast}(-\varepsilon)\Big]
\bigg\}\tanh(\varepsilon\beta).
\end{eqnarray}

The $y$ component of the supercurrent density, $J_y(x,y)$, takes the same form as
$J_x(x,y)$, with the partial-$x$
derivatives replaced by derivatives with respect to $y$, and
also the inclusion of the appropriate component of the vector potential.

\begin{figure}[b]
\includegraphics[width=7.50cm,height=4.5cm]{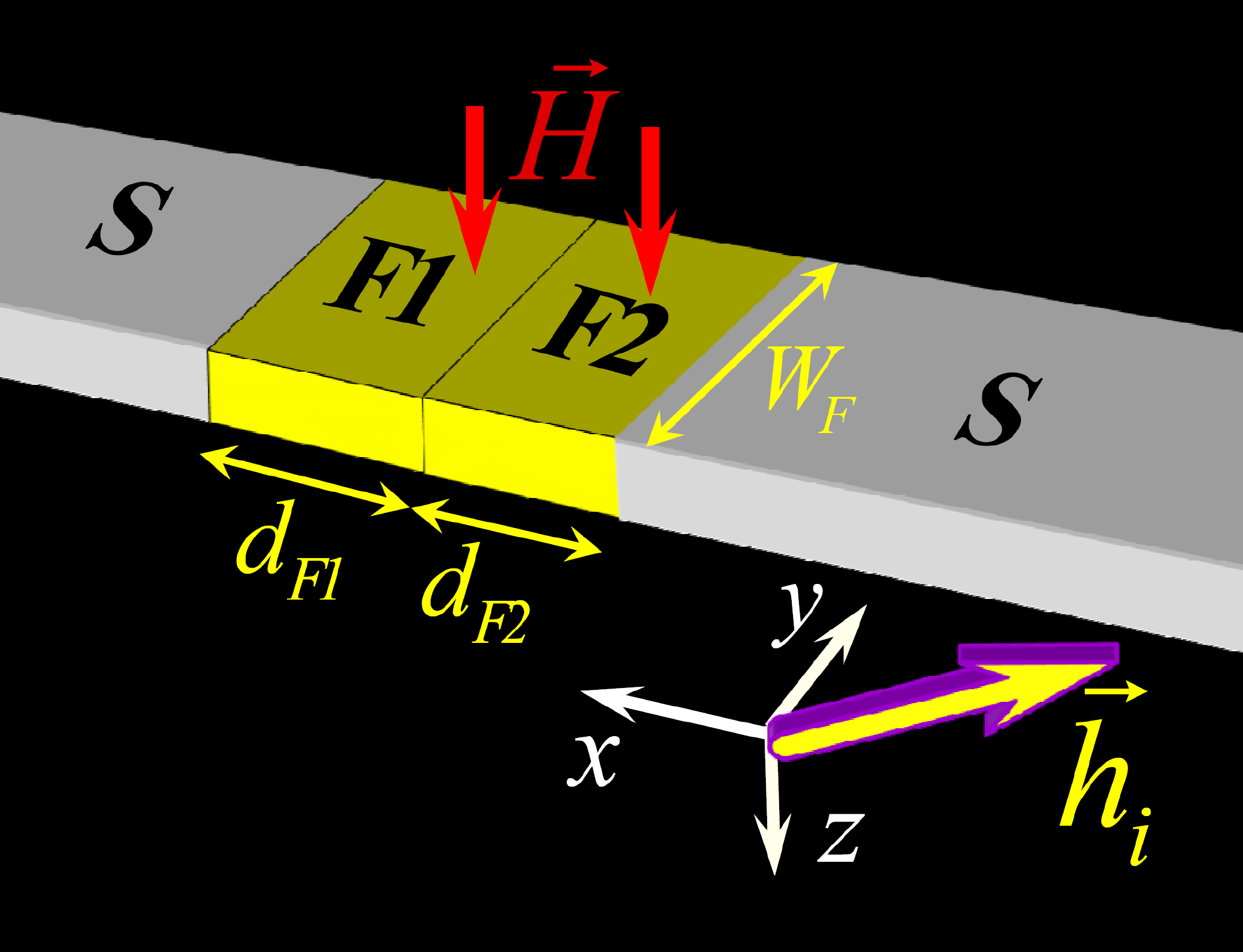}
\caption{\label{fig:model2} Proposed setup of an \sffs
Josephson junction where the double layer $F$ region has its interfaces
parallel with the $S$/$F$ interfaces.
The
ferromagnetic layers and $s$-wave superconducting terminals are
labeled by $F_1$, $F_2$, and $S$, respectively. Our theoretical
technique allows consideration of exchange fields with arbitrary
orientations. We thus
denote the magnetization of each $F$ layer by
$\vec{h}_i=(h_i^{x},h_i^{y},h_i^{z})$, for  $i=1,2$. The width of
the rectangular $F$ strips are equal and denoted by $W_F$, while
the length of each magnetic layer can differ and are denoted $d_{F1}$, and $d_{F2}$.
The junction is located in the
$xy$ plane and all interfaces are along the $y$ axis. An external
magnetic field $\vec{H}$ points along the $z$ axis, normal to
the junction plane.}
\end{figure}
\begin{figure*}[]
\includegraphics[width=18cm,height=10.50cm]{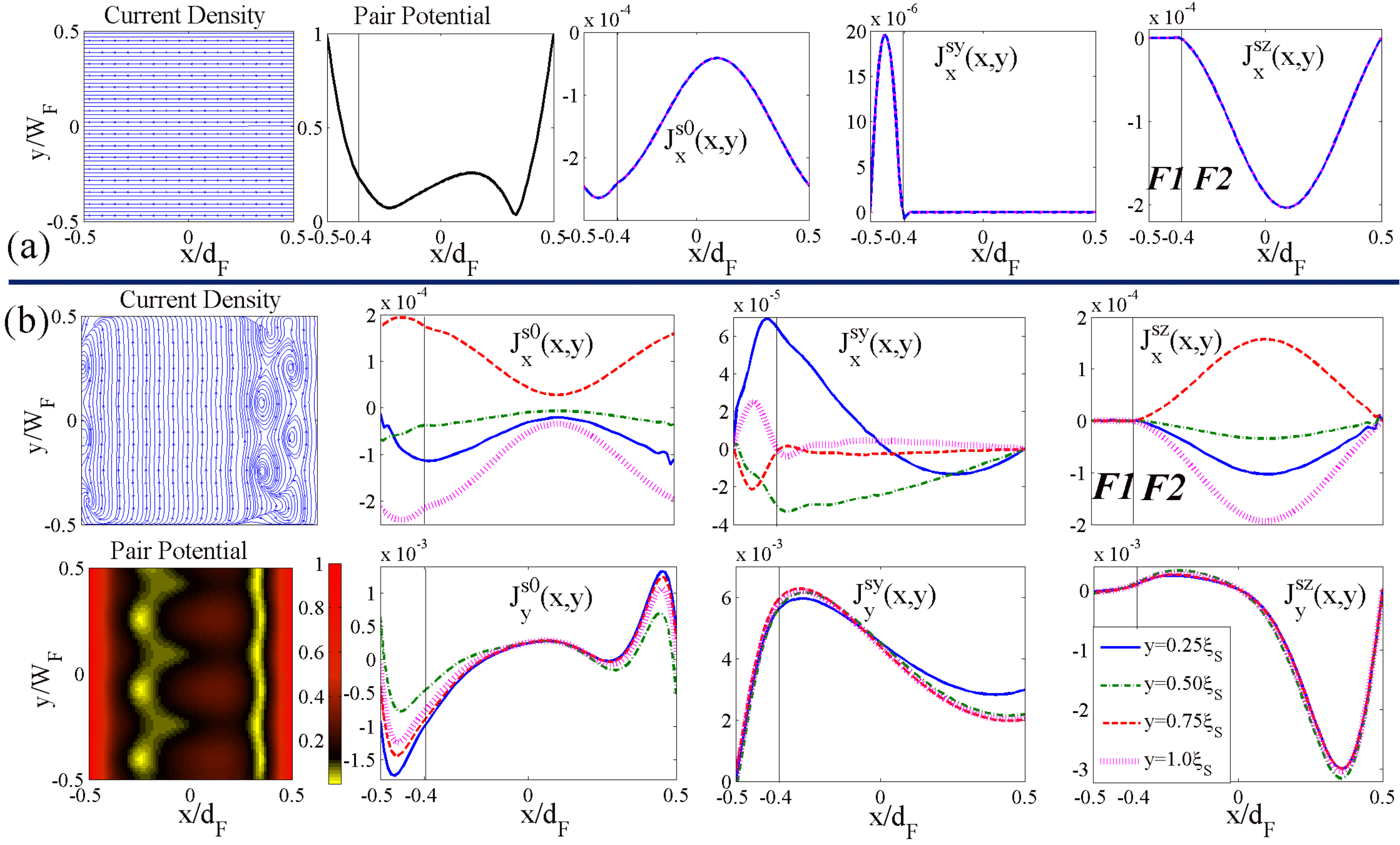} 
\caption{\label{fig:Jxy_37_Phi3pi_x} Spatial maps of
the total critical
current density, pair potential, and supercurrent
components in an \sffs Josephson junction for the configuration
depicted in Fig.~\ref{fig:model2}. The supercurrent is decomposed into
its even ($J^{s0}_{x,y}(x,y)$) and odd-frequency
($J^{sy}_{x,y}(x,y), J^{sz}_{x,y}(x,y)$) components.
The
two ferromagnetic layers have unequal lengths: $d_{F1}=0.2\xi_S$, and
$d_{F2}=1.8\xi_S$, while their widths are equal:
$W_{F1}=W_{F2}=2.0\xi_S$.
The vertical lines
identify the two ferromagnetic regions labeled $F_1$
and $F_2$.
In part (a), there is no applied magnetic field,
while in part (b) the external magnetic field
corresponds to a flux of $\Phi=3\Phi_0$.
The critical current components in both
cases, (a) and (b), are plotted as a function of $x$-position
at four different locations
along the
junction width: $y=0.25\xi_S, 0.50\xi_S, 0.75\xi_S, 1.0\xi_S$. }
\end{figure*}
Incorporating the spin-parametrization, Eq.~(\ref{eq:decomp}),
into the expression for the pair potential, Eq.~(\ref{eq:full_pair}),
we end up with the following compact equation:
\begin{equation}\label{eq:pair_prmtrz}
\tilde{U}_{\text{pair}}=\int_{-\infty}^{+\infty}d\varepsilon \Big\{
\mathbb{S}(\varepsilon)-\mathbb{S}(-\varepsilon)\Big\}\tanh(\varepsilon\beta).
\end{equation}
\end{widetext}
Thus, the pair potential involves only even-frequency
components of the parameterized Green's function, consistent
with the presence of a singlet order parameter in the left and right
superconducting banks.
In the next subsections, we solve the above
equations for two different magnetization configurations,
with and without an external magnetic field. The
spin-parametrization outlined here will
then delineate the various contributions
the singlet and triplet correlations make
to the supercurrent.

\subsection{\sffs heterostructures: parallel \ff and \fs
interfaces}\label{subsec:2D-sffs-pral}

Employing the spin-parametrization with the Usadel equations
together with the boundary conditions given in the previous section,
we numerically study the
proximity induced triplet correlations in a two-dimensional magnetic \sffs
Josephson junction. Here the \ff interface is
parallel with the outer $S$ interfaces. The
configuration is depicted in Fig.~\ref{fig:model2}.
The system is subject to an external
magnetic field $\vec{H}$ and we study the behavior of the corresponding
spin-triplet correlations.
We also compare these results
with the case of no external magnetic field.
As exhibited in Fig.~\ref{fig:model2}, the
external magnetic flux is directed along the $z$ direction, normal to
the junction face which resides in the $xy$ plane
so that the \fs and \ff interfaces are parallel to the $y$
axis.
We consider a rather general situation for the lengths and
magnetization directions of the $F$ strips. The lengths of the $F$ strips
can be unequal and are labeled by $d_{F1}$ and $d_{F2}$ ($d_{F1}\neq
d_{F2}$). The strength of the exchange fields in both $F$ regions are
equal $|\vec{h_1}|=|\vec{h_2}|=5.0\Delta_0$, while their orientations take arbitrary directions
$\vec{h}_{1,2}=(h^x_{1,2},h^y_{1,2},h^z_{1,2})$.
Also, the junction width
is equal to the width of the $F$ stripes, i.e., $W_{F1}=W_{F2}\equiv W_F$.

\begin{figure*}[]
\includegraphics[width=18cm,height=6.0cm]{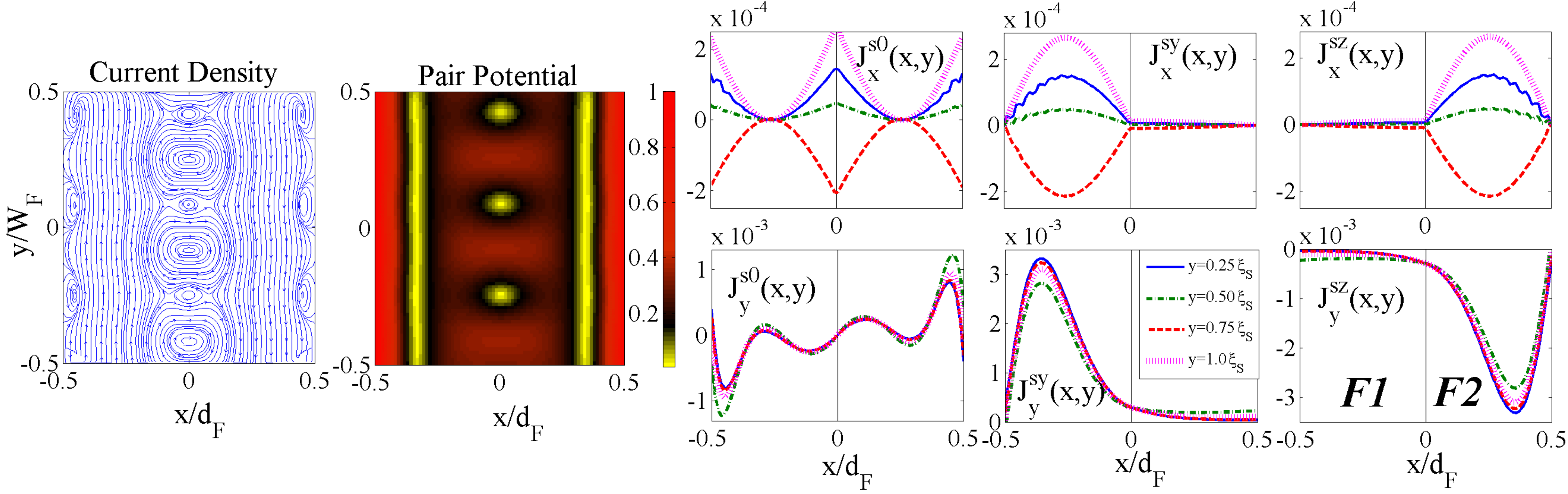}
\caption{\label{fig:jxy_11_Phi3pi_x} Spatial maps of
the critical
current density and pair potential as well as the critical charge
current components for a junction configuration
shown in Fig.~\ref{fig:model2}.
The same investigation as Fig.~\ref{fig:Jxy_37_Phi3pi_x}(b) is done here except we set
$d_{F1}=d_{F2}=1.0\xi_S$.
The junction width and external magnetic
flux are $W_{F}=2.0\xi_S$ and $\Phi=3\Phi_0$, respectively. The
vertical lines in the panels for the critical charge current components
show the spatial separation of the two magnetic regions labeled $F_1$ and
$F_2$.
The quantities
$J_{x}^{\beta}(x,y)$ show the current components for flow
along the $x$ direction whereas $J_{y}^{\beta}(x,y)$ represents
the components for flow along the $y$ direction.
Here $\beta$ denotes a specific component of the decomposed
charge supercurrent namely, $s0$, $sx$, $sy$, and $sz$.}
\end{figure*}

The charge supercurrent given by Eq.~(\ref{eq:current_param}) is
comprised of even-frequency singlet terms, $\mathbb{S}$, and the triplet
odd-frequency components,
$\vec{\mathbb{T}}$.
To study exactly the behavior of each
component of the Josephson charge current, we introduce the following
decomposition scheme which is
based on the discussions and notations in the previous sections:
\begin{subequations}
\begin{eqnarray}
&&\text{terms involving $\mathbb{S}$} \;\;\;\;\;\; \Rightarrow   \;\;\;\; J^{s0}_{x,y}, \\
&&\text{terms involving $\mathbb{T}_x$} \;\;\;\; \Rightarrow   \;\;\;\; J^{sx}_{x,y}, \\
&&\text{terms involving $\mathbb{T}_y$} \;\;\;\; \Rightarrow   \;\;\;\; J^{sy}_{x,y}, \\
&&\text{terms involving $\mathbb{T}_z$} \;\;\;\; \Rightarrow   \;\;\;\; J^{sz}_{x,y}.
\end{eqnarray}
\end{subequations}
In order to explicitly study
the influence of each component of the decomposed charge
supercurrent, we fix the magnetization of the
$F_1$ wire to be in the $y$-direction: $\vec{h}_1=(0,h^y,0)$.
Similarly, the magnetization of $F_2$ is orthogonal to
that of $F_1$, and oriented in the $z$ direction:
$\vec{h}_2=(0,0,h^z)$.
This orthogonal magnetization configuration is
an inhomogeneous magnetic state that results in effective
generation of equal-spin triplet correlations\cite{bergeret1}.

We show the spatial behavior
of the total charge supercurrent
density, pair potential, and the supercurrent
components for a phase difference of
$|\phi|=\pi/2$ in Fig.~\ref{fig:Jxy_37_Phi3pi_x} for the
\sffs junction depicted in Fig.~\ref{fig:model2}.
The geometric dimensions correspond to
$d_{F1}=0.2\xi_S$, $d_{F2}=1.8\xi_S$,
and the junction width $W_F$ is
set equal to $2.0\xi_S$.
The selected values of $d_F=d_{F1}+d_{F2}$
and $W_F$ are useful towards
understanding and analyzing
the spin-triplet
correlations in the different \sffs Josephson junctions
considered here.
For the panels found in (a),
the external magnetic flux is absent, while for those in (b),
a magnetic flux
$\Phi=3\Phi_0$ is
applied to the system.
In panels
(a), the charge current density is conserved
as exhibited by its spatial uniformity throughout the
ferromagnetic wire regions.
There is also no $y$ variations due the layered system
exhibiting
translational invariance in that direction.
Without
an external magnetic field, the charge supercurrent (and pair potential)
thus have no components along
the $y$ direction.
The pair potential is shown to be
an asymmetric function of
position along the junction, which is simply
due to the unequal length of each
$F$ wire and their
differing magnetizations.
The charge supercurrent density
components, $J_x^{s0}$, $J_x^{sy}$, and $J_x^{sz}$, are
also shown.
These quantities can only vary spatially
in the $x$-direction too since
in the absence of an external field,
the total supercurrent is generated from the phase difference $\phi$
between the $S$ terminals, which vary only along the junction length ($x$ direction).
The odd-frequency
triplet
components of the
supercurrent are localized in the $F_1$ or $F_2$
regions depending on the specific magnetization orientation in
each region (see also the discussion in Sec.~\ref{sec:1D}). It is evident that
$J_x^{sy}$ disappears in the right $F$ while $J_x^{sz}$ is
zero inside $F_1$ segment where $\vec{h}_1=(0,h^y_1,0),
\vec{h}_2=(0,0,h^z_2)$, respectively ($\vec{h}_1\rightarrow J^{sy}$,
$\vec{h}_2\rightarrow J^{sz}$). Thus $J^{sy}$ is
generated in $F_1$ and at the \ff interface is converted into $J^{sz}$ in
$F_2$, or vice versa. The two components are generated in one $F$
and do not penetrate into the other $F$ with orthogonal
magnetization.
Turning now to Fig.~\ref{fig:Jxy_37_Phi3pi_x}(b),
a finite magnetic flux
of $\Phi=3\Phi_0$
results now in a nonuniform supercurrent response that varies
in both the $x$ and $y$ directions.
Examining now the singlet correlations,
the
pair potential asymmetry
is again present
due to the spatially asymmetric magnetization regions, while
the external magnetic field induces vortices
with normal state cores along the junction
width.
The singlet and triplet components of
the supercurrent are plotted as
well.
The external magnetic field makes the
problem effectively two dimensional, as it induces
nonzero
current densities in the $y$ direction.
The
amplitude of the
equal-spin triplet component in the $x$-direction,
$J^{sy}_{x}(x,y)$, is
somewhat smaller
than the other components due in part to the small $F_1$ width,
$d_{F1}=0.2\xi_S$.
However,
comparing $J^{sy}_{x}(x,y)$ in panels (a) and (b),
we see that the presence of a magnetic field can result in
this triplet component
now existing in $F_2$, and it
can also be more prominent in $F_1$.

\begin{figure*}[]
\includegraphics[width=18cm,height=6.0cm]{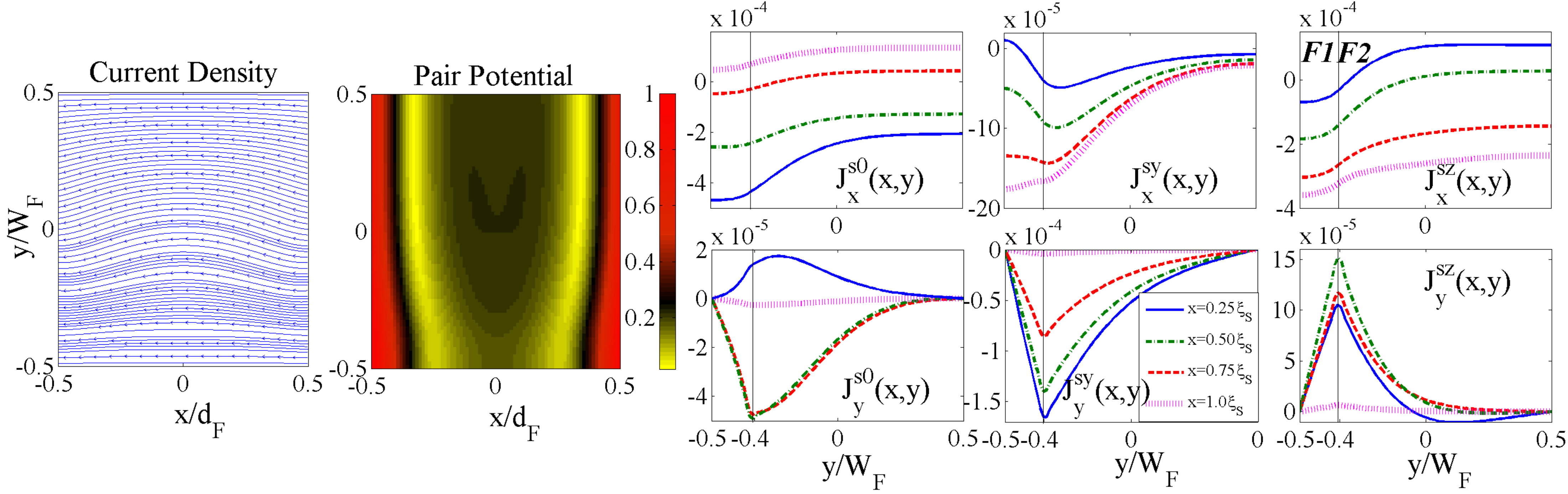} 
\caption{\label{fig:jxy_37_Phi0pi_y} Spatial maps of the
maximum total
charge current, pair potential, and the even-
($J^{s0}_{x,y}(x,y)$) and odd-frequency ($J^{sx}_{x,y}(x,y),
J^{sy}_{x,y}(x,y), J^{sz}_{x,y}(x,y)$) components of the charge current.
The configuration of the \sffs Josephson junction considered here is
shown in Fig. \ref{fig:model3}. The $F$/$F$ interface is now
perpendicular to the $S$/$F$ interfaces. The external magnetic flux
is zero, $\Phi=0$, and the junction length $d_F$ is equal to
$2.0\xi_S$.
The width of each ferromagnetic strip
are unequal, i.e., $W_{F1}=0.2\xi_S$ and $W_{F2}=1.8\xi_S$.
The current
components
are plotted against the lateral $y$
coordinate, i.e., along the
junction width.
Four various locations along the
junction length are considered: $x=0.25\xi_S$, $0.50\xi_S$,
$0.75\xi_S$, and $1.0\xi_S$. Vertical lines in the panels for the current
components separate the two ferromagnetic regions
along the $y$ direction (see Fig.~\ref{fig:model3}).}
\end{figure*}
Turning now to the $y$ components of the currents,
we see that
the magnitudes
have increased in
most instances by almost an
order of magnitude or more.
Although $J^{s0}_{y}(x,y),
J^{sy}_{y}(x,y)$, and $J^{sz}_{y}(x,y)$
vary little at the different $y$ locations,
the overall spatial
behavior is different for the odd
and even frequency triplet
components:
The equal spin $J^{sy}_{y}(x,y)$ penetrates considerably more
into the ferromagnetic regions compared with its $x$-component counterpart
and compared with the spin-0 triplet $J^{sz}_{y}(x,y)$.
These
results are consistent with the generation of a Meissner
supercurrent in $S$/$F$/$F$/$N$ structures.\cite{alidoust_missner}.

To study the effects of
varying the system geometry and its corresponding effects on
the triplet supercurrents, we consider more symmetric $F$
regions
with equal lengths
$d_{F1}=d_{F2}=1.0\xi_S$.
All other parameters are
identical to the previous \sffs junction in Fig.~\ref{fig:Jxy_37_Phi3pi_x}(b).
The first two panels on the left of Fig.~\ref{fig:jxy_11_Phi3pi_x} illustrate
2D spatial maps of the total supercurrent density and pair potential.
The remaining panels contain the spatial behavior of
the singlet and triplet contributions to the total current.
The effect of equal layer widths is reflected in the
regular array of vortex patterns and circulating currents in the
junction.
The pair potential at $x=0$ exhibits three zeros
when $\Phi=3\Phi_0$. This is
in accordance with the analytic expression in Eq.~(\ref{eq:pair_analytic}) and Eq.~(\ref{eq:pair_prmtrz}) which demonstrates
the actual behavior of singlet correlations, constituting the spatial profile of proximity pair potential.
To reveal the even and odd frequency contributions
to the contour plots, we
examine in the remaining panels the
supercurrent components as a function of $x$.
The vertical lines separate the two
ferromagnetic regions labeled by $F_1$ and $F_2$, as
shown in the schematic of Fig.~\ref{fig:model2}.
We see that the formerly
long-ranged $J_{y}^{sy}(x,y)$ and $J_{y}^{sz}(x,y)$, in
Fig.~\ref{fig:Jxy_37_Phi3pi_x}
now vanish here when $d_{F1}=d_{F2}=1.0\xi_S$.
\begin{figure}[b!]
\includegraphics[width=7.50cm,height=4.5cm]{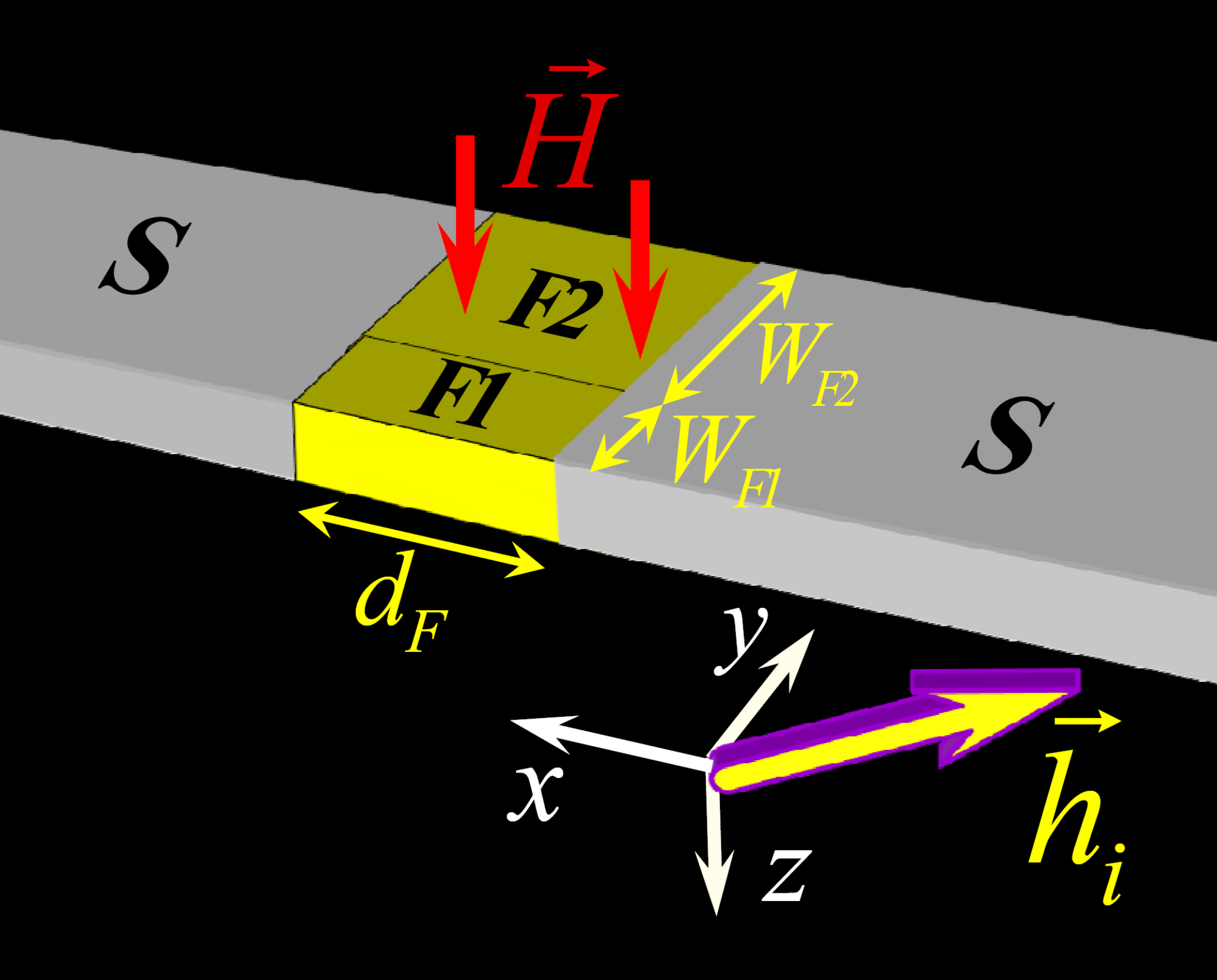}
\caption{\label{fig:model3}
Schematic of the \sffs
Josephson junction where the interface of the
double ferromagnet region is oriented
perpendicular to the $S$/$F$ interfaces.
The junction is located in
the $xy$ plane and the $S$/$F$ interfaces are along the $y$ axis while
the $F$/$F$ interface is along the $x$ axis.
The width of
the rectangular $F$ strips $W_{F1}$, and $W_{F2}$
are generally unequal, i.e., $W_{F1}\neq W_{F2}$.
The
lengths of the $F$ strips, $d_F$, however, are the same
and correspond to the junction length.
An external magnetic field
$\vec{H}$ is applied normal
to the junction plane along the $z$ axis. }
\end{figure}
This
demonstrates
the crucial role
that geometry can play in these type of junctions,
in particular the existence of odd-frequency correlations
tend to favor
configurations where the length
of the ferromagnetic layers are unequal, for instance, $d_{F2}\gg
d_{F1}$ can further enhance the effect.
This finding is also
consistent with
the results of a
$S$/$F$/$F$/$N$ configuration subjected to an external magnetic
field,\cite{alidoust_missner} where the $J_{x}^{sy}$ component of the
Meissner current is optimally anomalous if $d_{F2}\gg d_{F1}$.
The behavior of the singlet and triplet correlations
discussed here are typically highly dependent on the
magnetization of the double $F$ layer,
as well as the presence of an external magnetic field.
In the next section,
we therefore proceed to investigate
another
type of \sffs Josephson junction with
a different double $F$ layer arrangement.
\begin{figure*}[]
\includegraphics[width=18cm,height=6.0cm]{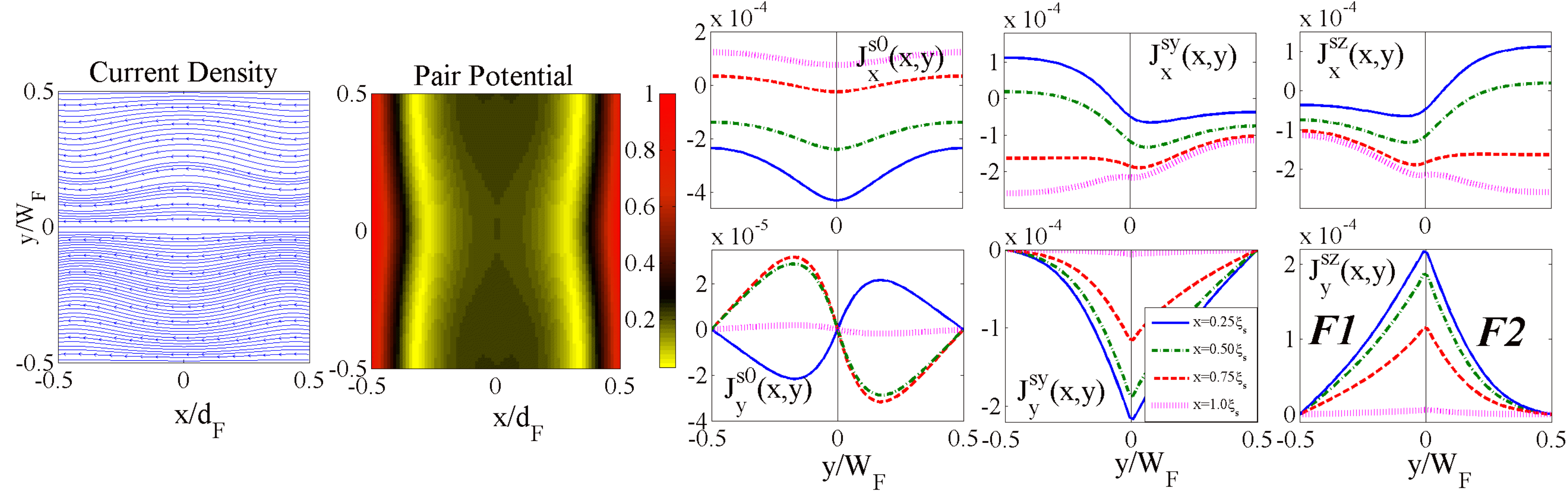} 
\caption{\label{fig:jxy_11_Phi0pi_y}
Spatial maps depicting a vector plot of the total
charge current density and density plot
of the pair potential for the
system
shown in Fig.~\ref{fig:model3}. The
remaining line plots correspond to the charge supercurrent
components: the singlet, $J^{s0}_{x,y}(x,y)$,
and triplets, $J^{sy}_{x,y}(x,y), J^{sz}_{x,y}(x,y)$. Here the
system parameters are identical
to those used previously in Fig.~\ref{fig:jxy_37_Phi0pi_y},
except the width of the ferromagnetic strips here are
equal: $W_{F1}=W_{F2}=1.0\xi_S$.
The junction length $d_F$
is equal to $2.0\xi_S$ and there is no external magnetic flux, $\Phi=0$.
The vertical lines indicate the
locations of the
$F$/$F$ interfaces
in between
the two ferromagnetic regions $F_1$ and $F_2$. }
\end{figure*}

\subsection{\sffs heterostructures: \ff interface
perpendicular to \fs interfaces}\label{subsec:2D-sffs-perp}

Here we consider a type of \sffs Josephson junction where the
previous ferromagnet double layer system is rotated by $\pi/2$,
so that the
\ff interface is normal to the outer \fs interfaces.
This layout is
shown in Fig.~\ref{fig:model3}.
As before,
we take the length of the ferromagnetic strips to be equal,
$d_{F1}=d_{F2}\equiv d_F$.
We first consider ferromagnets where
the width of the $F_1$ and $F_2$ regions are unequal:
$W_{F1}\neq W_{F2}$.
When
an applied
magnetic field, $\vec{H}$, is present, it is directed
along the $z$-axis, normal to the
plane of the system. To be consistent
with the previous subsection, we study two
regimes of ferromagnetic widths namely, $W_{F1}\ll W_{F2}$ and
$W_{F1}=W_{F2}$, and investigate
how the results vary for a finite external
magnetic flux $\Phi$.

\begin{figure*}[]
\includegraphics[width=18cm,height=6.0cm]{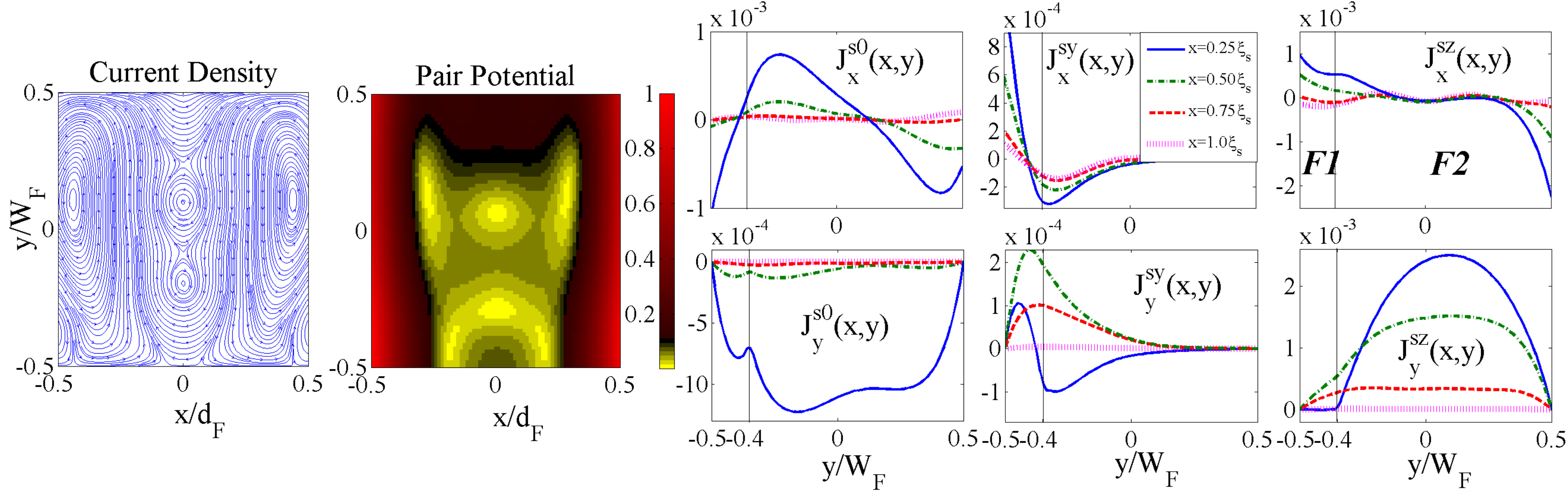} 
\caption{\label{fig:jxy_37_Phi3pi_y} Spatial maps of total critical
charge current density and pair potential and the decomposed
components of charge supercurrent $J^{s0}_{x,y}(x,y),
J^{sx}_{x,y}(x,y), J^{sy}_{x,y}(x,y), J^{sz}_{x,y}(x,y)$. The same
parameter set as Fig.~\ref{fig:jxy_37_Phi0pi_y} are used except we
now assume the system is subject to an external magnetic flux
$\Phi=3\Phi_0$. The vertical lines in the current components' panels
separate the two ferromagnetic segments, labeled by $F_1$, and
$F_2$, with orthogonal magnetization orientations. The current
components are plotted as a function of positions along the junction
width i.e. $y$ direction at four locations along the junction length
i.e. $x=0.25\xi_S$, $0.50\xi_S$, $0.75\xi_S$, and $1.0\xi_S$.}
\end{figure*}

In Fig.~\ref{fig:jxy_37_Phi0pi_y},
we show
the results for $W_{F1}=0.2\xi_S$,
and $W_{F2}=1.8\xi_S$, in the
absence of an applied magnetic field ($\Phi=0$).
First, the spatial
map of the
total maximum charge supercurrent shows that the current near
the \ff junction has a non-zero $y$-component. This is in
contrast to the uniform supercurrent shown in
Fig. \ref{fig:Jxy_37_Phi3pi_x}(a) for a
``parallel" \ff segment.
The induced $y$ component is
present over the entire junction width as exhibited by the
distorted
quasiparticle paths. The pair potential shows a symmetric behavior
along the junction length (the $x$ direction). However, similar to
the previous case, the pair potential is an asymmetric function of
$y$ coordinate. This again arises from the unequal
ferromagnetic wire widths ($W_{F1}\neq W_{F2}$) and their different
magnetization orientations.
The spatial behavior of the triplet and singlet contributions to the
supercurrent is
also shown
as a function of position along the junction width (the $y$ direction)
at four $x$ locations: $x=0.25\xi_S, 0.5\xi_S, 0.75\xi_S, 1.0\xi_S$.
The
amplitude of $J_{y}^{s0}$ is
around $\approx 1/10$ of $J_{y}^{sy}$
and $J_{y}^{sz}$ but comparable with $|J_{y}^{sy}+J_{y}^{sz}|$.
These components of $J_y$, and hence $J_y$ itself,
vanish
at $y=\pm 0.5$, corresponding to the vacuum boundary,
and consistent with the boundary conditions given by Eq.~(\ref{eq:bc}).
The vector plot
of the charge supercurrent
in Fig.~\ref{fig:jxy_37_Phi0pi_y} reveals
no current flow along $y$ in the middle
of the junction, $x=0$, throughout the junction width.
The $x$ component, $J_{x}^{s0}$, is most negative at
$x=0$ along the $y$ axis
while $J_{x}^{sy}$ and $J_{x}^{sz}$ are maximal. Therefore, the
singlet-triplet conversion  with
$W_{F2}\gg W_{F1}$ is maximal at $x=0$. Another important aspect of
this \sffs junction is seen in the behavior of $J_{x}^{sy}$
and $J_{x}^{sz}$ as a function of $y$: the two
components are generated in one $F$ region and penetrate
deeply into
the adjacent $F$ segment.
By comparing these plots with those of a
\sffs configuration where \ff interface is parallel with the \fs
interface subjected to an external magnetic field, presented in Fig.
\ref{fig:Jxy_37_Phi3pi_x}(b), one concludes that the spin-1 triplet
components of charge supercurrent can penetrate
the ferromagnetic regions when they flow parallel to
the junction interfaces.

In Fig.~\ref{fig:jxy_11_Phi0pi_y},
we investigate the geometrical effects
on the superconducting properties
for a system with
$F_1$
and
$F_2$ of equal width.
We have
$W_{F1}=W_{F2}=1.0\xi_S$, and
all previous parameters remain intact.
The current density map shows
that the
the induced $y$-component is equally distributed with respect to
$y=0$, the location of the \ff interface.
The pair potential
is a symmetric function of both the $x$ and $y$ coordinates
about the origin,
reflecting
the symmetric geometric configuration.
The panels representing the current components
reveal that the $y$-component to the current density
originates mainly from the singlet term, $J_{y}^{s0}$,
since it is evident that the odd-frequency components
$J_{y}^{sy}$ and $J_{y}^{sz}$ are
of opposite sign so that
$J_{y}^{sy}+J_{y}^{sz}\approx 0$.
Note that $J_{y}^{s0}$ is an odd function of $y$ for each fixed $x$-location.
This reflects the fact that the junction is symmetric along
the $y$ direction (see Fig. \ref{fig:jxy_37_Phi0pi_y}). From the
current density spatial map the same functionality appears for the
$x$ coordinate. The integration $\int_{-0.5}^{0.5}dx J_{y}^{s0}$
gives the total charge supercurrent flowing in the $y$ direction
which is therefore zero due since $J_{y}^{s0}$ is antisymmetric in $y$.
For the previous case shown in Fig.~\ref{fig:jxy_37_Phi0pi_y},
both triplet components
$J_{y}^{sy}$ and $J_{y}^{sz}$ contributed to the induced $y$
component in the total current density.
Considering now the flow of current along the $x$ direction,
the top panels display the
behavior of the odd-frequency
triplet components with spin-1 and spin-0 projections
on the $z$ quantization axes ($J_{x}^{sy}$ and
$J_{x}^{sz}$, respectively). Interestingly,
the generation of each of the two triplet
components tends to
mirror each others behavior.
Since the ferromagnetic regions are
symmetric, the singlet-triplet conversion of charge supercurrent
components at the middle of junction ($x=0$) across the junction is
not as extensive
compared to the previous case where $W_{F2}\gg W_{F1}$. It is
clear that the total current passing through the junction must be
conserved along the junction length ($I_{tot}(x)=const.$). However,
the singlet-triplet conversion in the charge supercurrent
density is maximal at $x=0$ and increases as the condition $W_{F2}\gg
W_{F1}$ is fulfilled.

In the previous two figures, the source of the driving current was the
macroscopic
phase differences
between the $S$ terminals in the Josephson structures.
We now introduce an additional source and
associated triplet correlations by
applying an external magnetic field (with corresponding
flux $\Phi=3\Phi_0$),
normal to the junction plane along the $z$-axis (see schematic, Fig.~\ref{fig:model3}).
Therefore,
Fig.~\ref{fig:jxy_37_Phi3pi_y}
exhibits our results for a junction with the same parameters
used in Fig.~\ref{fig:jxy_37_Phi0pi_y} but
now, $\Phi=3\Phi_0$.
The magnetic field causes
the quasiparticles to undergo circular motion
as shown in the vector plot
for the supercurrent density.
The pair potential also vanishes
at particular locations,
which now form a nontrivial pattern.
The charge
supercurrent components are clearly
modified by the external magnetic
field compared to $\Phi=0$ (Fig.~\ref{fig:jxy_37_Phi0pi_y}).
We have found that the behavior of the supercurrent
components
along $y$ are
similar to the corresponding components
along
$x$ when the \ff
junction is parallel with the \fs interfaces (Fig.~\ref{fig:Jxy_37_Phi3pi_x}(b)).
In both cases, the supercurrent
flows normal to the \ff junction.
Likewise, the $J_x$ component
as a function of
$y$ in Fig.~\ref{fig:jxy_37_Phi0pi_y} is similar  to
the $J_y$ component vs. $x$ as seen in Fig.~\ref{fig:Jxy_37_Phi3pi_x}(b).
Recall
that in Fig. \ref{fig:jxy_37_Phi3pi_y}, we have $\Phi=3\Phi_0$, while in Fig.
\ref{fig:Jxy_37_Phi3pi_x}(b) $\Phi=0$.
Here we find also that the
triplet components of
the charge supercurrent generated in one $F$
strip, flowing parallel with the \ff interface deeply penetrate
the adjacent $F$
compared to when the current
flows normal to the \ff interface.
It is also evident that the total net
supercurrent in the $y$ direction is zero, i.e.,
$\int_{-0.5}^{+0.5}dx(J_{y}^{s0}+J_{y}^{sy}+J_{y}^{sz})=0$.

\begin{figure*}[]
\includegraphics[width=18cm,height=6.0cm]{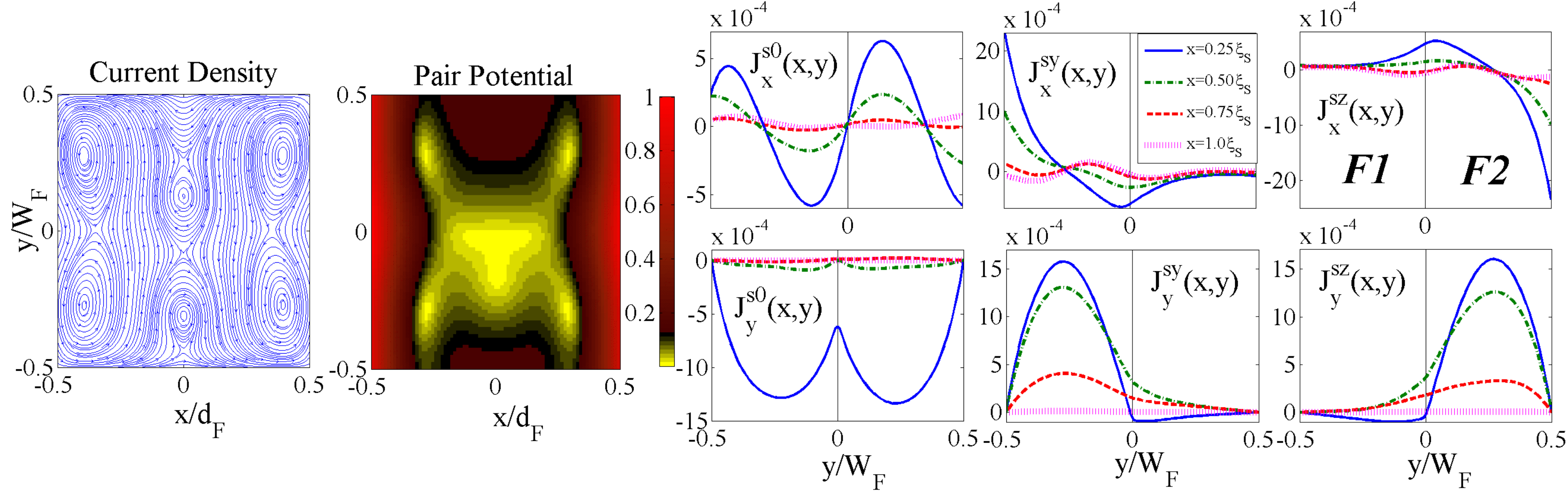}
\caption{\label{fig:jxy_11_Phi3pi_y} Spatial maps of the total charge
supercurrent and pair potential in addition to the current
components $J^{s0}_{x,y}(x,y), J^{sx}_{x,y}(x,y), J^{sy}_{x,y}(x,y),
J^{sz}_{x,y}(x,y)$, at maximum charge current
flowing across the junction. The same investigations as Fig.~\ref{fig:jxy_11_Phi0pi_y}
are presented here. The junction length is
fixed at $d_F=2.0\xi_S$ while the width of each ferromagnetic strip,
labeled by $F_1$ and $F_2$, are $W_{F1}=W_{F2}=1.0\xi_S$. Unlike
Fig.~\ref{fig:jxy_11_Phi0pi_y}, the external magnetic flux is
$\Phi=3\Phi_0$. The two ferromagnetic regions with orthogonal
magnetization orientations are separated by vertical lines and
labeled $F_1$ and $F_2$.}
\end{figure*}

Upon
changing
the width
of the
$F$ layers to $W_{F1}=W_{F2}=1.0\xi_S$,
Fig.~\ref{fig:jxy_11_Phi3pi_y} shows
that the circular paths observed in the
current density and zeroes in the spatial map of the pair
potential have reverted back to
a more symmetric configuration compared to
Fig.~\ref{fig:jxy_37_Phi3pi_y}.
Here the
magnetization orientations in the $F$ regions are orthogonal:
$\vec{h}_{1}=(0,h^y,0)$, and $\vec{h}_{2}=(0,0,h^z)$.
The panels containing the current
components $J_{x}^{sy}$ and $J_{x}^{sz}$ as a function of $y$
are now much more localized in the
$F$ regions, in
contrast to the previous case where $W_{F2}\gg W_{F1}$
(Fig.~\ref{fig:jxy_37_Phi0pi_y}).
Considering the current density along the $y$ direction,
the triplet components are also localized in the $F$ regions.
Similarities are observed
when comparing their $y$ dependence with the $x$ dependence of
the triplets
$J_{x}^{sy}$ and $J_{x}^{sz}$ in Fig.~\ref{fig:jxy_11_Phi3pi_x}.
This comprehensive
investigation
into the different
\sffs structures
has shown that
an applied magnetic field
can result
in the appearance
of odd-frequency triplet components to
the charge
transport
when the
supercurrent is parallel to the \ff interface.
These odd-frequency correlations
can exhibit extensive penetration into ferromagnetic regions
with
orthogonal magnetizations.
These findings are in stark contrast to those cases where the supercurrent
flows normal to uniform \ff double layers with orthogonal
magnetization orientations (see also the discussions of one-dimensional systems in Sec.~\ref{sec:1D}).

\subsection{Spin valve structure probing of
equal-spin triplet supercurrent} \label{subsec:2D-snpvlv}

\begin{figure}[b]
\centerline{\includegraphics[width=8.50cm,height=4.cm]{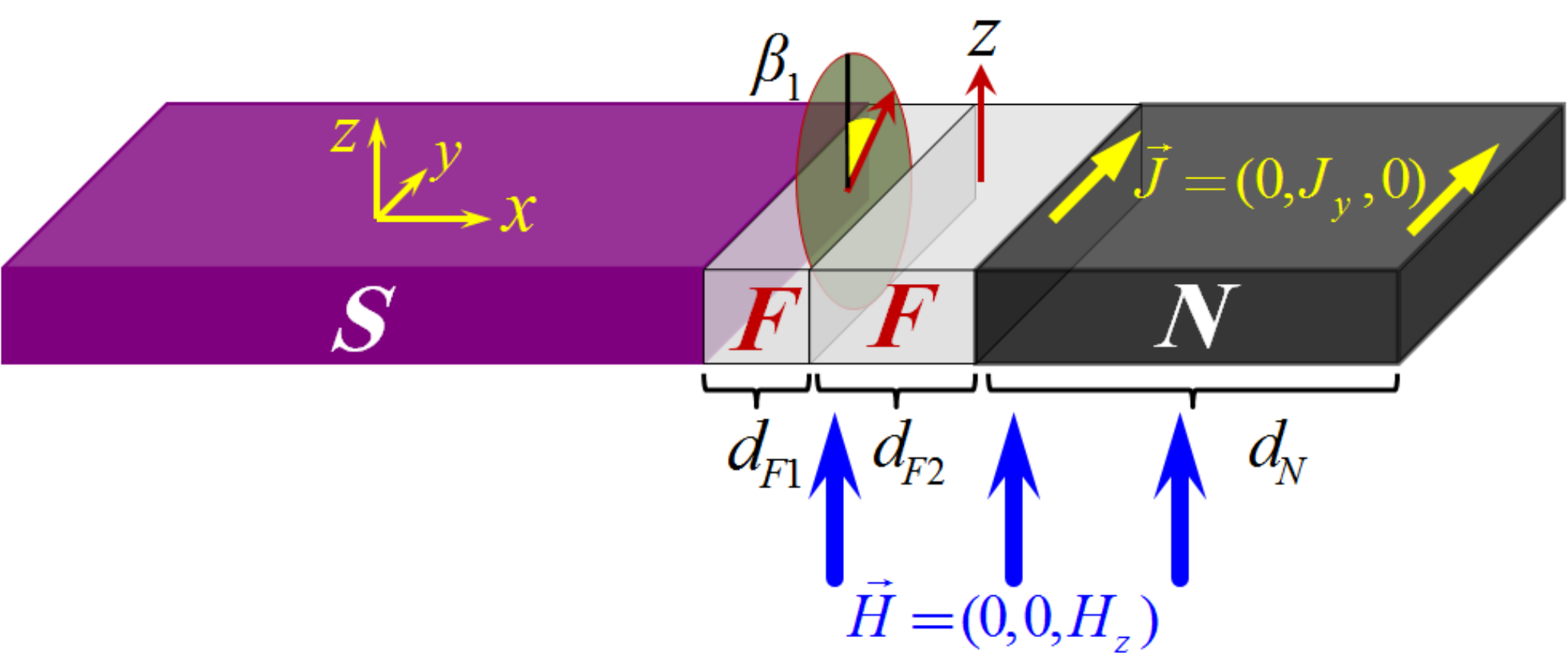}}
\caption{\label{fig:model4}(Color online) Schematic of the proposed
$S/F/F/N$ junction. The magnetization orientation of the middle $F$
layer with width $d_{F2}$ is fixed along the $z$ direction. The left
$F$ layer, with width $d_{F1}$, has a misaligned magnetization
direction corresponding to
$\vec{h}_1=h_0(\cos\pi/4,\sin\pi/4\sin\beta_1,\cos\beta_1)$. The
normal metal layer has width $d_N$. An external magnetic field is
applied to the system in the $z$ direction parallel to the
interfaces: $\vec{H}=(0, 0, H_z)$. Due to the external magnetic
field, a diamagnetic supercurrent flows in the $y$ direction
$\vec{J}(x)=(0, J_y(x), 0)$ and varies along $x$. The junction
interfaces lie in the $yz$ plane.}
\end{figure}

Having analyzed in detail the singlet-triplet contributions
to the supercurrent in various situations, we now proceed
to demonstrate an experimentally accessible structure that can directly detect
our predictions involving parallel transport in \ff bilayers.
We convert one of the outer $S$ terminals to
a finite sized normal layer, so that in effect  we consider
a simple
$S/F/F/N$ spin valve (Fig.~\ref{fig:model4}).
This structure
generates
pure odd-frequency spin-1 triplet correlations, and can be a
more experimentally
accessible system for generating and controlling triplet
correlations in supercurrent transport.\cite{al1}
The basic structure is made of two uniform
ferromagnetic layers with thicknesses $d_{F1}$, and $d_{F2}$, and a
relatively thick normal layer $d_{N}$, which is connected to a
superconducting terminal.
The $N$ layer assists in probing the equal-spin component of the
supercurrent efficiently since the exchange field there is zero, and this component of
supercurrent decays very slowly in the
normal metal.
\begin{figure*}[]
\centerline{\includegraphics[width=18.0cm,height=5.cm]{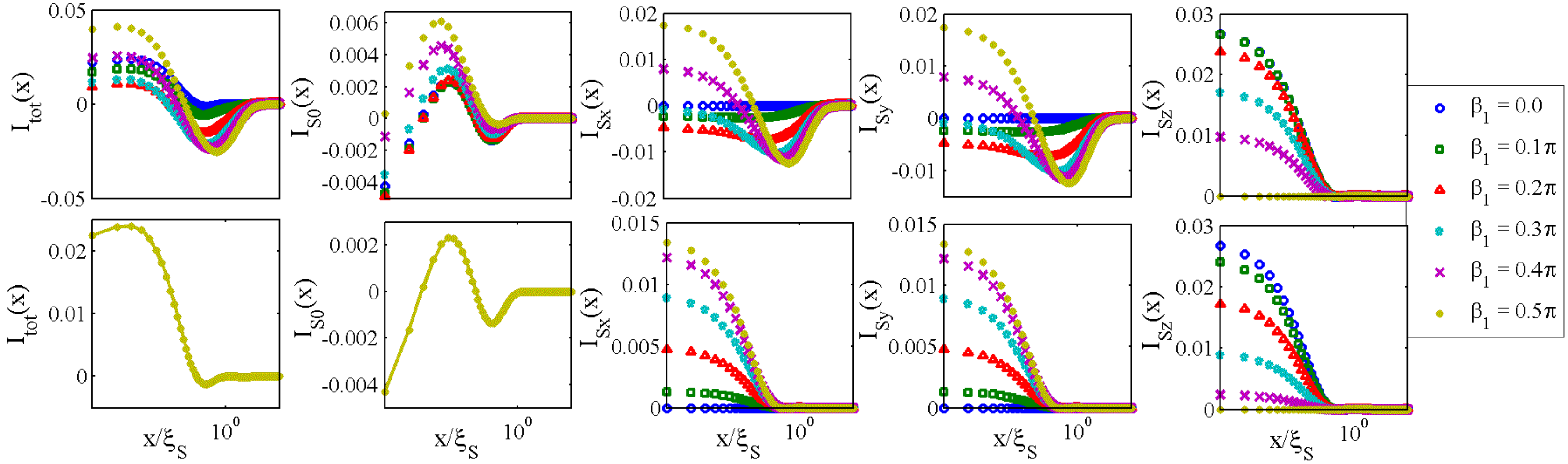}}
\caption{\label{fig:s_f_f_n}(Color online) \textit{Top row}: Spatial
behavior of the total supercurrent and its components as a function
of position inside the $S/F/F/N$ junction for various magnetization
orientations in $F_1$:
$\vec{h}_1=h_0(\cos\pi/4\sin\beta_1,\sin\pi/4\sin\beta_1,\cos\beta_1)$
for $\beta_1=0, 0.1\pi, 0.2\pi, 0.3\pi, 0.4\pi$, and $0.5\pi$. The
thickness of the left $F$ layer is fixed at $d_{F1}=0.15\xi_S$,
while the thickness of the right $F$ layer is equal to
$d_{F2}=1.85\xi_S$. \textit{Bottom row}: The left and right
thicknesses in the $F$ layers are now equal to
$d_{F1}=d_{F2}=\xi_S$. The sum of $d_{F1}$ and $d_{F2}$ is
identical for both cases investigated in the top and bottom rows.
The normal layer thickness is also fixed at $2.5\xi_S$ in both
cases. The logarithmic scales on the $x$-axes permits easier
visualization of the long-ranged spatial behavior of the odd-triplet
components that have spin projections normal to the $z$ quantization
axis.}
\end{figure*}
The magnetization of the left $F$ is determined by a rotation angle
$\beta_1$ with respect to the $z$ axis
($\vec{h}_1=h_0(\cos\pi/4\sin\beta_1,\sin\pi/4\sin\beta_1,\cos\beta_1)$)
while the magnetization of right $F$ layer is assumed to be fixed
along the $z$ axis. The non-superconducting part of the spin-valve is
subject to an external magnetic field $\vec{H}$ oriented along the
$z$ axis, $H_z$. The external magnetic field leads to a diamagnetic
supercurrent $\vec{J}$, which depends on $x$ and flows along the $y$
direction [denoted by $J_y(x)$] parallel to the interfaces. In the
presence of an external magnetic field, we make the usual
substitution $\hat{\partial}\rightarrow \vec{\nabla}+e\vec{A}/c$,
where $\vec{A}$ is the vector potential, related to the external
magnetic field ($\vec{H}$) via  $\vec{\nabla}\times\vec{A}=\vec{H}$.
In the linear response
regime\cite{sus1,sus2,sus3,sus4,sus5,sus6,Yokoyama_missner}, the supercurrent
density can be expressed by:
\begin{eqnarray}\label{eq:currentdensity_y}
&&\nonumber {J}_y(x) =-J_08ieA_y(x)\times\\\nonumber &&
\int_0^{\infty} d\varepsilon \tanh\varepsilon\beta\left\{
\mathbb{S}(\varepsilon,x)\mathbb{S}^{\ast}(-\varepsilon,x) -
\mathbb{S}(-\varepsilon,x)\mathbb{S}^{\ast}(\varepsilon,x)
+\right.\\\nonumber && \left.
\mathbb{T}_x(-\varepsilon,x)\mathbb{T}_x^{\ast}(\varepsilon,x) -
\mathbb{T}_x(\varepsilon,x)\mathbb{T}_x^{\ast}(-\varepsilon,x)+\right.\\\nonumber
&&
\left.\mathbb{T}_y(-\varepsilon,x)\mathbb{T}_y^{\ast}(\varepsilon,x)
-\mathbb{T}_y(\varepsilon,x)\mathbb{T}_y^{\ast}(-\varepsilon,x)+\right.\\
&& \left.
\mathbb{T}_z(-\varepsilon,x)\mathbb{T}_z^{\ast}(\varepsilon,x) -
\mathbb{T}_z(\varepsilon,x)\mathbb{T}_z^{\ast}(-\varepsilon,x)
\right\}.
\end{eqnarray}
Strictly speaking, to determine $A_y(x)$, it is necessary to solve
Maxwell's equation incorporating the Coulomb gauge in
conjunction with the expression for $J_y(x)$ and appropriate
boundary conditions for $H$
\cite{sus1,sus2,sus3,sus4,sus5,sus6,Yokoyama_missner}. We have found however
that through our extensive numerical investigations, $A_y(x)$ is
typically a linear function of $x$ and weakly varies
with the magnetization alignment, $\beta_1$.
Therefore, it is the energy and spatial dependence of the Green's
function components in Eq.~(\ref{eq:currentdensity_y}) which governs
the supercurrent density.
To study the supercurrent behavior, and similar to what was done
above for Josephson structures, we separate out the even- and
odd-frequency components of the total supercurrent the same as Sec.~\ref{subsec:1D-spin}.
We achieve this by first defining
$I_{tot}(x)$=$I_{S0}(x)$+$I_{Sx}(x)$+$I_{Sy}(x)$+$I_{Sz}(x)$, where
$I_{S0}(x)$, $I_{Sx}(x)$, $I_{Sy}(x)$, and $I_{Sz}(x)$ are given by,
\begin{subequations}\label{eq:currentdensity_comps}
\begin{eqnarray}
&& I_{S0}(x) =\nonumber\\&&\int_0^{\infty} d\varepsilon \left\{
\mathbb{S}(\varepsilon)\mathbb{S}^{\ast}(-\varepsilon) -
\mathbb{S}(-\varepsilon)\mathbb{S}^{\ast}(\varepsilon)\right\}\tanh\varepsilon\beta,
\\ && I_{Sx}(x) =\nonumber\\&&\int_0^{\infty} d\varepsilon \left\{
\mathbb{T}_x(-\varepsilon)\mathbb{T}_x^{\ast}(\varepsilon) -
\mathbb{T}_x(\varepsilon)\mathbb{T}_x^{\ast}(-\varepsilon)\right\}\tanh\varepsilon\beta,
\\ && I_{Sy}(x) =\nonumber\\&&\int_0^{\infty} d\varepsilon \left\{
\mathbb{T}_y(-\varepsilon)\mathbb{T}_y^{\ast}(\varepsilon)
-\mathbb{T}_y(\varepsilon)\mathbb{T}_y^{\ast}(-\varepsilon)\right\}\tanh\varepsilon\beta,
\\ && I_{Sz}(x) = \nonumber\\&&\int_0^{\infty} d\varepsilon \left\{
\mathbb{T}_z(-\varepsilon)\mathbb{T}_z^{\ast}(\varepsilon) -
\mathbb{T}_z(\varepsilon)\mathbb{T}_z^{\ast}(-\varepsilon) \right\}
\tanh\varepsilon\beta.
\end{eqnarray}
\end{subequations}
Also the boundary conditions at the right far end interface of the
normal metal should be modified accordingly (see
Fig.~\ref{fig:model3}):
\begin{subequations}
\begin{eqnarray}\label{bc_N}
&&\partial_x  (\mp \mathbb{T}_{x}(-\varepsilon)+i \mathbb{T}_{y}(-\varepsilon)) =0,\\
&&\partial_x  (\mp\mathbb{S}(-\varepsilon)+\mathbb{T}_z(-\varepsilon)) =0,\\
&&\partial_x  (\mp\mathbb{T}_{x}^{\ast}(\varepsilon)-i\mathbb{T}_{y}^{\ast}(\varepsilon))=0,\\
&&\partial_x
(\mp\mathbb{S}^{\ast}(\varepsilon)+\mathbb{T}_{z}^{\ast}(\varepsilon))
=0.
\end{eqnarray}
\end{subequations}

Having now established the method in which to determine the spatial
behavior of the supercurrent and its decompositions, we present in
Fig.~\ref{fig:s_f_f_n} our findings for the proposed nanovalve
subject to an external magnetic field. Several values of the left
magnetization orientation $\beta_1$ are investigated (the right $F$
magnetization is fixed along the $z$ axis), corresponding to
$\beta_1/\pi=0, 0.1, 0.2, 0.3, 0.4, 0.5$. In the top row, the
thickness of the left $F$ layer is much smaller than the thickness
of the adjacent $F$ layer: $d_{F1}(=0.15\xi_S)\ll
d_{F2}(=1.85\xi_S)$. The thickness of the normal metal is of
moderate size, set equal to $d_{N}=2.5\xi_S$. The total
non-superconducting region thus has a thickness of $4.5\xi_S$. To
demonstrate the long-ranged nature of the spin-1 triplet
components $I_{Sx}(x)$ and $I_{Sy}(x)$ in this simple structure with
only two magnetic regions with uniform and non-collinear
magnetization orientations, we introduce a three dimensional
magnetization in the left $F$ region:
$\vec{h}_1=h_0(\cos\pi/4\sin\beta_1,\sin\pi/4\sin\beta_1,\cos\beta_1)$.
To clearly show the spatially long-ranged components, we use a
logarithmic scale for each $x$-axis. It is seen that the triplet
component with zero spin projection on the $z$-axis, $I_{Sz}(x)$, is
short-ranged and becomes suppressed at positions approximately
greater than $x\sim 0.5\xi_S$. This is in drastic contrast to the
$I_{Sx}(x)$, and $I_{Sy}(x)$ components, which  propagate over the
entire non-superconducting region. The singlet component,
$I_{S0}(x)$, vanishes around $x\sim \xi_S$, which is about half the
 thickness of the magnetic layers
($d_{F1}+d_{F2}=2 \xi_S$). Thus, the long-ranged components can
strongly influence the total supercurrent, $I_{tot}(x)$, and be the
primary source of nonzero supercurrent in the normal metal. These
findings are therefore suggestive of an experimentally accessible
$S/F/F/N$ spin valve with long-ranged triplet correlations
controllable by magnetization rotation in one of the $F$ layers.

It turns out that the the pure spin-1 odd-frequency spin-valve
effect is optimal for asymmetric $F$ layer thicknesses. To
illustrate this, we consider a symmetric configuration with
$d_{F1}=d_{F2}=\xi_S$, which ensures the same total thickness of
the ferromagnet regions (the normal metal thickness is unchanged at
$d_{N}=2.5\xi_S$).
The results of this configuration are represented in the bottom row
of Fig.~\ref{fig:s_f_f_n}. The total current and its components are
plotted as a function of position inside the non-superconducting
regions for the same magnetization orientations in the asymmetric
case (top row).
We find that in contrast to the panels in the top row, the triplet
components $I_{Sx}(x)$, $I_{Sy}(x)$, and $I_{Sz}(x)$ are now all
short-ranged in the sense that they vanish near $x \sim 0.5\xi_S$.
In particular, the spin-1 components, $I_{Sx}(x)$, and $I_{Sy}(x)$
do not propagate in $F_2$, where the magnetization is
orthogonal to the correlations' spin orientations. The singlet
contribution, $I_{S0}(x)$, for both the $d_{F2}\gg d_{F1}$ and
$d_{F2}=d_{F1}$ cases shows similar behavior, where it vanishes at
around $x \sim \xi_S$. This can be understood by noting that
the spin-splitting effects of ferromagnetism destroy opposite-spin
paring correlations.
The total current also vanishes at $x \sim \xi_S$ for all
magnetization orientations, $\beta_1$, in contrast to what was
observed in the $d_{F2}\gg d_{F1}$ structure above.

We are now able to compare the behavior of various supercurrent
components in the \sffn and the \sffs Josephson structures presented in
Sec.~\ref{sec:1D}. As seen, our investigations for a wide
range of parameters, including misaligned \sffs structures with
$d_{F2}\gg d_{F1}$,  illustrate localized spin-1 components,
$I_{Sx}(x)$, $I_{Sy}(x)$, whereby their spin orientation follows the
local magnetization orientation. These components of
supercurrent are unable to propagate throughout the bilayer \ff
junctions where there is a relative orthogonal magnetization.
This comparison immediately reveals
the efficiency and advantages of a \sffn nanovalve
for detecting
proximity spin-1 triplet superconducting correlations
experimentally.
The pure spin-1 triplet diamagnetic response
discussed here for the \sffn spin valves might be measured using a local probe
such as
the density of states or current density.
As is well understood,
the triplet correlations can alter the induced minigap in the local density of states
in the nonsuperconducting regions.
For example, the traditional $s$-wave minigap profile can become
peaked due to the emergence of triplet correlations
which cause a resonance near the Fermi surface. Also, as seen, our
findings demonstrate that only  spin-1 triplet components of the
diamagnetic current survive in the $N$ region if $d_{F2}\gg d_{F1}$,
for sufficient magnetization misalignment.
Therefore, these results suggest that currents measured
experimentally within the $N$ wire for such systems should be comprised of
purely equal-spin
triplet correlations that are odd-frequency in character.

Therefore, the spin-parameterized technique
reveals the spatial profile of the odd and even frequency
components of the total supercurrent and provides additional 
insight into their behaviors. The general three-dimensional approach
which we outlined 
can then be employed for two-dimensional finite-size magnetic/superconducting  proximity hybrids
with arbitrary magnetization patterns.
In this paper we categorized
\fs hybrids with layered \f regions into two
classes: configurations hosting supercurrent transport 1) parallel and 2) perpendicular to
the \ff interface. The one-dimensional structures studied
in Sec.~\ref{sec:1D} belong to class 2), while the two-dimensional
configurations discussed in this section are examples of class 1).
For the fairly wide range of configurations considered
in this paper, we demonstrated that singlet
supercurrent flow ``parallel'' to
uniformly magnetized
$F/F$ interfaces can  generate substantial  long-range triplet supercurrents
{\it regardless} of the specific configuration and geometry
\cite{pugach_paral,Fominov,al1}. This is of course in stark contrast to
their counterparts where the singlet supercurrent flows ``perpendicular'' to
the \ff cross sections and fully discussed in Sec. \ref{sec:1D}.

\section{conclusion}\label{sec:conclusion}
In conclusion, we have employed
a quasiclassical method and
spin-parameterization 
technique in both the Usadel equations and
associated
boundary conditions.
This approach provides a suitable
computational and analytical
framework to pinpoint the spatial behaviors
of the decomposed
odd and even frequency {\it supercurrent} components in
layered ferromagnetic junctions. We have studied the transport characteristics of
the spin decomposed supercurrent components for two generic scenarios and several
experimentally relevant structures: systems with 1) parallel or 2) perpendicular
charge supercurrent flow relative to the magnetic interfaces. 
Two types of finite-size
two-dimensional magnetic \sffs Josephson junctions subject
to an
external magnetic field, and supporting supercurrents
flow parallel to the \ff interface, are considered.
In one type, we assume the \ff interface is
parallel to the \fs interfaces, while in the other type, the \ff
interface is perpendicular to the \fs interfaces.
Our studies revealed that when a
supercurrent flows parallel to the \ff junction,
the long-ranged spin-triplet supercurrent will be effectively generated and
it will propagate deeply into the ferromagnetic regions independent
of junction geometries. This phenomenon
is more pronounced when the thickness of ferromagnetic strips
are unequal ($d_{F2}\gg d_{F1}$ or $W_{F2}\gg W_{F1}$), depending on
the junction type under consideration. This effect disappears
when the supercurrent direction is perpendicular to the \ff
interface. To gain more insight and to have
explicit comparisons,
we also studied the various aspects of the
spin-decomposed supercurrent components in one-dimensional $S/F/F/S$,
$S/F/F/F/S$, and $S/Ho/F/Ho/S$ hybrids using the spin-parametrization scheme.
We studied the spatial maps of each supercurrent component and found
that the total supercurrent is dominated by the long-range triplet
component, which also governs the behavior of total supercurrent.
Our results demonstrated that
structures with inhomogeneous magnetization patterns generate
long-ranged supercurrents stronger than their uniform counterparts
and that the component of supercurrent corresponding
to the rotating
magnetization component  is long-ranged in such structures.
Finally, we proposed an \sffn spin-valve,
which represents an experimentally accessible platform
to probe the predicted phenomena in the parallel supercurrent transport scenario. We
have studied the spatial maps of the singlet and triplet supercurrent components of this spin-valve
when subjected to an external magnetic field. Our
findings demonstrated that for misaligned magnetizations
and different thicknesses of the $F$ layers, $d_{F2}\gg d_{F1}$, a
long-ranged odd-triplet component of the supercurrent response
arises and populates the adjacent normal metal $N$.  

\acknowledgments

We would like to thank G. Sewell for his valuable instructions in
the numerical parts of this work. We also appreciate N.O. Birge and
F.S. Bergeret 
for useful discussions and comments. K.H. is supported in part by ONR
and by a grant of supercomputer resources provided by the DOD HPCMP.


\begin{thebibliography}{000}


\bibitem{zutic} I. Zutic, J. Fabian, and S.D. Sarma,  \rmp \textbf{76}, 323 (2004).


\bibitem{golubov1} A.A. Golubov, M.Yu. Kupriyanov, E. Ilichev, Rev. Mod. Phys. {\bf 76}, 411
(2004).

\bibitem{buzdin1} A. Buzdin, Rev. Mod. Phys. {\bf 77}, 935 (2005).

\bibitem{bergeret1} F.S. Bergeret, A.F. Volkov, K.B. Efetov, Rev. Mod. Phys. {\bf 77}, 1321
(2005).

\bibitem{efetov1} K.B. Efetov, I.A. Garifullin, A.F. Volkov,K. Westerholt, \textit{Magnetic
Heterostructures,Advances and Perspectives in Spinstructures and
Spintransport.} ed. by H. Zabel, S.D. Bader, Series. Springer Tracts
in Mod. Phys. vol {\bf 227} (Springer, New York, 2007), P. 252.

\bibitem{eschrigh1} M. Eschrig, Phys. Today \textbf{64}, 43
(2011).

\bibitem{al1} M. Alidoust and K. Halterman, Appl. Phys. Lett. {\bf 105}, 202601 (2014). 

\bibitem{al2} M. Alidoust and K. Halterman, Phys. Rev. B {\bf 89}, 195111 (2014). 
 
 \bibitem{al3} K. Halterman, O. T. Valls, and M. Alidoust, Phys. Rev. Lett. {\bf 111}, 046602 1694 (2013). 
 
 
\bibitem{Bobkov} A.M. Bobkov and I.V. Bobkova, Phys. Rev. B \textbf{84}, 054533
(2011); Phys. Rev. Lett. {\bf 108}, 197002 (2012).

\bibitem{brataas1} A.G. Malshukov, and A. Brataas, \prb \textbf{86}, 094517 (2012).


\bibitem{Ioffe} L.B. Ioffe, V.B. Geshkenbein, M.V. Feigelman, A.L. Fauchere,
and G. Blatter, Nature (London) {\bf 398}, 679 (1999).

\bibitem{Barash} Y.S. Barash, I.V. Bobkova, and T. Kopp, Phys. Rev. B \textbf{66}, 140503(R)
(2002).

\bibitem{Cottet} A. Cottet, T. Kontos, W. Belzig, C. Schonenberger, C. Bruder,
Europhys. Lett. \textbf{74}, 320 (2006).

\bibitem{Crouzy} B. Crouzy, S. Tollis, and D.A. Ivanov, Phys. Rev. B \textbf{76},
134502 (2007).

\bibitem{Fominov} Ya.V. Fominov, A.F. Volkov, and K.B. Efetov, Phys. Rev. B \textbf{75}, 104509 (2007).

\bibitem{Radovic1} Z. Pajovic, M. Bozovic, Z. Radovic, J. Cayssol, and A. Buzdin, Phys.
Rev. B \textbf{74}, 184509 (2006).

\bibitem{Sellier} H. Sellier, C. Baraduc, F. Lefloch, and R. Calemczuk, Phys. Rev.
Lett. \textbf{92}, 257005 (2004).

\bibitem{Pugach2} N.G. Pugach, M.Yu. Kupriyanov, A.V. Vedyayev, C. Lacroix, E.
Goldobin, D. Koelle, R. Kleiner, and A.S. Sidorenko Phys. Rev. B
\textbf{80}, 134516 (2009).

\bibitem{Jin} L.-J. Jin, Y. Wang, L. Wen, G.-Q. Zha, and S.-P. Zhou
Physics Letters A \textbf{376}, 2435 (2012).


\bibitem{eschrigh3} M. Eschrig, J. Kopu, J.C. Cuevas, and G. Schon, Phys. Rev. Lett.
{\bf 90}, 137003 (2003).

\bibitem{Baibich} M.N. Baibich, J.M. Broto, A. Fert, F. Nguyen Van Dau, F. Petroff,
P. Eitenne, G. Creuzet, A. Friederich, and J. Chazelas, Phys. Rev.
Lett. {\bf 61}, 2472 (1988).

\bibitem{Grunberg} G.
Binasch, P. Grunberg, F. Saurenbach, and W. Zinn, Phys. Rev. B {\bf 39},
4828 (1989).

\bibitem{alidoust2} M. Alidoust, G. Sewell, and J. Linder
Phys. Rev. B \textbf{85}, 144520 (2012).

\bibitem{alidoust1} M. Alidoust, J. Linder, G. Rashedi, T. Yokoyama, and A. Sodbo, \prb
\textbf{81}, 014512 (2010).



\bibitem{Zdravkov} V.I. Zdravkov, J. Kehrle, G. Obermeier, D. Lenk, H.-A. Krug von Nidda, C. Muller,
M.Yu. Kupriyanov, A.S. Sidorenko, S. Horn, R. Tidecks, and L.R.
Tagirov, Phys. Rev. B {\bf 87}, 144507 (2013).


\bibitem{Bakurskiy} S.V. Bakurskiy, N.V. Klenov, I.I. Soloviev, M.Yu. Kupriyanov, and A.A.
Golubov, Phys. Rev. B {\bf 88}, 144519 (2013).

\bibitem{makhlin} Y. Makhlin, G. Schoen, A. Shnirman, Rev. Mod. Phys. \textbf{73}, 357 (2001).



\bibitem{ryaz0} V.V. Ryazanov, V.A. Oboznov, A.Yu. Rusanov, A.V. Veretennikov,
A.A. Golubov, and J. Aarts, Phys. Rev. Lett. \textbf{86}, 2427
(2001).

\bibitem{Bulaevskii} L.N. Bulaevskii, M.J. Graf, C.D. Batista, and V.G. Kogan, Phys. Rev. B \textbf{83}, 144526 (2011).

\bibitem{buzdin2} A.I. Buzdin, L.N. Bulaevskii, S.V. Panyukov, JETP Lett. \textbf{35}, 178
(1982).


\bibitem{kh_0pi} K. Halterman, and O.T. Valls, Phys. Rev. B 69, 014517 (2004).


\bibitem{bergeret2} F.S. Bergeret, A.F. Volkov, K.B. Efetov, Phys. Rev. Lett. \textbf{86}, 4096
(2001).

\bibitem{Asano2} Y. Asano, Y. Tanaka, T. Yokoyama, and S. Kashiwaya, Phys.
Rev. B \textbf{74}, 064507 (2006).

\bibitem{Lofwander1} M. Eschrig, T. Lofwander, Nat. Phys. \textbf{4}, 138 (2008).

\bibitem{Lofwander2} T. Lofwander, T. Champel, J. Durst, M. Eschrig, Phys. Rev. Lett. \textbf{95}, 187003 (2005).

\bibitem{Kontos} T. Kontos, M. Aprili, J. Lesueur, X. Grison, Phys. Rev. Lett. \textbf{86}, 304 (2001).

\bibitem{Sosnin} I. Sosnin, H. Cho, V.T. Petrashov, and A.F. Volkov, Phys. Rev.
Lett. \textbf{96}, 157002 (2006).

\bibitem{halterman1} K. Halterman, P.H. Barsic, and O.T. Valls
Phys. Rev. Lett. \textbf{99}, 127002 (2007).

\bibitem{Birge} T.S. Khaire, M.A. Khasawneh, W.P. Pratt Jr., N.O. Birge, Phys. Rev. Lett. \textbf{104}, 137002 (2010).

\bibitem{Keizer}  R.S. Keizer, S.T.B. Goennenwein, T.M. Klapwijk, G. Miao,
G. Xiao and A. Gupta, Nature \textbf{439}, 825 (2006).


\bibitem{Houzet3} M. Houzet and A.I. Buzdin, Phys. Rev. B {\bf 76}, 060504(R)
(2007).

\bibitem{rob2} M. Alidoust, and J. Linder, \prb \textbf{82}, 224504 (2010).

\bibitem{alidoust_missner} M. Alidoust, K. Halterman, J. Linder,  Phys. Rev. B \textbf{89}, 054508 (2014).

\bibitem{Houzet1} M. Houzet, A.I. Buzdin, \prb  \textbf{74}, 214507 (2006).

\bibitem{Trifunovic3} M. Knezevic, L. Trifunovic, Z. Radovic, Phys. Rev. B \textbf{85}, 094517
(2012).

\bibitem{Trifunovic2} L. Trifunovic, Phys. Rev. Lett. \textbf{107}, 047001 (2011).

\bibitem{pugach_paral} T.Y. Karminskaya, M.Y. Kupriyanov, and A.A. Golubov, JETP Lett. {\bf 87}, 570 (2008); A.I. Buzdin, A.S. Melnikov, and N.G. Pugach,
Phys. Rev. B {\bf 83}, 144515 (2011).


\bibitem{Hikino} S. Hikino and S. Yunoki, Phys. Rev. Lett. \textbf{110}, 237003
(2013).


\bibitem{rob1} J.W.A. Robinson, J.D.S. Witt, M.G. Blamire, Science \textbf{329}, 5987
(2010).

\bibitem{rob4} C. Wu, O.T. Valls, and K. Halterman
\prl \textbf{108}, 117005 (2012).


\bibitem{halterman2} K. Halterman and O. T. Valls
Phys. Rev. B \textbf{80}, 104502 (2009).


\bibitem{Oboznov} V.A. Oboznov, V.V. Bolginov, A.K. Feofanov, V.V. Ryazanov, and A.I.
Buzdin, Phys. Rev. Lett. {\bf 96}, 197003 (2006).

\bibitem{Ovsyannikov} G.A. Ovsyannikov, A.E. Sheyerman, A.V. Shadrin, Yu.V.
Kislinskii, K.Y. Constantinian, and A. Kalabukhov JETP Letters {\bf 97},
145 (2013)

\bibitem{Karminskaya} T.Yu. Karminskaya, A.A. Golubov, and M.Yu. Kupriyanov,
Phys. Rev. B {\bf 84}, 064531 (2011).

\bibitem{kh_sv} C. Wu, O.T. Valls, and K. Halterman, Phys. Rev. B {\bf 86}, 014523
(2012).

\bibitem{buzdin3} A.S. Melnikov, A.V. Samokhvalov, S.M. Kuznetsova, and A.I.
Buzdin, Phys. Rev. Lett. 109, 237006 (2012).

\bibitem{Beasley} S. Oh, D. Youm and M.R. Beasley, Appl. Phys. Lett. {\bf 71}, 2376
(1997).

\bibitem{Fominov2} Y.V. Fominov  \etal, JETP Lett. \textbf{91}, 308 (2010).


\bibitem{exper1} P.V. Leksin, N.N. Garifyanov, I.A. Garifullin, Ya.V. Fominov,
J. Schumann, Y. Krupskaya, V. Kataev, O.G. Schmidt, and B.
B�chner, Phys. Rev. Lett. {\bf 109}, 057005 (2012).

\bibitem{exper2} P.V. Leksin, N.N. Garifyanov, I.A. Garifullin, J. Schumann,
V. Kataev, O.G. Schmidt, and B.
B�chner, Phys. Rev. Lett. {\bf 106}, 067005 (2011).

\bibitem{exper3} P.V. Leksin, N.N. Garifyanov, I.A. Garifullin,
J. Schumann, V. Kataev, O.G. Schmidt, and B.
B�chner, Phys. Rev. B {\bf 85}, 024502 (2012).

\bibitem{exper4} B. Li, N. Roschewsky, B.A. Assaf, M. Eich, M. Epstein-Martin, D. Heiman,
M. M�nzenberg, and J. S.
Moodera, Phys. Rev. Lett. {\bf 110}, 097001 (2013).

\bibitem{exper5} P.V. Leksin, N.N. Garifyanov, I.A. Garifullin, J. Schumann,
H. Vinzelberg, V. Kataev, R. Klingeler, O.G. Schmidt and B.
B�chner, Appl. Phys. Lett. {\bf 97}, 102505 (2010).

\bibitem{exper6} A.A. Jara, C. Safranski, I.N. Krivorotov, C. Wu, A.N. Malmi-Kakkada,
O.T. Valls, and K. Halterman, \prb \textbf{89}, 184502 (2014).

\bibitem{exper7} M.G. Flokstra, T.C.Cunningham, J. Kim, N. Satchell,
G. Burnell, S.J. Bending, P.J. Curran, S.J. Langridge,
C. Kinane, J.F.K. Cooper, N. Pugach, M. Eschrig, S.L. Lee, arXiv:1404.2950.

\bibitem{nazarov} V. Braude, Yu.V. Nazarov, Phys. Rev. Lett. {\bf 98}, 077003
(2007).

\bibitem{Cuevas_frh2} F.S. Bergeret and J.C. Cuevas, J. Low Temp. Phys., {\bf 153} 304 (2008).



\bibitem{Rowell} J.M. Rowell, Phys. Rev. Lett. \textbf{11}, 200 (1963).

\bibitem{Josephson} B.D. Josephson, Phys. Lett. \textbf{1}, 251 (1962).

\bibitem{Cuevas_frh1} J.C. Cuevas and F.S. Bergeret, Phys. Rev. Lett., {\bf 99} 217002 (2007).

\bibitem{alidoust_nfrh1} M. Alidoust, G. Sewell, and J. Linder, Phys. Rev. Lett. \textbf{108}, 037001 (2012).

\bibitem{alidoust_nfrh2} M. Alidoust, and J. Linder, Phys. Rev. B \textbf{87}, 060503(R) (2013).

\bibitem{Ledermann} U. Ledermann, A.L. Fauche`re, and G. Blatter, Phys. Rev. B \textbf{59},
R9027 (1999).


\bibitem{Usadel} K. Usadel, Phys. Rev. Lett. \textbf{25}, 507 (1970); A.I. Larkin and Y.N. Ovchinnikov, in Nonequilibrium Superconductivity, edited by D.
Langenberg and A. Larkin (Elsevier, Amesterdam, 1986), P. 493.

\bibitem{morten} J.P. Morten, M.S. thesis, Norwegian University of Science and
Technology, 2003.

\bibitem{Belzig_solid_state} W. Belzig, F.K. Wilhelm, C. Bruder, G. Sch�n, A.
D.Zaikin, \textit{Superlattices and Microstructures}, Vol. {\bf 25}, Iss.
5�6, Pg. 1251 (1999).



\bibitem{cite:zaitsev} A. V. Zaitsev, Zh. Eksp. Teor. Fiz. \textbf{86}, 1742
(1984) (Sov. Phys. JETP \textbf{59}, 1015 (1984); M. Y. Kuprianov
\etal, Sov. Phys. JETP \textbf{67}, 1163 (1988).

\bibitem{Angers} L. Angers, F. Chiodi, G. Montambaux, M. Ferrier, S. Gu\`{e}ron,
H. Bouchiat, and J.C. Cuevas, Phys. Rev. B \textbf{77}, 165408 (2008).

\bibitem{Chiodi} F. Chiodi, M. Ferrier, Gu\`{e}ron, J.C. Cuevas, G. Montambaux, F.
Fortuna, A. Kasumov, H. Bouchiat, Phys. Rev. B \textbf{86}, 064510
(2012).

\bibitem{Clem1} J.R. Clem, Phys. Rev. B \textbf{81}, 144515 (2010).



\bibitem{efetov2} A.F. Volkov and K.B. Efetov, Phys. Rev. B \textbf{81}, 144522 (2010).

\bibitem{Champel1} T. Champel, M. Eschrig, Phys. Rev. B \textbf{72}, 054523 (2005).

\bibitem{Eschrig2} M. Eschrig, T. Lofwander, T. Champel, J.C. Cuevas, J. Kopu,
G. Schon, J. Low Temp. Phys. \textbf{147}, 457, (2007).

\bibitem{robinson3} G.B. Halsz, J.W.A. Robinson, J.F. Annett, and M.G. Blamire,
Phys. Rev. B {\bf 79}, 224505 (2009)

\bibitem{annett} D. Fritsch and J.F. Annett, New J. Phys. {\bf 16}, 055005
(2014).


\bibitem{Suderow} H. Suderow, I. Guillamon, J.G. Rodrigo, S. Vieira,
arXiv:1403.5514.

\bibitem{Abrikosov1} A.A. Abrikosov, Fundamentals of the Theory of Metals (North-Holland,
Amsterdam, 1988).

\bibitem{Abrikosov2} T. Yamashita, L. Rindereer, J. Low Temp. Phys. {\bf 21}, 153
(1975).

\bibitem{Abrikosov3} A.A. Abrikosov, Sov. Phys. JETP {\bf 5}, 1174 (1957).

\bibitem{Tinkham} M. Tinkham, Introduction to Superconductivity (McGraw-Hill, New
York, 1996).

\bibitem{Champel2} T. Lofwander, T. Champel, M. Eschrig, Phys. Rev. B \textbf{75}, 014512
(2007).


\bibitem{sus1} Higashitani and K. Nagai, Physica B 194, 1385 (1994).

\bibitem{sus2} W. Belzig, C. Bruder, and G. Sch\"{o}n, Phys. Rev. B {\bf 53}, 5727 (1996).

\bibitem{sus3} P. Visani et al., Phys. Rev. Lett. {\bf 65}, 1514 (1990).

\bibitem{sus4} F. Bernd, Muller-Allinger and A. Celia Mota, Phys. Rev. Lett. {\bf 84}, 3161 (2000).

\bibitem{sus5} F. Bernd Muller-Allinger and A. Celia Mota, Phys. Rev. B {\bf 62}, 6120 (2000).

\bibitem{sus6} C. Bruder and Y. Imry, Phys. Rev. Lett. {\bf 80}, 5782 (1998).

\bibitem{Yokoyama_missner} T. Yokoyama, Y. Tanaka, and N. Nagaosa, Phys. Rev. Lett. \textbf{106}, 246601 (2011).

\bibitem{al33} K. Halterman, O. T. Valls, and M. Alidoust, Phys. Rev. Lett. {\bf 111}, 046602 1694 (2013).




%
%
%
%

%
%
%
%
%
%
%
%
%
%
%
%
%
%
%
%







\end{thebibliography}
\end{document}